\documentclass[onecolumn,notitlepage,pra,showpacs,superscriptaddress,amssymb,amsmath,amsmath]{revtex4-1}
\usepackage{graphicx}
\usepackage{epstopdf}
\usepackage{bm}
\usepackage{hyperref}
\usepackage{cleveref}
\usepackage{comment}
\usepackage{color}
\usepackage{physics}
\usepackage{amsmath}
\usepackage{tikz}
\usepackage{enumitem}
\usepackage{bbold}
\usepackage{ulem}

\newcommand{\relu}[1]{\text{ReLU}\left(#1\right)}

\definecolor{orangish}{rgb}{0.976, 0.616, 0.243}
\definecolor{alex}{rgb}{0.514107, 0.281697, 0.998596}
\definecolor{CW}{rgb}{0.164385, 0.69482, 0.192477}

\begin{document}

\title{Unsupervised machine learning of topological phase transitions \\ from experimental data}

\author{Niklas K\"{a}ming}
\affiliation{ILP - Institut f\"{u}r Laserphysik, Universit\"{a}t Hamburg, Luruper Chaussee 149, 22761 Hamburg, Germany}

\author{Anna Dawid}
\affiliation{Faculty of Physics,  University of Warsaw, Pasteura 5, 02-093 Warsaw, Poland}
\affiliation{ICFO - Institut de Ci\`encies Fot\`oniques, The Barcelona Institute of Science and Technology, Av. Carl Friedrich Gauss 3, 08860 Castelldefels (Barcelona), Spain}

\author{Korbinian Kottmann}
\affiliation{ICFO - Institut de Ci\`encies Fot\`oniques, The Barcelona Institute of Science and Technology, Av. Carl Friedrich Gauss 3, 08860 Castelldefels (Barcelona), Spain}

\author{Maciej Lewenstein}
\affiliation{ICFO - Institut de Ci\`encies Fot\`oniques, The Barcelona Institute of Science and Technology, Av. Carl Friedrich Gauss 3, 08860 Castelldefels (Barcelona), Spain}
\affiliation{ICREA, Pg.~Llu\'is Campanys 23, 08010 Barcelona, Spain}

\author{Klaus Sengstock}
\affiliation{ILP - Institut f\"{u}r Laserphysik, Universit\"{a}t Hamburg, Luruper Chaussee 149, 22761 Hamburg, Germany}
\affiliation{The Hamburg Centre for Ultrafast Imaging, Luruper Chaussee 149, 22761 Hamburg, Germany}
\affiliation{ZOQ – Zentrum f\"{u}r Optische Quantentechnologien, Universit\"{a}t Hamburg, Luruper Chaussee 149, 22761 Hamburg, Germany}

\author{Alexandre Dauphin}
\affiliation{ICFO - Institut de Ci\`encies Fot\`oniques, The Barcelona Institute of Science and Technology, Av. Carl Friedrich Gauss 3, 08860 Castelldefels (Barcelona), Spain}

\author{Christof Weitenberg}
\affiliation{ILP - Institut f\"{u}r Laserphysik, Universit\"{a}t Hamburg, Luruper Chaussee 149, 22761 Hamburg, Germany}
\affiliation{The Hamburg Centre for Ultrafast Imaging, Luruper Chaussee 149, 22761 Hamburg, Germany}

\date{\today}

\begin{abstract}
Identifying phase transitions is one of the key challenges in quantum many-body physics. Recently, machine learning methods have been shown to be an alternative way of localising phase boundaries also from noisy and imperfect data and without the knowledge of the order parameter. Here we apply different unsupervised machine learning techniques including anomaly detection and influence functions to experimental data from ultracold atoms. In this way we obtain the topological phase diagram of the Haldane model in a completely unbiased fashion. We show that the methods can successfully be applied to experimental data at finite temperature and to data of Floquet systems, when postprocessing the data to a single micromotion phase. Our work provides a benchmark for unsupervised detection of new exotic phases in complex many-body systems. 

\end{abstract}

\maketitle
\section{Introduction}

Machine learning techniques have recently achieved remarkable successes in analysing large data sets in various areas. These developments lead also to promising applications in quantum physics \cite{Carleo19RevMod, Carrasquilla20AdvPhys}. Examples include the efficient representation of quantum many-body states \cite{Carleo17Science}, efficient state tomography from restricted experimental data \cite{Torlai18NatPhys, Torlai19PRL, Neugebauer20PRA}, optimisation of experimental preparation \cite{Wigley16SciRep, Tranter18NatComm, Bukov18PRX, Davletov20PRA} and identification of phases of matter \cite{Carrasquilla17NatPhys, Chng17PRX, Broecker17SciRep, Nieuwenburg17NatPhys, Wang16PRB, Wang17PRB, Ohtsuki16JPSJ, kottmann2020unsupervised,Huembeli18PRB,Huembeli19}. 
For the latter, machine learning methods were applied to data both from numerical simulations and from experiments such as scanning tunneling microscopy images of condensed matter systems \cite{Zhang19Nature, Ziatdinov16Nanotechnology}, neutron scattering data from spin ice systems \cite{Samarakoon20NatComm} as well as real and momentum-space images of ultracold atom systems \cite{Rem19NatPhys, Bohrdt19NatPhys, Khatami20PRA}. When analysing experimental data, machine learning can unfold its full potential by identifying the relevant information despite noise and other imperfections such as finite temperature or restricted access to relevant observables. Another prospect of the machine learning analysis is to identify novel phases and order parameters in exotic regimes \cite{Khatami20PRA, Miles20arXiv}. While supervised machine learning methods, i.e. with labelled training data, have been broadly applied, unsupervised methods dealing with unlabelled data were so far mainly restricted to numerical studies \cite{Nieuwenburg17NatPhys, Broecker17arxiv, Wetzel17PRE, Chng18PRE, Greplova20NJP, kottmann2020unsupervised, Lidiak20PRL, Arnold20arXiv}. 

Machine learning techniques can be employed for various tasks related to phase transitions including the comparison of experimental data to competing theory descriptions in an unbiased way \cite{Bohrdt19NatPhys}, the analysis of patterns in the trained filters of convolutional neural networks \cite{Khatami20PRA,Miles20arXiv}, the generation of new images \cite{Casert20arXiv}, or the extraction of physical parameters and concepts \cite{Lu20PRX, Iten20PRL}. Finally, the question of interpretability of neural networks has obtained a new stimulus by the application to physical problems \cite{Wang16PRB, Ponte17PRB, Greitemann19PRB, Dawid20NJP, Zhang20PRR, Wetzel20PRR, Miles20arXiv}.

Quantum simulators based on ultracold atoms in optical lattice allow engineering a variety of quantum many-body systems and probing them with detection methods complementary to solid-state systems \cite{Lewenstein12Oxford}. In particular, topological systems can be created by adding artificial gauge fields \cite{Dalibard11RMP, Cooper19RMP} using periodic driving, i.e. so-called Floquet engineering \cite{Bukov15AdvPhys, Eckardt17RMP}. Topological phases of matter are an active field of study, but the absence of a local order parameter generically poses a challenge to their detection \cite{Cooper19}. Therefore, the classification of topological phases receives particular attention with machine learning techniques \cite{Zhang17PRL, Deng17PRB, Zhang18PRL, Carvalho18PRB, Beach2018PRB, Rodriguez-Nieva19NatPhys, Greplova20NJP, Holanda20PRB, Long20PRL, Scheurer20PRL}. With cold atoms, many detection methods have been demonstrated including the transverse Hall drift \cite{price12,dauphin13,Jotzu14Nature, Aidelsburger15NatPhys}, Berry phase measurements \cite{Duca15Science}, quantized circular dichroism \cite{Tran17SciAdv, Asteria19NatPhys} as well as Bloch state tomography \cite{Alba11PRL, Hauke14PRL, Flaschner16Science, Flaschner18NatPhys, Tarnowski19NatComm}. The latter is based on momentum-space images after quench dynamics, which also form the basis of the machine learning analysis in this article.

Here, we apply unsupervised machine learning techniques to experimental data from topological phases of a Haldane-like \cite{Haldane88PRL} model realised in ultracold atom quantum simulators. We also address the problem of dealing with the micromotion inherently arising in Floquet systems by using machine learning for data postprocessing, which allows to effectively change the micromotion phase of all data to a desired value. While supervised machine learning can deal with data of different micromotion phases \cite{Rem19NatPhys}, unsupervised techniques we applied within this work cannot.

Unsupervised learning of phase transitions can roughly be divided in two categories: clustering-based methods \cite{Wang16PRB, Wetzel17PRE, Wang17PRB, Hu17, Chng18PRE, Ming19npj, Rosson20PRA, Rodriguez-Nieva19NatPhys, Lidiak20PRL} and learning-success-based methods \cite{Nieuwenburg17NatPhys, Broecker17arxiv, Huembeli18PRB, Greplova20NJP, kottmann2020unsupervised}. In this work we apply methods of both categories to the data, which we postprocess to a single micromotion phase. Clustering-based methods identify the phases by clustering the data in a suitably chosen space and associating each cluster with a different phase, employing concepts such as PCA, t-SNE, autoencoder or diffusion maps. In this category, we find that a k-means cluster analysis in the latent space of an autoencoder does identify the phase transitions, when it is applied to cuts through the phase diagram separately. This approach, however, cannot distinguish between the different signs of the Chern number. Learning success-based methods use the success of the training process for different trial classifications to judge on the similarity of the data. In this context, we use anomaly detection \cite{kottmann2020unsupervised} and influence functions \cite{Dawid20NJP}. By carefully combining the information from these techniques we can uncover the full phase diagram from noisy experimental data in a completely unsupervised way. Our results provide an important benchmark for unsupervised machine learning of phases of matter and evaluate methods that might be useful for revealing new exotic order in complex systems.

The structure of the article is as follows. We start with the description of the used methods (Sec. \ref{sec:methods}). In section \ref{sec:experimental_setup}, we describe the experimental setup and the protocols to obtain the data. Section \ref{ssec:ML_methods} gives an overview over the different machine learning methods we use within this work. In the result section \ref{sec:latent-space-with-micromotion}, we first employ a latent space analysis to detect the different topological phases. Afterwards, we describe how to postprocess the data to a desired micromotion phase in section \ref{sec:data-postprocessing} and check the validity of this approach with influence functions in section \ref{ssec:confirming_removal_IF}. We use the postprocessed data to detect the different topological phases with the method of latent space analysis (Sec. \ref{sec:kmeansclustering}) again. Section \ref{sec:anomaly_detection} describes an anomaly detection scheme to separate different topological phases. We finally use the influence functions in section \ref{ssec:similarity-analysis-IF} to distinguish between the two topologically non-trivial phases.

\section{Methods}
\label{sec:methods}
\subsection{Experimental setup and data acquisition}
\label{sec:experimental_setup}
\begin{figure}
\begin{center}
\includegraphics[width=0.9\columnwidth]{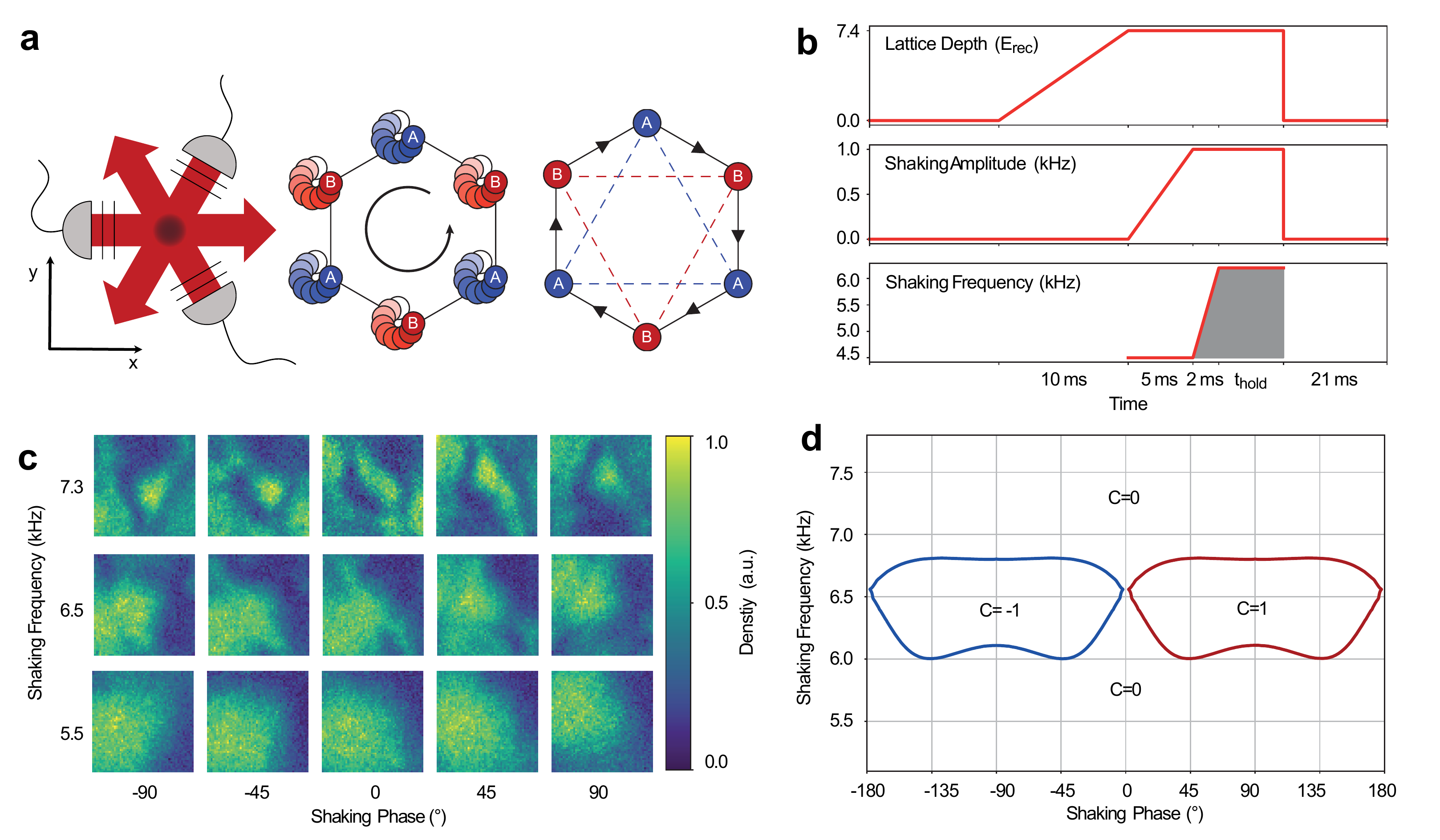}
\end{center}
\vspace{-0.7cm}
\caption{Experimental setup. (\textbf{a}) Three laserbeams forming a hexagonal optical lattice by interfering under $120^\circ$ (left). The polarisation and thus the geometry can be tuned via the two waveplates indicated by the black lines. The hexagonal lattice consists of A and B sites with an energy offset, $\Delta_{AB}$. We periodically drive the lattice on an elliptical trajectory (center) in order to obtain an effective Floquet Hamiltonian with Peierls phases on the nearest-neighbor tunneling elements (right) giving rise to topological bands. (\textbf{b}) The atoms are adiabatically prepared in the lowest Floquet band by ramping the lattice depth, the shaking amplitude, and the shaking frequency to different final values. Different hold times allow sampling different micromotion phases given by the grey area under the curve. The atoms are imaged after a time-of-flight expansion of 21 ms. (\textbf{c}) Typical momentum-space images for different shaking frequencies and shaking phases. The images are centered around zero momentum and have a width of one reciprocal lattice vector. (\textbf{d}) The topological phase diagram for the lowest band as predicted by a numerical Floquet calculation featuring the two lobes of non-trivial Chern number, $C=\pm 1$, characteristic of the Haldane model.}
\label{fig:exp_setup}
\vspace{-0.55cm}
\end{figure}

The data is taken in experiments performed with ultracold atoms in optical lattices \cite{Lewenstein12Oxford}, which are established as a well-controllable system for studying solid-state physics in general and topological phases in particular \cite{Dalibard11RMP, Cooper19RMP}. The topological Haldane model \cite{Haldane88PRL} is realised by Floquet-driving of a honeycomb lattice \cite{Oka09PRB, Rechtsman13Nature, Jotzu14Nature, Flaschner16Science}. In this specific configuration, the experiments start with a hexagonal lattice with a large offset $\Delta_{\mathrm{AB}}=2\pi\cdot 6.1$\,kHz between the two sublattices, realised by a suitable polarisation of the three interfering laser beams that form the optical lattice \cite{Flaschner16Science} (Fig.\,\ref{fig:exp_setup}a). The lattice is then accelerated on elliptical trajectories by a phase modulation of the lattice beams characterised by the shaking phase $\varphi$ between the modulation along the $x$ and $y$ direction. The resulting effective Floquet Hamiltonian features non-trivial Chern numbers and gives rise to a topological phase diagram closely related to the original Haldane model (Fig.\,\ref{fig:exp_setup}d) \cite{Tarnowski19NatComm, Rem19NatPhys}. The control parameters are the shaking phase $\varphi$, which gives rise to the time-reversal-symmetry-breaking, and the shaking frequency $f_{\mathrm{sh}}$, which gives rise to non-trivial Chern numbers $C=\pm1$ for near-resonant shaking with the sublattice offset $f_{\mathrm{sh}}\approx \Delta_{\mathrm{AB}}/2\pi$.

The numerical prediction for the phase boundary (Fig.\,\ref{fig:exp_setup}d) results from a Floquet calculation for a tight-binding model of the hexagonal lattice based on the shaking parameters and the calibrated parameters of the static lattice. It has been shown to agree well with previous measurements of topological properties in the system \cite{Flaschner16Science, Tarnowski19NatComm, Asteria19NatPhys, Rem19NatPhys}, except for a slight shift of the topological region towards higher frequencies for the experimental data. This shift might be due to the uncertainty in the calibration of the static lattice or to contributions of higher bands, which where neglected in the two-band tight-binding model. Note that the calibration uncertainty of the polarisation of the lattice beams of $0.2^\circ$ leads to an uncertainty of the expected phase transition points of around 200 Hz \cite{Rem19NatPhys}.

The experiments are performed with ultracold spin-polarised fermionic atoms of $^{40}$K with mass $m=40 \mathrm{u}$ prepared in the lowest band of the optical lattice formed by laser beams with a wavelength of $\lambda=1064\,$nm as in the earlier work \cite{Flaschner16Science, Rem19NatPhys}. The characteristic energy scale is the recoil energy $E_{\mathrm{rec}}=h^2/(2 m \lambda^2)$. In the transverse direction the cloud is weakly harmonically confined. In order to adiabatically prepare the lowest band of the Floquet system, we gradually ramp up the Floquet drive in two steps (Fig.\,\ref{fig:exp_setup}b): 1) we ramp up the shaking amplitude to 1\,kHz within 5\,ms at the far off-resonant shaking frequency of $f_{\mathrm{sh}}^{\mathrm{ini}}=4.5\,$kHz, 2) we ramp the shaking frequency to the final values $f_{\mathrm{sh}}^{\mathrm{fin}}$ within $t_{\mathrm{ramp}}=2\,$ms at the fixed shaking amplitude. Due to the Floquet heating, this procedure leads to a population of the lowest band of typically 50-75\%. Previous work on supervised machine learning has shown that the Chern number of the lowest band can be faithfully obtained despite the non-zero temperature \cite{Rem19NatPhys}. 

For the detection of the state, all potentials are switched off leading to a free expansion of the system known as time-of-flight imaging. The expansion maps the original momentum distribution on the real-space density, which is then imaged by absorption imaging. This procedure can be related to the Bloch-state tomography \cite{Hauke14PRL,Flaschner16Science, Tarnowski19NatComm}, which is based on the quench dynamics after projection onto the static lattice with large $\Delta_{\mathrm{AB}}$, realizing the special case of zero hold time in the static lattice. This connection motivates the usefulness of the experimental images for detecting topology, although the proper Bloch state tomography explicitly relies on the full quench dynamics for disentangling the parameters \cite{Hauke14PRL}. 

In the experimental protocol, we hold the atoms in the Floquet system for different hold times $t_{\mathrm{hold}}$ at the final shaking frequency in steps smaller than the Floquet period, in order to sample different instances of the Floquet micromotion $\phi$. The micromotion phase is then given by $\phi=f_{\mathrm{sh}}^{\mathrm{fin}}(t_{\mathrm{ramp}}/2+t_{\mathrm{hold}})+f_{\mathrm{sh}}^{\mathrm{ini}}t_{\mathrm{ramp}}$. This convention traces the micromotion back to the start of the driving with a kick in a fixed direction and allow relating the micromotion phases of data with different shaking frequencies. The micromotion is an intrinsic property of Floquet systems and while it can give rise to new physics \cite{Kitagawa_2010,Rudner13PRX}, it is often a nuisance when studying the effective Floquet Hamiltonian \cite{Dauphin_2017,Kumar_2020}. 

For the analysis, we restrict the images to a square region of 56x56 pixels centered around zero momentum, $k=0$, where 56 pixel correspond to the length of a reciprocal lattice vector (Fig.\,\ref{fig:exp_setup}c). The images are furthermore individually normalized to the interval [0,1]. In total we use 10,436 images with varying shaking phase, shaking frequency and micromotion phase with just a few images per parameter. While supervised learning often requires an additional large training data set at parameters, which allow a labelling, the unsupervised methods discussed below can identify the phase transitions with the data homogeneously sampled across the parameter space alone.

\subsection{Machine learning methods}\label{ssec:ML_methods}

In the various machine learning applications of this article  we use deep neural networks (NNs) composed of combinations of fully-connected and convolutional layers \cite{Goodfellow16}. After each layer, the output is processed by a non-linear activation function, which in this work is mainly the so-called rectified unit function $\relu{x} = \left(0 \text{ if } x \leq 0; x \text{ if } x \ge 0\right)$ or its variations (e.g. leaky ReLU \cite{Maas2013}).
The archetypical task of a NN is supervised learning, where the network output $y^\text{out}_i$ is trained to approximate a desired label $y_i$ for every input $x_i$ of the dataset $\mathcal{D}=\{x,y\}$. In order to achieve this task, we define a loss function $l_i=l(y^\text{out}_i,y_i)$ that captures the success of this endeavor. Training then comes down to minimising the total loss function $L=1/N\sum_{i=1}^{N} l_i$ with respect to the trainable parameters $\{\omega\}$ of the deep NN. Commonly used loss functions are mean square error and binary cross-entropy. This high-dimensional optimisation problem can be tackled with gradient descent, where the parameters $\{\omega\}$ are iteratively shifted in the direction of the negative gradient, i.e. $\omega \rightarrow \omega - \alpha \nabla_\omega L$, where $\alpha$ is the so-called learning rate and a hyperparameter. We here mainly use more involved gradient-based optimisation strategies like ADAM \cite{Kingma2015} in order to speed-up the training process.

In this work, we employ a special NN architecture called autoencoder (AE) \cite{LeCun1987PhD,Bourlard1988,Hinton1993}. An AE is composed of two subsequent deep NNs, called encoder and decoder. The neurons $z$ at the output of the encoder are called bottleneck neurons and the dimension of this space is typically chosen to be smaller than the input space. The task of an AE is to find an efficient compression of the input data through the encoder at the bottleneck, from which the decoder is able to reproduce the original input $x_i$ at the output stage $y^\text{out}_i$. The loss function $l(x_i,y^\text{out}_i)$ is therefore defined between the input data and the output of the AE and there is no necessity for labelled data.
In this work, we found it sufficient to use the mean squared error total loss function, $L_\text{MSE} (x_i, y_i^\text{out}) = \frac{1}{N}\sum_i |x_i - y_i^\text{out}|^2.$
AEs are typically used in the context of unsupervised learning for tasks such as dimensionality reduction or anomaly detection. We will later use the success of this compression by looking at the loss $L$ to differentiate between phases of a phase diagram in \cref{sec:anomaly_detection}. Finally, an autoencoder can also be trained in a supervised way given a series of inputs associated to a corresponding output. Such supervised method has applications in image denoising or colorization \cite{Vincent2008ICML, Xie2012NIPS, Baldassarre2017arXiv}.

To improve stability for data transformation, we also employ so-called variational autoencoders (VAEs) \cite{kingma2014autoencoding, rezende2014stochastic}. The key difference with respect to the previous architecture is the introduction of an engineered regularisation at the bottleneck. Instead of encoding the input $x$ into a feature $z(x)$ in the latent space, it is encoded into a probability distribution $p(z|x)$. A feature $z$ is then sampled from $p(z|x)$ to be passed to the decoder. This introduces two main advantages. First, the feature space is regularised such that neuron activations at the bottleneck are more interpretable \cite{doersch2016tutorial}. Second, this way we can generate new data after training by sampling at the bottleneck. However, we use the variational autoencoder here to gain more stability in the transformation of experimental data.
Additionally, one can introduce a question neuron, that is an extra input neuron feeding directly into the bottleneck. The information provided there can be for example a physical parameter corresponding to the image we provide. We will later use VAEs to postprocess the data to a fixed micromotion phase.

The influence function \cite{Cook77, Koh17} is an interpretability method that can be understood as a numerically-feasible approximation of leave-one-out (LOO) training. LOO training consists of retraining the model after removing a single training point and checking how it changed the test loss connected to the prediction on the chosen test point. If the prediction got worse (better), i.e. the test loss got larger (smaller), the removed training point was a helpful (harmful) one. However, such an analysis with LOO trainings is prohibitively expensive because, for the full picture, it requires the number of training procedures equal to the training dataset size multiplied by the number of chosen test points. Instead, the most complicated step in the numerical approximation, i.e. influence functions, consists of a single computation of the Hessian's inverse of the training loss with respect to the model parameters.

An influence function estimates the influence that a single training point has on a prediction made on a single test point. In this paper, we apply it to the convolutional neural network (CNN) trained in a supervised way. Analysis of the most influential points indicates which data features are dominant in the model's predictions. We use this property in section \ref{ssec:confirming_removal_IF}. Moreover, we interpret the training points which are similarly influential to a particular prediction as similar. In this way, the influence functions' values, $\mathcal{I}$, provide a notion of similarity found by a trained network in a given problem. Thanks to this property, they allow finding distinctive similarity regions within the same class \cite{Dawid20NJP}, which indicates improper labelling and proves useful in section \ref{ssec:similarity-analysis-IF}. Analysis of what training points the model regards as similar and which data features are dominant in the model's predictions increases the model's interpretability. Influence function values, $\mathcal{I}(x_{\text{train}}, x_{\text{test}})$, can only be compared for fixed test point and various training points, and within the same model. Therefore, we need to fix a test point whenever we calculate a set of $\mathcal{I}$ for the similarity analysis.

All machine learning techniques were implemented using NumPy, PyTorch and Tensorflow \cite{NumPy,pytorch, tensorflow15}. The specifics of the architectures with reproducible code for all performed tasks can be found in our notebooks \cite{notebooks}.

\section{Results}
\label{sec:results}
\subsection{Latent-space interpretation of autoencoders}
\label{sec:latent-space-with-micromotion}

\begin{figure}[b]
\begin{center}
\includegraphics[width=\columnwidth]{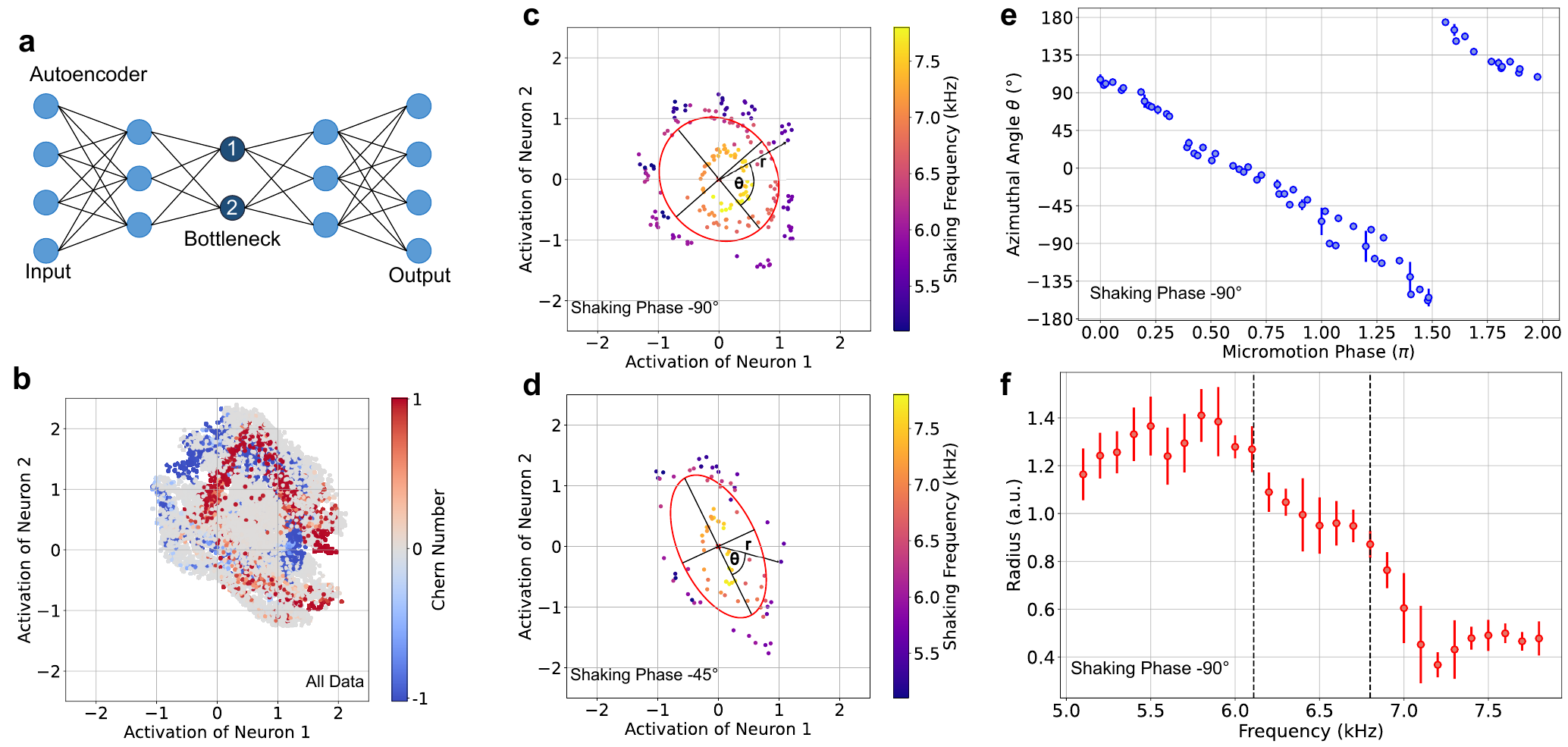}
\end{center}
\vspace{-0.7cm}
\caption{Bottleneck analysis. (\textbf{a}) Sketch of the autoencoder with two bottleneck neurons, which is trained to reproduce the entire dataset. (\textbf{b}) Activations of the two bottleneck neurons form the latent space, in which each point corresponds to one image. In the scatter plot of all data, the Chern number is color coded according to the numerics, illustrating that the data does not cluster according to the Chern number in latent space. (\textbf{c}-\textbf{d}) The analysis of single cuts through the phase diagram, i.e. of data with fixed shaking phase of (\textbf{c}) $\varphi=-90^\circ$ and (\textbf{d}) $\varphi=-45^\circ$. The data forms rings in latent space and is fitted by an ellipse (red line), which forms a new coordinate system with the azimuthal angle $\theta$ measured relative to the longer half axis and the radius $r$ as sketched with the arrow. (\textbf{e}-\textbf{f}) Analysis of latent space in elliptical coordinates. (\textbf{e}) The azimuthal angle $\theta$ is linearly related to the micromotion phase for $\varphi=90^\circ$ of the individual images independent of their shaking frequency. (\textbf{f}) The mean radial coordinate for a given shaking frequency traces out a monotonously decreasing curve with no clear signature of the phase transition and indicate three plateaus in accordance with the phase boundaries. The error bar is the standard deviation from averaging over the images with a given shaking frequency. The plots for other shaking frequencies look similar. This association with the micromotion means that latent space can be interpreted, but also that micromotion is the dominant feature hiding possible signatures of the topological phase transitions.}
\label{fig:latent_space}
\vspace{-0.55cm}
\end{figure}

As a first step, we produce and analyse a low-dimensional representation of the data in the latent space of an autoencoder formed by the activations of the bottleneck neurons. Autoencoders are important tools for unsupervised learning \cite{Hinton06Science}. The autoencoder consists of several convolutional layers and a fully connected bottleneck formed by two neurons (Fig.~\ref{fig:latent_space}a). We choose a 2D latent space to create an easy-to-understand visual representation of the given samples. The complete implementation details can be found in our notebooks \cite{notebooks}. We checked that choosing more dimensions in latent space does not lead to an improvement. The autoencoder is trained on the complete dataset.

The two dimensional latent-space representation of all images yields a dense cloud of data points without an apparent clustering (Fig.~\ref{fig:latent_space}b). The picture becomes clearer, when we restrict the data to fixed shaking phases, i.e. vertical cuts through the phase diagram (Fig.~\ref{fig:latent_space}c-d). The data then lies on elliptical structures with the radius related to the shaking frequency. For further analysis, we fit an ellipse using direct least square fitting \cite{FitzgibbonIEEE} and perform a coordinate transformation to extract the elliptical coordinates radius $r$ and azimuthal angle $\theta$ measured from the major axis of the fitted ellipse. 

The azimuthal angle can be clearly connected to the micromotion phase showing a linear dependence (Fig.~\ref{fig:latent_space}e) for a shaking phase of $\varphi=90^\circ$. The same dependence can also be seen with the azimuthal coordinate of the center of mass of the raw images, which provides a direct connection between the time of flight images and latent space. See appendix \ref{app:com_micromotion} for further details. We furthermore explore possible information hidden in the radial coordinate (Fig.~\ref{fig:latent_space}f). The mean radius decreases with shaking frequency with some signs of plateaus, but without a sufficiently clear separation with shaking frequency for making a prediction of phase transitions. The latent space representation can thus be physically interpreted via the micromotion, but it cannot provide an identification of the topological phases. We attribute this to the dominance of the micromotion, which we try to eliminate in the following section \ref{sec:data-postprocessing}.

\subsection{Data postprocessing to desired micromotion phase}
\label{sec:data-postprocessing}
In Floquet systems, the micromotion poses an additional challenge for identifying phase transitions. We find that the unsupervised machine learning only works if all images have the same micromotion phase, i.e. the center of mass displacement is in the same direction. Because the data was taken with a sampling of different micromotion phases, we need to post process it to the desired sampling of a fixed micromotion phase. We show that this can be accomplished by machine learning techniques based on VAEs, which are a powerful tool for data transformation and generation \cite{kingma2014autoencoding,rezende2014stochastic}. Our architecture uses an additional question neuron in the bottleneck, which has previously been proven successful in identifying relevant physical properties \cite{Iten20PRL}. Here we use the additional question neuron for supervised training of the VAE.
The given challenge is related to tasks such as fringe removal in absorption imaging \cite{Oeckeloen10PRA} or removal of timing jitter in pump-probe experiments \cite{Fung16Nature}, but we believe that our method based on VAE is very broadly applicable to postprocessing to desired sampling in different experimental scenarios.

\begin{figure}[b]
\begin{center}
\includegraphics[width=0.9\columnwidth]{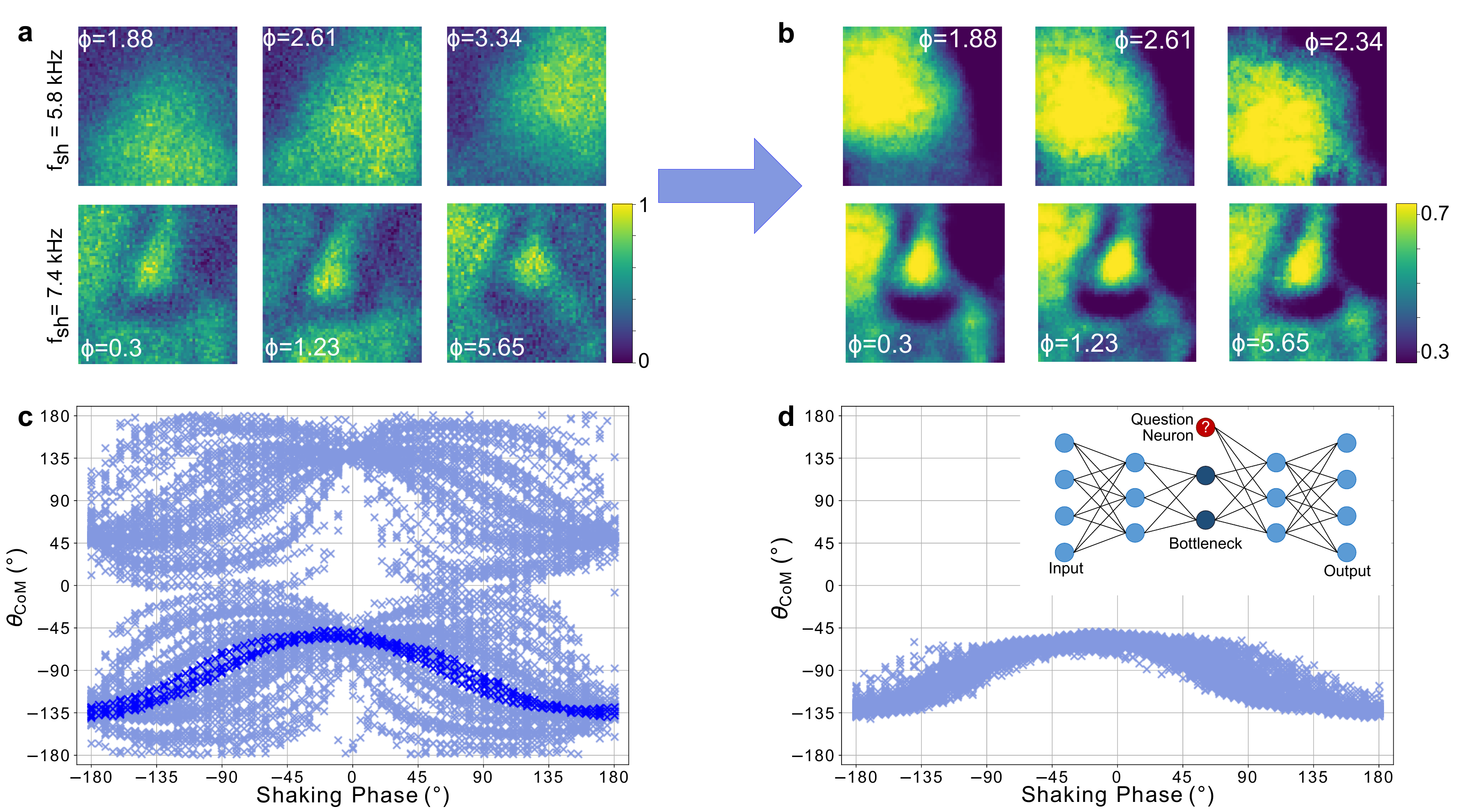}
\end{center}
\vspace{-0.7cm}
\caption{Postprocessing data to desired micromotion phase using a variational autoencoder with an additional question neuron. (\textbf{a}) Example images at shaking frequencies of 5.8\,kHz and 7.4\,kHz and shaking phase $\varphi=90^\circ$ for different micromotion phases $\phi$ illustrating the motion of the center of mass with the micromotion phase. (\textbf{b}) The same example images as in \textbf{a} rephased by the variational autoencoder to a micromotion phase of $\phi_0=0.0$ show a center of mass position only weakly dependent on the original micromotion phase $\phi$. (\textbf{c}) The azimuthal coordinate of the center of mass of the original images as a function of the shaking phase shows a complicated dependence. (\textbf{d}) The azimuthal coordinate of the center of mass ($\theta_{\mathrm{CoM}}$) of all images forms a narrow band and is only dependent on the shaking phase, but not the original micromotion phase. This dependence on the shaking phase is similar to the dependence of the data for actual micromotion phase $\phi=0.0$ before rephasing (dark blue points in \textbf{c}). The comparison of the distribution of azimuthal coordinates of the center of mass in \textbf{c} and \textbf{d} illustrates that the rephasing of the micromotion was successful. The micromotion phase of the processed data $\phi_0=0.0$ is controlled by an additional question neuron in the bottleneck of the VAE (inset in \textbf{d}).}
\label{fig:micromotion_removal}
\vspace{-0.55cm}
\end{figure}

The encoder of the VAE consists of several convolutional stages and several layers of fully connected neurons. The last layer of the encoder has 26 fully connected neurons thus the latent space covers 13 probability distributions. The decoder has again several fully connected layers followed by a few transposed convolutional stages. The first fully connected layer is also attached to the input of the question neuron. In total the autoencoder has over 3 million trainable parameters. The complete implementation details can be found in our notebooks \cite{notebooks}. To optimise the hyperparameters of our autoencoder, we use the hyperparameter optimisation library optuna \cite{optuna19} and train over 60,000 different network architectures. To identify the best working network we use the structural similarity index \cite{wang14ssim} as a measurement for performance.
For each point in the phase diagram with a fixed shaking frequency and shaking phase we took several images with different micromotion phases by varying the hold time. As a new dataset, we select all combinations and permutations of images for a fixed shaking frequency and shaking phase and calculated their micromotion phase difference $\Delta\phi=\phi_\mathrm{output}-\phi_\mathrm{input}$. Thus the dataset includes 63,050 image pairs with a given $\Delta\phi$. Thus in contrast to the other autoencoders we employed in context of this paper, the input and output are different for the VAE. We randomly choose 10\% of the dataset for validation purposes and hide them from the network during training. We train with one image as input, which we refer to as input image and one image with the same shaking frequency and shaking phase but a different micromotion phase as output and the micromotion phase difference as input for the question neuron.
Samples of original images are given in figure \ref{fig:micromotion_removal}a.  
After training, we use the variational autoencoder to transform all original images in the dataset to a micromotion phase of $\phi_0=0.0$ by choosing their micromotion phase with an opposite sign as input for the question neuron. The postprocessed images with a single micromotion phase are similar to the original data except for some noise removal and a squeezing of the distribution of pixel values to a range of 0.3 to 0.7, which we attribute to the non-linear activations in the network (Fig.~\ref{fig:micromotion_removal}b). 

Because the micromotion is directly related to the center of mass of the images, one can see the success of the postprocessing in the example images in Fig.~\ref{fig:micromotion_removal}b: after micromotion is set to a fixed value, the images with different micromotion phases look very similar and have the same direction of the displacement of the center of mass. This is further confirmed by a comparison of the distributions of the azimuthal center of mass coordinates for all images before and after rephasing (Fig.~\ref{fig:micromotion_removal}c,d). The highlighted blue data in (Fig.~\ref{fig:micromotion_removal}c) are the center of masses for the micromotion phase $\phi=0.0$. The narrow distribution in (Fig.~\ref{fig:micromotion_removal}d) shows that the rephasing was successful.
For the identification of the phase transitions with unsupervised machine learning discussed below, it is important that we use the postprocessed data with a single micromotion phase.

\subsection{Confirming the postprocessing to a desired micromotion phase with influence functions}
\label{ssec:confirming_removal_IF}

\begin{figure}[b]
\begin{center}
\includegraphics[width=\columnwidth]{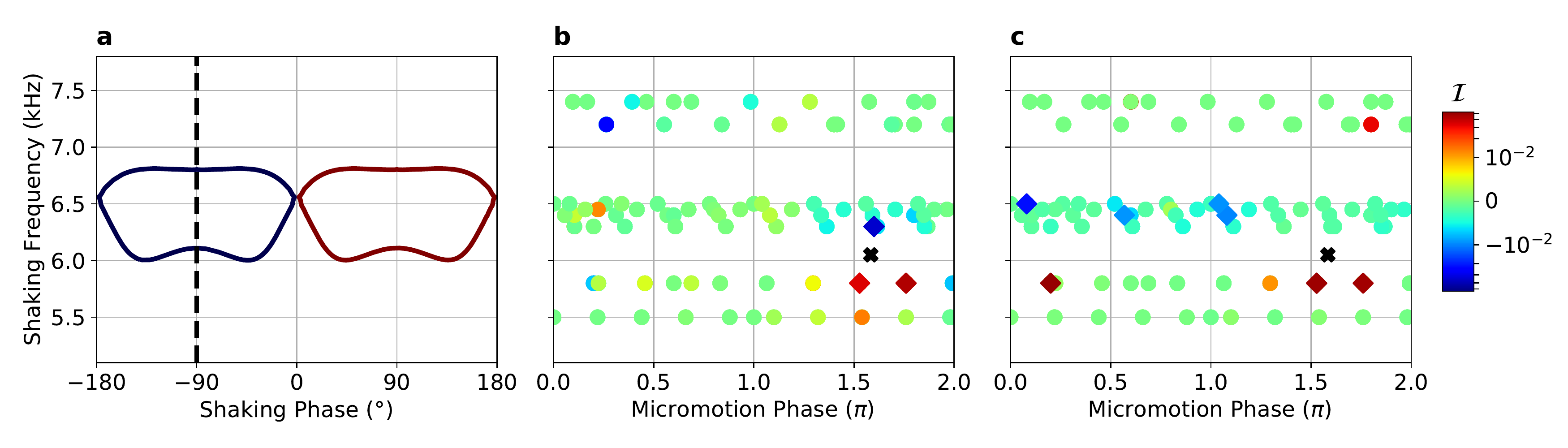}
\end{center}
\caption{Confirming the postprocessing to a desired micromotion phase with influence functions. Panel \textbf{a} presents the phase diagram with theoretical boundaries between phases and with the cut at a shaking phase of $-90^{\circ}$ marked with black line, which is analysed on the right. Panels \textbf{b}-\textbf{c} show the colour-coded influence function values for the all training points marked with dots and for the same test point (marked with black cross) for a single cut at phase $-90^{\circ}$ across the shaking frequency before and after the postprocessing. Five most helpful (harmful) training points are marked with red (blue) diamonds and they can overlap. (\textbf{b}) The most influential (both helpful and harmful) training points are the points with the shaking frequency and the micromotion phase closest to the test point. (\textbf{c}) After fixing the micromotion phase, the most influential points are distributed across various original micromotion phase values indicating the successful removal of this property which now is ignored by the CNN.}
\label{fig:IF_micromotion}
\end{figure}

To get further evidence that the postprocessing described in section \ref{sec:data-postprocessing} was successful, we confirm it with influence functions introduced in section \ref{ssec:ML_methods}. Influence functions provide an interpretation of the machine learning model by indicating which training points are influential for a chosen prediction. Analysis of the most influential examples can reveal the characteristics which impacts the machine learning predictions.

Firstly, we train a convolutional neural network (CNN) in a supervised way to classify original images with micromotion phase. Instead of analyzing the whole 2D diagram, we consider only a single cut at the fixed shaking phase $-90^{\circ}$ which simplifies the visualization of the results without changing them. Within this single cut, the Chern number of the system changes from 0 to -1 and back to 0 with increasing shaking frequency. The labelled training data contains then only two phases (C=0 and C=-1). To avoid influence from experimental imperfections, we exclude data close to the theoretically-predicted phase transitions. With the trained CNN, we calculate the influence functions determining how influential the whole training dataset is for the prediction chosen to be in the transition region. The results are presented in figure \ref{fig:IF_micromotion}b,c with the black cross indicating the test point and dots representing the training data and their color-coded influence function values. Colors vary from red for helpful training points, through green for least influential (ignored), to blue for harmful.

We see in panel b that the most influential (both helpful and harmful) data for the chosen prediction are those with both the most similar shaking frequency and micromotion phase. Learning the shaking frequency is expected as it is the parameter governing the phase transition. However, the CNN also regards the micromotion phase as influential when making a prediction, while we know that this property is physically irrelevant for the transition. Micromotion phase is an intrinsic property of the Floquet realization of the topological Hamiltonian, but does not change the topology of the effective Floquet Hamiltonian.

We do the analogous analysis for postprocessed training data, i.e. with the removed micromotion phase. Panel c in figure \ref{fig:IF_micromotion} shows that the most influential points are now randomly distributed along the original micromotion phase axis. It tells us that the CNN no longer sees this parameter as influential and confirms that the data was successfully postprocessed to a constant micromotion phase.

Let us note that when training on data with or without the micromotion, in both cases the validation and test accuracy of the trained CNN are similar. It means, interestingly, that in this set-up the predictive power of the network is not impacted by learning the quantity which is physically irrelevant.

\subsection{Clustering in latent space}
\label{sec:kmeansclustering}

\begin{figure}[b]
\begin{center}
\includegraphics[width=\columnwidth]{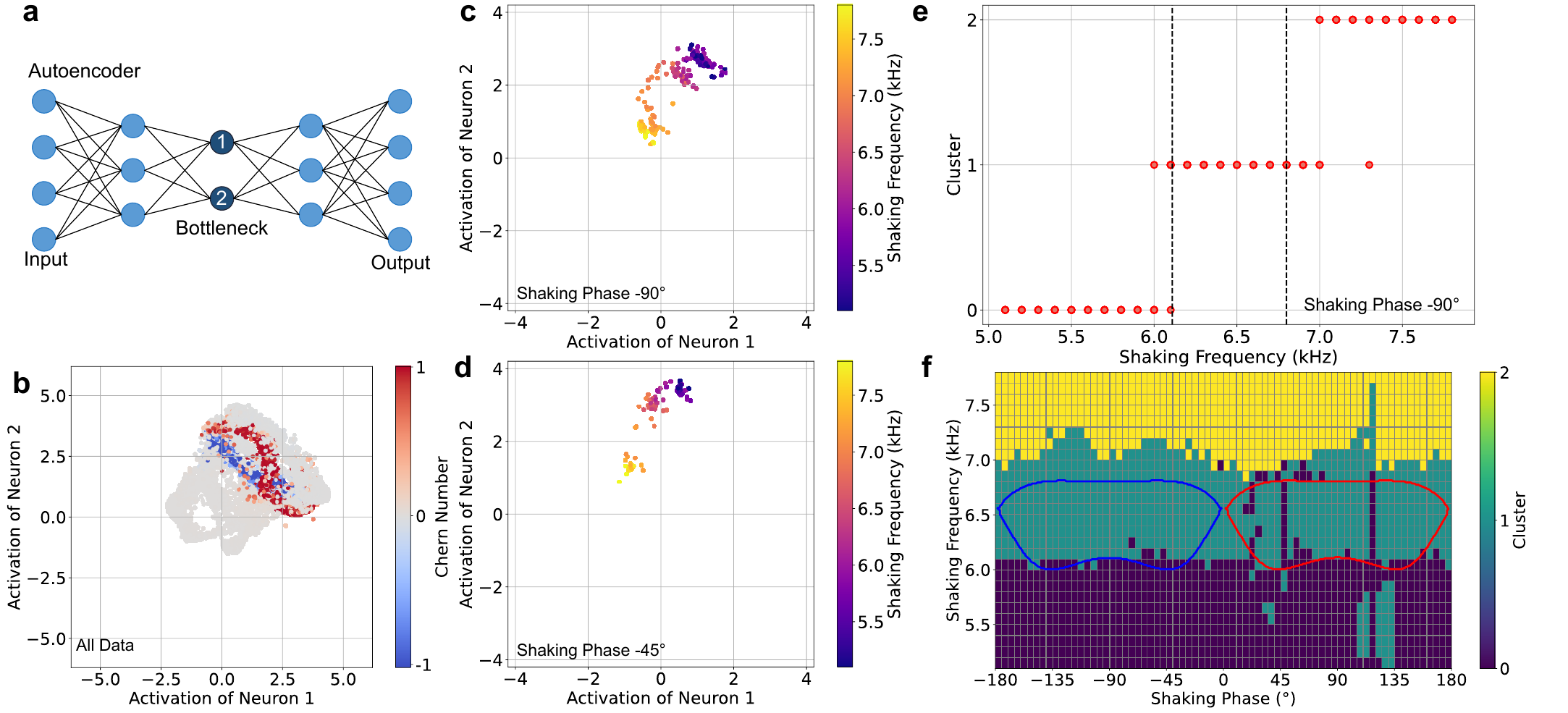}
\end{center}
\vspace{-0.7cm}
\caption{Bottleneck analysis of rephased data. The bottleneck analysis with a simple autoencoder (sketched in \textbf{a}) for the data after rephasing to sampling of a fixed micromotion phase. (\textbf{c})-(\textbf{d}) The analysis of single cuts through the phase diagram, i.e. of data with fixed shaking phase of $\varphi=-90^\circ$ in \textbf{b} and $\varphi=-45^\circ$ in \textbf{c}, shows a significant clustering according to the shaking frequency (color coded). (\textbf{e}) The cluster analysis of latent space. A k-means clustering analysis for $k=3$ clusters shows that the data is indeed clustered according to shaking frequency. There are several values for each shaking frequency for the different original micromotion phases, which mostly lie on top of each other in the plot. (\textbf{f}) Naming the three clusters increasing with frequency and combining the data from all cuts yields a topological phase diagram, which is reasonable agreement with the numerical prediction (blue and red lines), but does not distinguish between $C=1$ and $C=-1$.}
\label{fig:latent_space_rephased}
\vspace{-0.55cm}
\end{figure}

In the following sections we apply different unsupervised machine learning methods to the postprocessed data with a constant micromotion phase and compare their performance. We start with a method of the clustering-category, which identify clusters in some low-dimensional representation of the data as different phases. Specifically, we employ the same convolutional autoencoder as in section \ref{sec:latent-space-with-micromotion}, but now applied to the postprocessed data. 

The latent space representation of the complete dataset is shown in figure \ref{fig:latent_space_rephased}b. The data is color encoded by the theoretical predictions for the Chern number. It appears that the different topological classes tend to form ring structures in the 2D latent space. As in  section \ref{sec:latent-space-with-micromotion}, we now restrict ourselves to single shaking phase cuts.
Figure \ref{fig:latent_space_rephased}c and d show the latent space for two such cuts and we observe three main clusters related to three frequency regimes.

Choosing k-means clustering, a standard method to solve cluster problems, it is possible to automate the clustering process of the different latent space representations of the observed image data. We use the k-means functionality of Scikit-learn \cite{scikit-learn} and set the number of clusters to 3 and the number of max iteration to 500. All other parameters are set to standard according to the documentation. We tried different random seeds to prove stability. The results of the cluster analysis are shown in figure \ref{fig:latent_space_rephased}e and f.

This allows one to predict phase boundaries in good agreement with the theoretical predictions given by the dashed lines. The slight shift to higher frequencies is in accordance with all other methods and is explained in section \ref{sec:experimental_setup}. Evaluating the data for all shaking phases allows reconstructing the complete 2D Haldane phase diagram shown in figure \ref{fig:latent_space_rephased}f.

The procedure of separately analysing vertical cuts through the phase diagram can fundamentally not distinguish between the C=1 and C=-1 phases at positive and negative shaking phases. Furthermore, a similar analysis of horizontal cuts along constant shaking frequencies through the phase diagram does not produce clustering. Therefore further methods are required to fully identify the topological phases.

\subsection{Anomaly detection scheme}
\label{sec:anomaly_detection}
\begin{figure}[b]
\begin{center}
\includegraphics[width=\columnwidth]{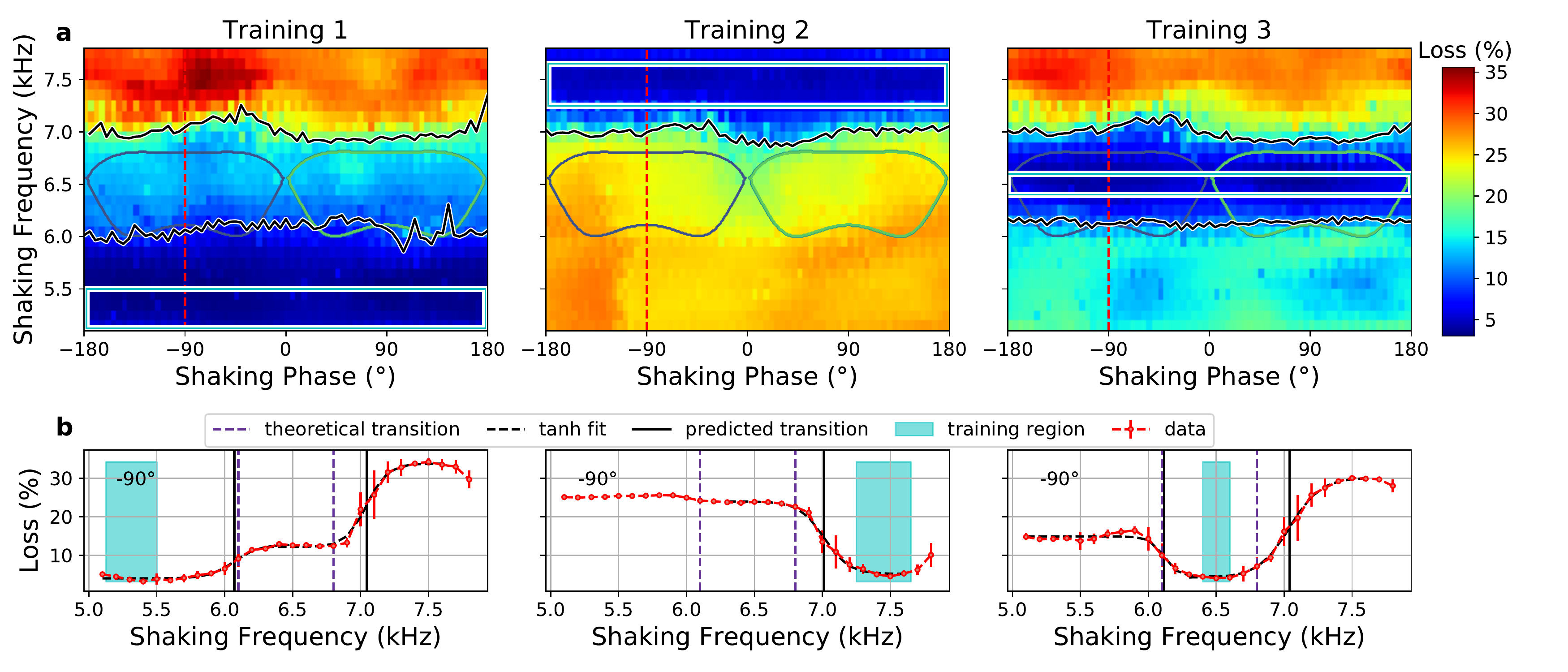}
\end{center}
\caption{Anomaly detection scheme: We start by training in the topologically trivial phase (Training 1). Due to the symmetry in the shaking phase we use all angles from $[-180^\circ,180^\circ]$ and small frequencies up to $5.5$ kHz as indicated by the light blue box in \textbf{a}. Panel \textbf{b} shows a single cut at shaking phase $-90^\circ$ where we can see two plateaus which we identify with the topological non-trivial and trivial phases, respectively. Training 2: We continue by training in the region of highest loss in the first iteration for high frequencies. From these two training iterations we can already narrow down the two boundaries in the phase diagram. Training 3 completes the overall picture and confirms the phases mapped in Training 1 and 2.}
\label{fig:KK_master}
\end{figure}

\begin{figure}
\begin{center}
\includegraphics[width=.55\columnwidth]{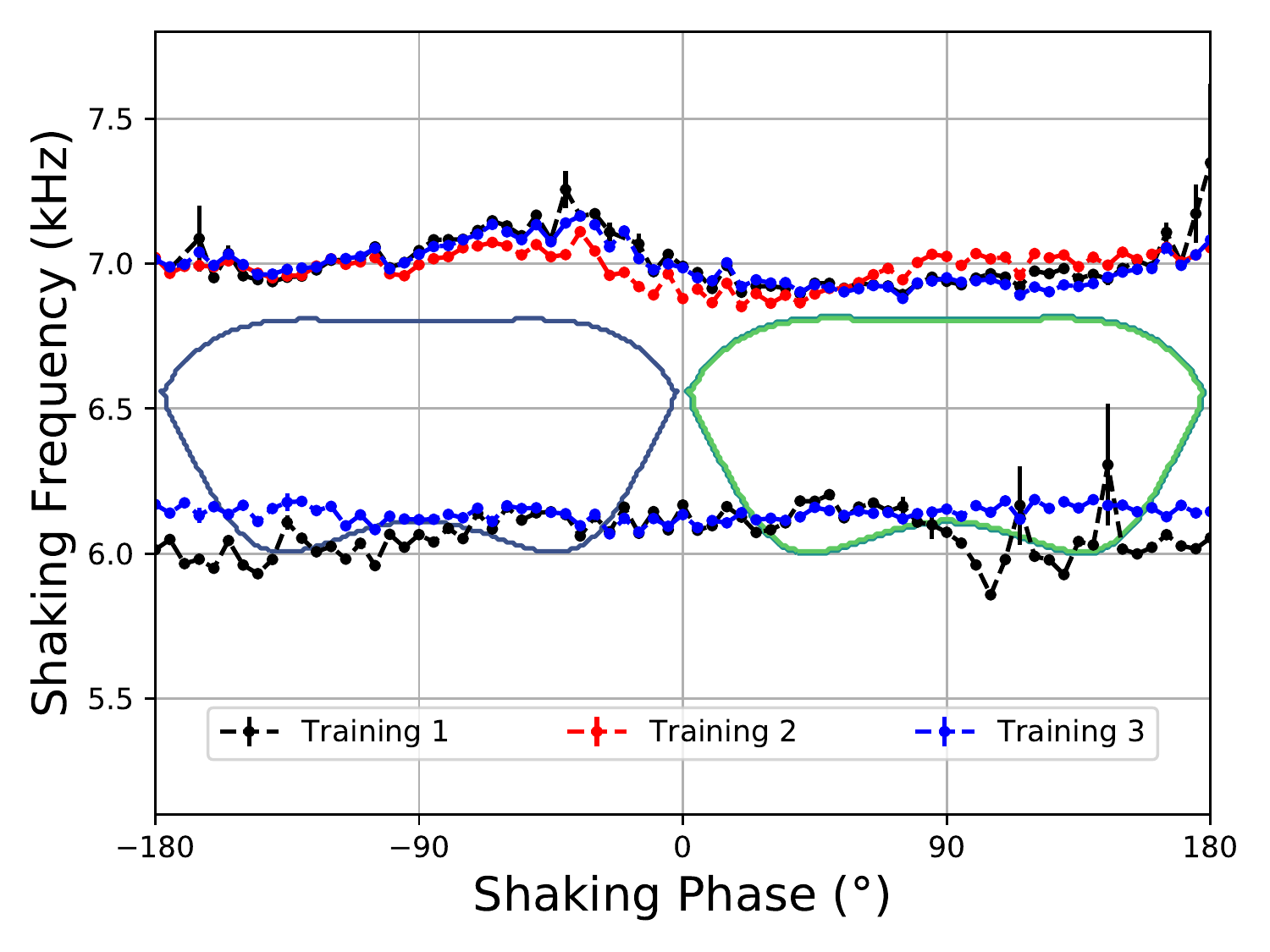}
\end{center}
\caption{Phase boundaries drawn from four training iterations of anomaly detection in figure \ref{fig:KK_master}, separating three different phases, which we identify with the topologically trivial and non-trivial phases. The obtained boundaries from the different training iterations are consistent with each other}
\label{fig:KK_boundaries}
\end{figure}

We followed the approach in \cite{kottmann2020unsupervised} and performed an unsupervised learning scheme called anomaly detection to map out the phase diagram in a few training iterations. We used a convolutional autoencoder, similar to the network described in \cref{sec:data-postprocessing}. The network consisted of an encoder and decoder made of two convolutional layers each, with a fully-connected bottleneck of $50$ units. For full details about the model and process, we refer the interested reader to \cite{notebooks}, where all steps described in this section can be exactly reproduced.
The idea is the following: We started by defining a region of the phase diagram in which we trained the autoencoder to encode and decode the images with low mean-squared error $L_\text{MSE} (A^\text{in},A^\text{out})$ between input and output images $A^\text{in},A^\text{out} \in \mathbb{R}^{56\times56}$. The network learns the characteristic features of the phase that the images were taken from and fails to reproduce images from the other phases. By looking at the loss for all images after training in only part of the phase diagram, we distinguished between the phase it has been trained on and the remaining phases via different plateaus of the loss function. Furthermore, by fitting a sum of tanh functions to the loss curve as a function of shaking frequency, we obtained a phase boundary. We then repeated this process by training in the region of high loss from the previous training round until we found all boundaries.

We show this process in figure \ref{fig:KK_master} where we started by training in the low-frequency regime (Training 1). We used all values of the shaking phase from $[-180^\circ,180^\circ]$ and small shaking frequencies up to $5.5$ kHz as indicated by the cyan rectangle. We identified three different plateaus in the loss, between which we obtained two boundaries. As seen in panel 1b), where we took a single cut of the phase diagram at $-90^\circ$, the different phases make up plateaus. So we fitted a $\tanh$ function to estimate the boundaries as indicated by the grey lines.

We continued the process and trained in the high-frequency regime, where the first iterations yielded the highest loss. As seen in Training 2 (Fig. \ref{fig:KK_master}), we find the reverse picture with a clear boundary slightly above the theoretically predicted transition. This boundary from Training 2 nicely coincides with the second boundary from Training 1.
To complete the picture, we also trained in the intermediate-frequency regime that yielded higher loss in the previous training iterations. Here the training region is significantly smaller, yet the results still match nicely with the previous training iterations. Note that generally, the results are robust concerning the size of the training region in frequencies. We show this in figure \ref{fig:boundary_consistency} in \cref{appendix:anomaly-detection}. The images provide sufficient information to separate the different phases and map out the phase diagram with this method. We present the predicted diagram in figure \ref{fig:KK_boundaries}. We notice that with this method, it is not possible to differentiate the non-trivial topological phases with Chern number $1$ and $-1$ because the trained compression does not generalize well in the shaking phase parameter. We provide further details in \cref{appendix:anomaly-detection}. In the following section, we overcome this shortcoming and complete the phase diagram, i.e. separating the two topological regions, employing the influence functions. We see that the transition between intermediate and high frequencies for all three training rounds is slightly above the theoretically predicted transition, which is due to a mismatch between theory and experiment as discussed in \cref{sec:experimental_setup}.

\subsection{Analysis of data similarity within three phases with influence functions}\label{ssec:similarity-analysis-IF}

After obtaining the phase boundaries from the anomaly detection scheme, as described in the previous section, we analyse how similar data are within the three distinguished phases. Such an analysis not only can confirm the predictions of unsupervised machine learning schemes but also reveal the existence of additional phase transitions. To this end, we train a CNN on the postprocessed experimental data, i.e. with the single micromotion phase, with labels assigned by the anomaly detection scheme. Therefore, we have three labels corresponding to the low-frequency, topological, and high-frequency phases. We employ influence functions, described in section \ref{ssec:ML_methods}, to analyse which training data are influential for a given prediction. Similarly influential training data, $\lbrace x_{\text{train}} \rbrace$, with a similar influence functions' values, $\mathcal{I}(x_{\text{train}}, x_{\text{test}})$, for a particular test point, $x_{\text{test}}$, can be then interpreted as similar from the machine learning model's point of view in a given problem.

\begin{figure}[b]
\begin{center}
\includegraphics[width=\columnwidth]{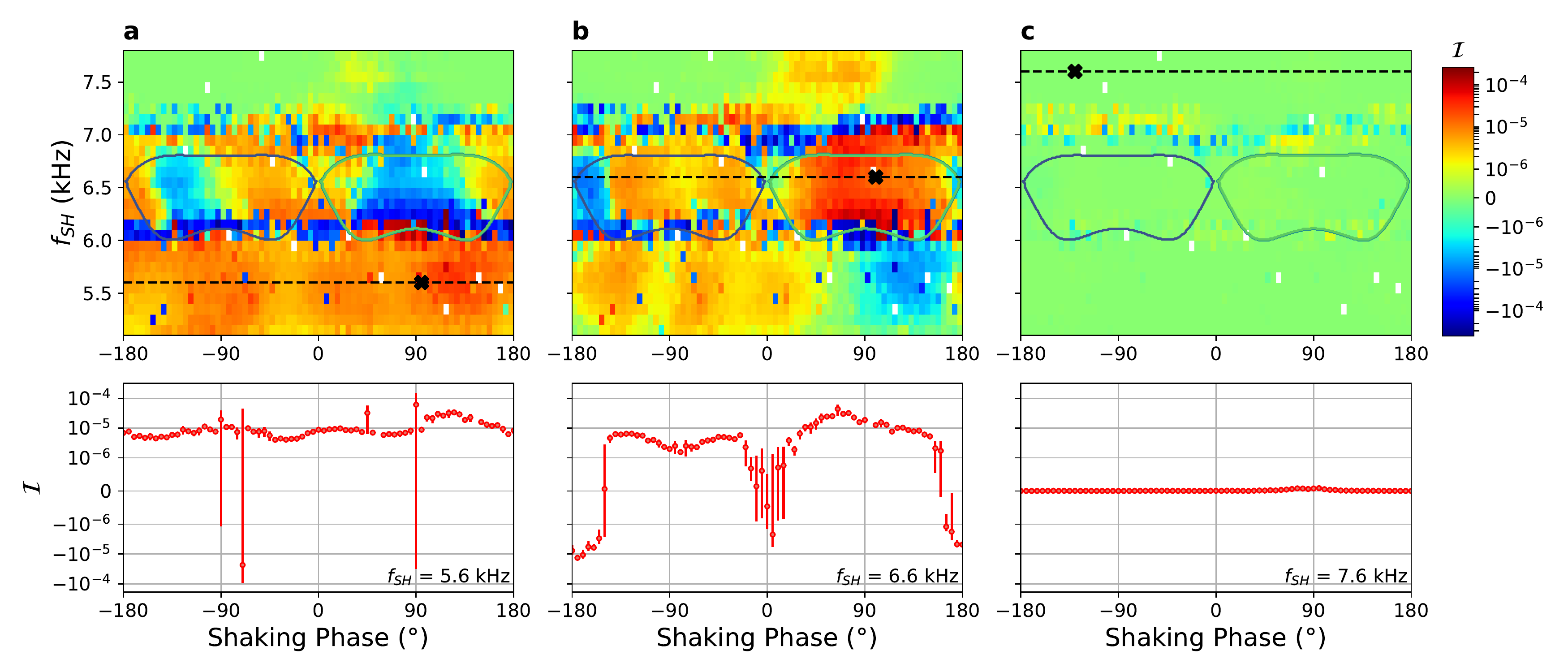}
\end{center}
\caption{Analysis of data similarity within three anomaly-detected phases with influence functions. The upper row shows the phase diagrams with the color-coded influence function values for all training data and the prediction for the test point marked with a black cross. Test point belongs to (\textbf{a}) low-frequency, (\textbf{b}) intermediate-frequency, and (\textbf{c}) high-frequency phase. The lower row presents the influence function values from the single cut through a corresponding phase diagram above at the fixed shaking frequency, $f_{\mathrm{sh}}$. Panels \textbf{a} and \textbf{c} show quite uniform similarity patterns, while \textbf{b} suggests the existence of additional phase. Note the use of a symmetric logarithmic scale. We mark theoretically-predicted boundaries with continuous blue and green lines. White pixels correspond to the lack of available training data.}
\label{fig:IF_3rd_phase}
\end{figure}

To analyse similarity of training data, we need to compare $\mathcal{I}(x_{\text{train}}, x_{\text{test}})$ and therefore fix the test point for which $\mathcal{I}(x_{\text{train}}, x_{\text{test}})$ is calculated. In figure \ref{fig:IF_3rd_phase}, we plot three sets of $\mathcal{I}(x_{\text{train}}, x_{\text{test}})$ calculated for all training data and three different test points: one is located in the low-frequency regime (panel a), the second in the intermediate-frequency regime (panel b), and final in the high-frequency regime (panel c). Each element of the phase diagrams in the upper row indicates color-coded $\mathcal{I}$ value for a corresponding test point marked with black cross. If an element corresponds to more than one training point (if more measurements were performed for a given frequency and shaking phase), then we plot the mean of $\mathcal{I}(x_{\text{train}}, x_{\text{test}})$. Red (blue) color indicates the most helpful (harmful) training points for a given prediction. White dots correspond to the lack of available training data. The lower row of figure \ref{fig:IF_3rd_phase} contains the mean of $\mathcal{I}(x_{\text{train}}, x_{\text{test}})$ for a single cut across the phase diagram for a fixed shaking frequency, $f_{\mathrm{sh}}$. The error bars represent the standard deviation, being non-zero for all shaking phases for which multiple measurements were taken.

In panel a, we see that the low-frequency training data are all quite similarly influential for the model while predicting that the black-cross test point belongs to the same low-frequency phase. The uniformity in question is also well visible in the cut through the phase diagram in the lower panel of figure \ref{fig:IF_3rd_phase}. 
Apart from single $\mathcal{I}(x_{\text{train}}, x_{\text{test}})$ with large variation indicating experimental outliers in the training set, the formed similarity pattern is quite uniform. Notice, however, the symmetric logarithmic scale for $\mathcal{I}$. When ignoring the outliers, $\mathcal{I}$ values span almost one order of magnitude. The lowest $\mathcal{I}$ values of around $5 \cdot 10^{-6}$ are located in the negative shaking phase, and the largest $\mathcal{I}$ values around $3 \cdot 10^{-5}$ are for training points which have positive shaking phase similar to $x_{\text{test}}$. It tells us that the shaking phase is an influential factor in the predictions in the low-frequency phase. However, it is not a determining one. Otherwise, the largest $\mathcal{I}$ values would be much more localized in the shaking phase axis. We also note that the $\mathcal{I}$ values always highlight the boundaries between phases for two reasons. Firstly, data around the phase transitions are usually the most confusing for the model. They are labelled as belonging to either of the phases, being at the same time non-representative of any phase. The second reason is of purely numerical nature. Regardless if boundaries are placed in accordance to physical ones, the data around the boundaries plays a unique role in the training, containing the most important information for the model. In general, we expect that the confusing phase transition regions, indicated by large $\mathcal{I}$ values, in experimental data should be broader as compared to numerical studies \cite{Dawid20NJP}. It is due to the fact that the experimental system is finite and inhomogeneous and therefore the phase transition is intrinsically broadened.

Panel c shows even more uniform behaviour. It contains $\mathcal{I}$ values for the test point localised in the high-frequency regime. What may seem surprising is that almost all $\mathcal{I}$ values are practically zero. It means that none of the training points is of significant influence when making the chosen prediction. The reason is that the prediction on the test point from panel c has an extremely high certainty and it has an impact on the $\mathcal{I}$ values. In fact, $|\mathcal{I}(x_{\text{train}},x_{\text{test}})|$ values are proportional to the uncertainty of the prediction made on $x_{\text{test}}$. When the prediction's uncertainty is very low, the $\mathcal{I}$ values are also very small.

Panel b is analogous to the previous panels, but this time the test point, for which $\mathcal{I}$ values are computed, is localized in the intermediate-frequency regime (which we know contains two topological phases). The striking feature of panel b is the lack of uniformity in the intermediate-frequency regime which is well visible in the lower plot of panel b, which contains $\mathcal{I}$ values for the single cut through the phase diagram for the fixed frequency of 6.6 kHz. In between two plateaus, i.e. around shaking phase of 0 and $180^\circ$, there are significant dips in the influence functions' values reaching negative values. They show that the training data in this part of the diagram is different enough to be harmful for the analysed prediction. They are misleading for the model as they are labelled the same (as belonging to the topological phase) while they are actually quite different. It is analogous to the reason why influence functions always highlight boundaries between phases. This leads to the conclusion that within the anomaly-detected intermediate-frequency regime there are additional boundaries, therefore more phases. Another observation supporting this conclusion are two similarity plateaus on the negative and positive sides of the shaking phase, separated by the detected boundaries. They are well visible in the lower plot of panel b. The average values of two plateaus differ by almost order of magnitude, indicating two distinct patterns. Simultaneously, these patterns are more similar to each other than to the low- or high-frequency phase, which suggests the similar character of two phases detected in the intermediate-frequency regime.

The similarity analysis described above reveals the existence of two phases within the anomaly-detected topological phase. We note that this analysis is vastly simplified by removing the micromotion phase from the time-of-flight images. Results from section \ref{ssec:confirming_removal_IF} show that the micromotion phase was a very influential factor for the trained CNN before the post-processing. Therefore, the analysis would need to include the impact of the micromotion phase on the CNN's predictions.

\section{Conclusion}
In this article, we have applied different unsupervised machine learning methods for identifying topological phase transitions in experimental data of Haldane-like model realised with ultracold atoms. The topological phase diagram of the elliptically shaken hexagonal lattice hosts topologically non-trivial phases at an intermediate shaking frequency and trivial phases both for low and high shaking frequencies. Furthermore, the sign of the Chern number changes with the sign of the shaking phase, i.e. the orientation of shaking, giving rise to two distinct non-trivial phases.

A necessary step for the successful unsupervised learning was fixing the micromotion phase inherent to the Floquet realization of the topological phases via a variational autoencoder with a question neuron. This postprocessing of the experimental data to the desired sampling demonstrated here is an exciting tool on its own.

 Both a clustering analysis in an appropriate low-dimensional representation of the data and anomaly detection in the loss function correctly identified the three regions as a function of shaking frequency. The correct identification of the two regions with the opposite sign of the Chern number was only possible by combining this information with the insights from an influence function on the supervised training on the incomplete phase diagram. In total, the full phase diagram, which can also be identified via supervised machine learning on labelled data, could be obtained in a fully unsupervised way by combining the different methods.
 
 The successful identification of the phase diagram demonstrates that unsupervised machine learning can correctly identify phases even for noisy data and despite the finite temperature of the system. In the future, these methods can be applied to exotic quantum many-body systems with unknown phase diagrams or hidden order \cite{Khatami20PRA, Miles20arXiv} to support the interpretation of the data and to guide the experimental exploration of the parameter space.

\begin{acknowledgments}

We thank Gorka Mu\~noz-Gil, Gabriel Fern\'andez-Fern\'andez, Patrick Huembeli and Ludwig Mathey for insightful discussions and Benno Rem, Matthias Tarnowski and Luca Asteria for providing the experimental data. The team in Hamburg thanks the PHYSnet for providing computational resources. 

The work in Hamburg was supported by the Deutsche Forschungsgemeinschaft (DFG, German Research Foundation) via Research Unit FOR 2414 under project number 277974659 and via the Cluster of Excellence ’CUI: Advanced Imaging of Matter’ - EXC 2056 - under project number 390715994.

The work at ICFO was supported by the Spanish Ministry of Economy and Competitiveness (Plan National FISICATEAMO and FIDEUA PID2019-106901GB-I00/10.13039 / 501100011033, “Severo Ochoa” program for Centres of Excellence in R\&D (CEX2019-000910-S), FPI, FIS2020-TRANQI), European Social Fund, Fundació Privada Cellex, Fundació Mir-Puig, Generalitat de Catalunya (AGAUR Grant No. 2017 SGR 1341, CERCA program, QuantumCAT U16-011424, co-funded by ERDF Operational Program of Catalonia 2014-2020), ERC AdG NOQIA, MINECO-EU QUANTERA MAQS (funded by State Research Agency (AEI) PCI2019-111828-2 / 10.13039/501100011033), and the National Science Centre, Poland-Symfonia Grant No. 2016/20/W/ST4/00314. 

An.D. acknowledges the financial support from the National Science Centre, Poland, within the Preludium grant No. 2019/33/N/ST2/03123 and the Etiuda grant No. 2020/36/T/ST2/00588 as well as the Foundation for Polish Science within the First Team programme co-financed by the EU Regional Development Fund. This project has received funding from the European Unions Horizon 2020 research and innovation programme under the Marie Skłodowska-Curie grant agreement No. 713729 (K.K.). Al.D. acknowledges the financial support from a fellowship granted by la Caixa Foundation (ID 100010434, fellowship code LCF/BQ/PR20/11770012).

\end{acknowledgments}

\subsection*{Author Contributions}
N.K., An.D., and K.K. performed the numerical calculation and evalulation of the different machine learning architectures and analysed the data. N.K. performed the bottleneck analysis of the AE and the data postprocessing via the VAE. An.D. performed the confirmation of the data postprocessing and the identification of the different topological phases via the influence function. K.K. performed the unsupervised identifiction of phases via anomaly detection. N.K., An.D., and K.K. contributed equally to this work. Al.D. and M.L. supervised the work at ICFO (An.D. and K.K.), C.W. and K.S. supervised the work in Hamburg (N. K.). All authors contributed to the interpretation of the results and to the writing of the manuscript.

------------------------------------------------------------------------------
\normalem

\begin{thebibliography}{103}%
\makeatletter
\providecommand \@ifxundefined [1]{%
 \@ifx{#1\undefined}
}%
\providecommand \@ifnum [1]{%
 \ifnum #1\expandafter \@firstoftwo
 \else \expandafter \@secondoftwo
 \fi
}%
\providecommand \@ifx [1]{%
 \ifx #1\expandafter \@firstoftwo
 \else \expandafter \@secondoftwo
 \fi
}%
\providecommand \natexlab [1]{#1}%
\providecommand \enquote  [1]{``#1''}%
\providecommand \bibnamefont  [1]{#1}%
\providecommand \bibfnamefont [1]{#1}%
\providecommand \citenamefont [1]{#1}%
\providecommand \href@noop [0]{\@secondoftwo}%
\providecommand \href [0]{\begingroup \@sanitize@url \@href}%
\providecommand \@href[1]{\@@startlink{#1}\@@href}%
\providecommand \@@href[1]{\endgroup#1\@@endlink}%
\providecommand \@sanitize@url [0]{\catcode `\\12\catcode `\$12\catcode
  `\&12\catcode `\#12\catcode `\^12\catcode `\_12\catcode `\%12\relax}%
\providecommand \@@startlink[1]{}%
\providecommand \@@endlink[0]{}%
\providecommand \url  [0]{\begingroup\@sanitize@url \@url }%
\providecommand \@url [1]{\endgroup\@href {#1}{\urlprefix }}%
\providecommand \urlprefix  [0]{URL }%
\providecommand \Eprint [0]{\href }%
\providecommand \doibase [0]{http://dx.doi.org/}%
\providecommand \selectlanguage [0]{\@gobble}%
\providecommand \bibinfo  [0]{\@secondoftwo}%
\providecommand \bibfield  [0]{\@secondoftwo}%
\providecommand \translation [1]{[#1]}%
\providecommand \BibitemOpen [0]{}%
\providecommand \bibitemStop [0]{}%
\providecommand \bibitemNoStop [0]{.\EOS\space}%
\providecommand \EOS [0]{\spacefactor3000\relax}%
\providecommand \BibitemShut  [1]{\csname bibitem#1\endcsname}%
\let\auto@bib@innerbib\@empty
\bibitem [{\citenamefont {Carleo}\ \emph {et~al.}(2019)\citenamefont {Carleo},
  \citenamefont {Cirac}, \citenamefont {Cranmer}, \citenamefont {Daudet},
  \citenamefont {Schuld}, \citenamefont {Tishby}, \citenamefont
  {Vogt-Maranto},\ and\ \citenamefont {Zdeborov\'{a}}}]{Carleo19RevMod}%
  \BibitemOpen
  \bibfield  {author} {\bibinfo {author} {\bibfnamefont {G.}~\bibnamefont
  {Carleo}}, \bibinfo {author} {\bibfnamefont {I.}~\bibnamefont {Cirac}},
  \bibinfo {author} {\bibfnamefont {K.}~\bibnamefont {Cranmer}}, \bibinfo
  {author} {\bibfnamefont {L.}~\bibnamefont {Daudet}}, \bibinfo {author}
  {\bibfnamefont {M.}~\bibnamefont {Schuld}}, \bibinfo {author} {\bibfnamefont
  {N.}~\bibnamefont {Tishby}}, \bibinfo {author} {\bibfnamefont
  {L.}~\bibnamefont {Vogt-Maranto}}, \ and\ \bibinfo {author} {\bibfnamefont
  {L.}~\bibnamefont {Zdeborov\'{a}}},\ }\bibfield  {title} {\bibinfo {title}
  {\emph {Machine learning and the physical sciences}},\ }\href {\doibase
  10.1103/RevModPhys.91.045002} {\bibfield  {journal} {\bibinfo  {journal}
  {Rev. Mod. Phys.}\ }\textbf {\bibinfo {volume} {91}},\ \bibinfo {pages}
  {045002} (\bibinfo {year} {2019})}\BibitemShut {NoStop}%
\bibitem [{\citenamefont {Carrasquilla}(2020)}]{Carrasquilla20AdvPhys}%
  \BibitemOpen
  \bibfield  {author} {\bibinfo {author} {\bibfnamefont {J.}~\bibnamefont
  {Carrasquilla}},\ }\bibfield  {title} {\bibinfo {title} {\emph {{Machine
  learning for quantum matter}}},\ }\href {\doibase
  10.1080/23746149.2020.1797528} {\bibfield  {journal} {\bibinfo  {journal}
  {Advances in Physics: X}\ }\textbf {\bibinfo {volume} {5}} (\bibinfo {year}
  {2020}),\ 10.1080/23746149.2020.1797528},\ \Eprint
  {http://arxiv.org/abs/2003.11040} {arXiv:2003.11040} \BibitemShut {NoStop}%
\bibitem [{\citenamefont {Carleo}\ and\ \citenamefont
  {Troyer}(2017)}]{Carleo17Science}%
  \BibitemOpen
  \bibfield  {author} {\bibinfo {author} {\bibfnamefont {G.}~\bibnamefont
  {Carleo}}\ and\ \bibinfo {author} {\bibfnamefont {M.}~\bibnamefont
  {Troyer}},\ }\bibfield  {title} {\bibinfo {title} {\emph {Solving the quantum
  many-body problem with artificial neural networks}},\ }\href {\doibase
  10.1126/science.aag2302} {\bibfield  {journal} {\bibinfo  {journal}
  {Science}\ }\textbf {\bibinfo {volume} {355}},\ \bibinfo {pages} {602}
  (\bibinfo {year} {2017})}\BibitemShut {NoStop}%
\bibitem [{\citenamefont {Torlai}\ \emph {et~al.}(2018)\citenamefont {Torlai},
  \citenamefont {Mazzola}, \citenamefont {Carrasquilla}, \citenamefont
  {Troyer}, \citenamefont {Melko},\ and\ \citenamefont
  {Carleo}}]{Torlai18NatPhys}%
  \BibitemOpen
  \bibfield  {author} {\bibinfo {author} {\bibfnamefont {G.}~\bibnamefont
  {Torlai}}, \bibinfo {author} {\bibfnamefont {G.}~\bibnamefont {Mazzola}},
  \bibinfo {author} {\bibfnamefont {J.}~\bibnamefont {Carrasquilla}}, \bibinfo
  {author} {\bibfnamefont {M.}~\bibnamefont {Troyer}}, \bibinfo {author}
  {\bibfnamefont {R.}~\bibnamefont {Melko}}, \ and\ \bibinfo {author}
  {\bibfnamefont {G.}~\bibnamefont {Carleo}},\ }\bibfield  {title} {\bibinfo
  {title} {\emph {Neural-network quantum state tomography}},\ }\href {\doibase
  10.1038/s41567-018-0048-5} {\bibfield  {journal} {\bibinfo  {journal} {Nat.
  Phys.}\ }\textbf {\bibinfo {volume} {14}},\ \bibinfo {pages} {447} (\bibinfo
  {year} {2018})}\BibitemShut {NoStop}%
\bibitem [{\citenamefont {Torlai}\ \emph {et~al.}(2019)\citenamefont {Torlai},
  \citenamefont {Timar}, \citenamefont {{Van Nieuwenburg}}, \citenamefont
  {Levine}, \citenamefont {Omran}, \citenamefont {Keesling}, \citenamefont
  {Bernien}, \citenamefont {Greiner}, \citenamefont {Vuleti{\'{c}}},
  \citenamefont {Lukin}, \citenamefont {Melko},\ and\ \citenamefont
  {Endres}}]{Torlai19PRL}%
  \BibitemOpen
  \bibfield  {author} {\bibinfo {author} {\bibfnamefont {G.}~\bibnamefont
  {Torlai}}, \bibinfo {author} {\bibfnamefont {B.}~\bibnamefont {Timar}},
  \bibinfo {author} {\bibfnamefont {E.~P.}\ \bibnamefont {{Van Nieuwenburg}}},
  \bibinfo {author} {\bibfnamefont {H.}~\bibnamefont {Levine}}, \bibinfo
  {author} {\bibfnamefont {A.}~\bibnamefont {Omran}}, \bibinfo {author}
  {\bibfnamefont {A.}~\bibnamefont {Keesling}}, \bibinfo {author}
  {\bibfnamefont {H.}~\bibnamefont {Bernien}}, \bibinfo {author} {\bibfnamefont
  {M.}~\bibnamefont {Greiner}}, \bibinfo {author} {\bibfnamefont
  {V.}~\bibnamefont {Vuleti{\'{c}}}}, \bibinfo {author} {\bibfnamefont {M.~D.}\
  \bibnamefont {Lukin}}, \bibinfo {author} {\bibfnamefont {R.~G.}\ \bibnamefont
  {Melko}}, \ and\ \bibinfo {author} {\bibfnamefont {M.}~\bibnamefont
  {Endres}},\ }\bibfield  {title} {\bibinfo {title} {\emph {{Integrating Neural
  Networks with a Quantum Simulator for State Reconstruction}}},\ }\href
  {\doibase 10.1103/PhysRevLett.123.230504} {\bibfield  {journal} {\bibinfo
  {journal} {Physical Review Letters}\ }\textbf {\bibinfo {volume} {123}},\
  \bibinfo {pages} {19} (\bibinfo {year} {2019})},\ \Eprint
  {http://arxiv.org/abs/1904.08441} {arXiv:1904.08441} \BibitemShut {NoStop}%
\bibitem [{\citenamefont {Neugebauer}\ \emph {et~al.}(2020)\citenamefont
  {Neugebauer}, \citenamefont {Fischer}, \citenamefont {J{\"{a}}ger},
  \citenamefont {Czischek}, \citenamefont {Jochim}, \citenamefont
  {Weidem{\"{u}}ller},\ and\ \citenamefont {G{\"{a}}rttner}}]{Neugebauer20PRA}%
  \BibitemOpen
  \bibfield  {author} {\bibinfo {author} {\bibfnamefont {M.}~\bibnamefont
  {Neugebauer}}, \bibinfo {author} {\bibfnamefont {L.}~\bibnamefont {Fischer}},
  \bibinfo {author} {\bibfnamefont {A.}~\bibnamefont {J{\"{a}}ger}}, \bibinfo
  {author} {\bibfnamefont {S.}~\bibnamefont {Czischek}}, \bibinfo {author}
  {\bibfnamefont {S.}~\bibnamefont {Jochim}}, \bibinfo {author} {\bibfnamefont
  {M.}~\bibnamefont {Weidem{\"{u}}ller}}, \ and\ \bibinfo {author}
  {\bibfnamefont {M.}~\bibnamefont {G{\"{a}}rttner}},\ }\bibfield  {title}
  {\bibinfo {title} {\emph {{Neural network quantum state tomography in a
  two-qubit experiment}}},\ }\href {\doibase 10.1103/PhysRevA.102.042604}
  {\bibfield  {journal} {\bibinfo  {journal} {Phys. Rev. A}\ }\textbf {\bibinfo
  {volume} {102}},\ \bibinfo {pages} {042604} (\bibinfo {year} {2020})},\
  \Eprint {http://arxiv.org/abs/2007.16185} {arXiv:2007.16185} \BibitemShut
  {NoStop}%
\bibitem [{\citenamefont {Wigley}\ \emph {et~al.}(2016)\citenamefont {Wigley},
  \citenamefont {Everitt}, \citenamefont {{Van Den Hengel}}, \citenamefont
  {Bastian}, \citenamefont {Sooriyabandara}, \citenamefont {Mcdonald},
  \citenamefont {Hardman}, \citenamefont {Quinlivan}, \citenamefont {Manju},
  \citenamefont {Kuhn}, \citenamefont {Petersen}, \citenamefont {Luiten},
  \citenamefont {Hope}, \citenamefont {Robins},\ and\ \citenamefont
  {Hush}}]{Wigley16SciRep}%
  \BibitemOpen
  \bibfield  {author} {\bibinfo {author} {\bibfnamefont {P.~B.}\ \bibnamefont
  {Wigley}}, \bibinfo {author} {\bibfnamefont {P.~J.}\ \bibnamefont {Everitt}},
  \bibinfo {author} {\bibfnamefont {A.}~\bibnamefont {{Van Den Hengel}}},
  \bibinfo {author} {\bibfnamefont {J.~W.}\ \bibnamefont {Bastian}}, \bibinfo
  {author} {\bibfnamefont {M.~A.}\ \bibnamefont {Sooriyabandara}}, \bibinfo
  {author} {\bibfnamefont {G.~D.}\ \bibnamefont {Mcdonald}}, \bibinfo {author}
  {\bibfnamefont {K.~S.}\ \bibnamefont {Hardman}}, \bibinfo {author}
  {\bibfnamefont {C.~D.}\ \bibnamefont {Quinlivan}}, \bibinfo {author}
  {\bibfnamefont {P.}~\bibnamefont {Manju}}, \bibinfo {author} {\bibfnamefont
  {C.~C.}\ \bibnamefont {Kuhn}}, \bibinfo {author} {\bibfnamefont {I.~R.}\
  \bibnamefont {Petersen}}, \bibinfo {author} {\bibfnamefont {A.~N.}\
  \bibnamefont {Luiten}}, \bibinfo {author} {\bibfnamefont {J.~J.}\
  \bibnamefont {Hope}}, \bibinfo {author} {\bibfnamefont {N.~P.}\ \bibnamefont
  {Robins}}, \ and\ \bibinfo {author} {\bibfnamefont {M.~R.}\ \bibnamefont
  {Hush}},\ }\bibfield  {title} {\bibinfo {title} {\emph {{Fast
  machine-learning online optimization of ultra-cold-atom experiments}}},\
  }\href {\doibase 10.1038/srep25890} {\bibfield  {journal} {\bibinfo
  {journal} {Scientific Reports}\ }\textbf {\bibinfo {volume} {6}},\ \bibinfo
  {pages} {25890} (\bibinfo {year} {2016})},\ \Eprint
  {http://arxiv.org/abs/1507.04964} {arXiv:1507.04964} \BibitemShut {NoStop}%
\bibitem [{\citenamefont {Tranter}\ \emph {et~al.}(2018)\citenamefont
  {Tranter}, \citenamefont {Slatyer}, \citenamefont {Hush}, \citenamefont
  {Leung}, \citenamefont {Everett}, \citenamefont {Paul}, \citenamefont
  {Vernaz-Gris}, \citenamefont {Lam}, \citenamefont {Buchler},\ and\
  \citenamefont {Campbell}}]{Tranter18NatComm}%
  \BibitemOpen
  \bibfield  {author} {\bibinfo {author} {\bibfnamefont {A.~D.}\ \bibnamefont
  {Tranter}}, \bibinfo {author} {\bibfnamefont {H.~J.}\ \bibnamefont
  {Slatyer}}, \bibinfo {author} {\bibfnamefont {M.~R.}\ \bibnamefont {Hush}},
  \bibinfo {author} {\bibfnamefont {A.~C.}\ \bibnamefont {Leung}}, \bibinfo
  {author} {\bibfnamefont {J.~L.}\ \bibnamefont {Everett}}, \bibinfo {author}
  {\bibfnamefont {K.~V.}\ \bibnamefont {Paul}}, \bibinfo {author}
  {\bibfnamefont {P.}~\bibnamefont {Vernaz-Gris}}, \bibinfo {author}
  {\bibfnamefont {P.~K.}\ \bibnamefont {Lam}}, \bibinfo {author} {\bibfnamefont
  {B.~C.}\ \bibnamefont {Buchler}}, \ and\ \bibinfo {author} {\bibfnamefont
  {G.~T.}\ \bibnamefont {Campbell}},\ }\bibfield  {title} {\bibinfo {title}
  {\emph {{Multiparameter optimisation of a magneto-optical trap using deep
  learning}}},\ }\href {\doibase 10.1038/s41467-018-06847-1} {\bibfield
  {journal} {\bibinfo  {journal} {Nat. Comm.}\ }\textbf {\bibinfo {volume} {9}}
  (\bibinfo {year} {2018}),\ 10.1038/s41467-018-06847-1},\ \Eprint
  {http://arxiv.org/abs/1805.00654} {arXiv:1805.00654} \BibitemShut {NoStop}%
\bibitem [{\citenamefont {Bukov}\ \emph {et~al.}(2018)\citenamefont {Bukov},
  \citenamefont {Day}, \citenamefont {Sels}, \citenamefont {Weinberg},
  \citenamefont {Polkovnikov},\ and\ \citenamefont {Mehta}}]{Bukov18PRX}%
  \BibitemOpen
  \bibfield  {author} {\bibinfo {author} {\bibfnamefont {M.}~\bibnamefont
  {Bukov}}, \bibinfo {author} {\bibfnamefont {A.~G.}\ \bibnamefont {Day}},
  \bibinfo {author} {\bibfnamefont {D.}~\bibnamefont {Sels}}, \bibinfo {author}
  {\bibfnamefont {P.}~\bibnamefont {Weinberg}}, \bibinfo {author}
  {\bibfnamefont {A.}~\bibnamefont {Polkovnikov}}, \ and\ \bibinfo {author}
  {\bibfnamefont {P.}~\bibnamefont {Mehta}},\ }\bibfield  {title} {\bibinfo
  {title} {\emph {{Reinforcement Learning in Different Phases of Quantum
  Control}}},\ }\href {\doibase 10.1103/PhysRevX.8.031086} {\bibfield
  {journal} {\bibinfo  {journal} {Physical Review X}\ }\textbf {\bibinfo
  {volume} {8}},\ \bibinfo {pages} {031086} (\bibinfo {year} {2018})},\ \Eprint
  {http://arxiv.org/abs/1705.00565} {arXiv:1705.00565} \BibitemShut {NoStop}%
\bibitem [{\citenamefont {Davletov}\ \emph {et~al.}(2020)\citenamefont
  {Davletov}, \citenamefont {Tsyganok}, \citenamefont {Khlebnikov},
  \citenamefont {Pershin}, \citenamefont {Shaykin},\ and\ \citenamefont
  {Akimov}}]{Davletov20PRA}%
  \BibitemOpen
  \bibfield  {author} {\bibinfo {author} {\bibfnamefont {E.~T.}\ \bibnamefont
  {Davletov}}, \bibinfo {author} {\bibfnamefont {V.~V.}\ \bibnamefont
  {Tsyganok}}, \bibinfo {author} {\bibfnamefont {V.~A.}\ \bibnamefont
  {Khlebnikov}}, \bibinfo {author} {\bibfnamefont {D.~A.}\ \bibnamefont
  {Pershin}}, \bibinfo {author} {\bibfnamefont {D.~V.}\ \bibnamefont
  {Shaykin}}, \ and\ \bibinfo {author} {\bibfnamefont {A.~V.}\ \bibnamefont
  {Akimov}},\ }\bibfield  {title} {\bibinfo {title} {\emph {{Machine learning
  for achieving Bose-Einstein condensation of thulium atoms}}},\ }\href
  {\doibase 10.1103/PhysRevA.102.011302} {\bibfield  {journal} {\bibinfo
  {journal} {Physical Review A}\ }\textbf {\bibinfo {volume} {102}},\ \bibinfo
  {pages} {011302(R)} (\bibinfo {year} {2020})},\ \Eprint
  {http://arxiv.org/abs/2003.00346} {arXiv:2003.00346} \BibitemShut {NoStop}%
\bibitem [{\citenamefont {Carrasquilla}\ and\ \citenamefont
  {Melko}(2017)}]{Carrasquilla17NatPhys}%
  \BibitemOpen
  \bibfield  {author} {\bibinfo {author} {\bibfnamefont {J.}~\bibnamefont
  {Carrasquilla}}\ and\ \bibinfo {author} {\bibfnamefont {R.~G.}\ \bibnamefont
  {Melko}},\ }\bibfield  {title} {\bibinfo {title} {\emph {Machine learning
  phases of matter}},\ }\href {\doibase 10.1038/nphys4035} {\bibfield
  {journal} {\bibinfo  {journal} {Nat. Phys.}\ }\textbf {\bibinfo {volume}
  {13}},\ \bibinfo {pages} {431} (\bibinfo {year} {2017})}\BibitemShut
  {NoStop}%
\bibitem [{\citenamefont {Ch'ng}\ \emph {et~al.}(2017)\citenamefont {Ch'ng},
  \citenamefont {Carrasquilla}, \citenamefont {Melko},\ and\ \citenamefont
  {Khatami}}]{Chng17PRX}%
  \BibitemOpen
  \bibfield  {author} {\bibinfo {author} {\bibfnamefont {K.}~\bibnamefont
  {Ch'ng}}, \bibinfo {author} {\bibfnamefont {J.}~\bibnamefont {Carrasquilla}},
  \bibinfo {author} {\bibfnamefont {R.~G.}\ \bibnamefont {Melko}}, \ and\
  \bibinfo {author} {\bibfnamefont {E.}~\bibnamefont {Khatami}},\ }\bibfield
  {title} {\bibinfo {title} {\emph {{Machine learning phases of strongly
  correlated fermions}}},\ }\href {\doibase 10.1103/PhysRevX.7.031038}
  {\bibfield  {journal} {\bibinfo  {journal} {Physical Review X}\ }\textbf
  {\bibinfo {volume} {7}},\ \bibinfo {pages} {031038} (\bibinfo {year}
  {2017})},\ \Eprint {http://arxiv.org/abs/1609.02552} {arXiv:1609.02552}
  \BibitemShut {NoStop}%
\bibitem [{\citenamefont {Broecker}\ \emph
  {et~al.}(2017{\natexlab{a}})\citenamefont {Broecker}, \citenamefont
  {Carrasquilla}, \citenamefont {Melko},\ and\ \citenamefont
  {Trebst}}]{Broecker17SciRep}%
  \BibitemOpen
  \bibfield  {author} {\bibinfo {author} {\bibfnamefont {P.}~\bibnamefont
  {Broecker}}, \bibinfo {author} {\bibfnamefont {J.}~\bibnamefont
  {Carrasquilla}}, \bibinfo {author} {\bibfnamefont {R.~G.}\ \bibnamefont
  {Melko}}, \ and\ \bibinfo {author} {\bibfnamefont {S.}~\bibnamefont
  {Trebst}},\ }\bibfield  {title} {\bibinfo {title} {\emph {Machine learning
  quantum phases of matter beyond the fermion sign problem}},\ }\href {\doibase
  10.1038/s41598-017-09098-0} {\bibfield  {journal} {\bibinfo  {journal} {Sci.
  Rep.}\ }\textbf {\bibinfo {volume} {7}},\ \bibinfo {pages} {8823} (\bibinfo
  {year} {2017}{\natexlab{a}})}\BibitemShut {NoStop}%
\bibitem [{\citenamefont {van Nieuwenburg}\ \emph {et~al.}(2017)\citenamefont
  {van Nieuwenburg}, \citenamefont {Liu},\ and\ \citenamefont
  {Huber}}]{Nieuwenburg17NatPhys}%
  \BibitemOpen
  \bibfield  {author} {\bibinfo {author} {\bibfnamefont {E.~P.~L.}\
  \bibnamefont {van Nieuwenburg}}, \bibinfo {author} {\bibfnamefont {Y.-H.}\
  \bibnamefont {Liu}}, \ and\ \bibinfo {author} {\bibfnamefont {S.~D.}\
  \bibnamefont {Huber}},\ }\bibfield  {title} {\bibinfo {title} {\emph
  {Learning phase transitions by confusion}},\ }\href {\doibase
  10.1038/nphys4037} {\bibfield  {journal} {\bibinfo  {journal} {Nat. Phys.}\
  }\textbf {\bibinfo {volume} {13}},\ \bibinfo {pages} {435} (\bibinfo {year}
  {2017})}\BibitemShut {NoStop}%
\bibitem [{\citenamefont {Wang}(2016)}]{Wang16PRB}%
  \BibitemOpen
  \bibfield  {author} {\bibinfo {author} {\bibfnamefont {L.}~\bibnamefont
  {Wang}},\ }\bibfield  {title} {\bibinfo {title} {\emph {Discovering phase
  transitions with unsupervised learning}},\ }\href {\doibase
  10.1103/PhysRevB.94.195105} {\bibfield  {journal} {\bibinfo  {journal} {Phys.
  Rev. B}\ }\textbf {\bibinfo {volume} {94}},\ \bibinfo {pages} {195105}
  (\bibinfo {year} {2016})}\BibitemShut {NoStop}%
\bibitem [{\citenamefont {Wang}\ and\ \citenamefont {Zhai}(2017)}]{Wang17PRB}%
  \BibitemOpen
  \bibfield  {author} {\bibinfo {author} {\bibfnamefont {C.}~\bibnamefont
  {Wang}}\ and\ \bibinfo {author} {\bibfnamefont {H.}~\bibnamefont {Zhai}},\
  }\bibfield  {title} {\bibinfo {title} {\emph {Machine learning of frustrated
  classical spin models. I. Principal component analysis}},\ }\href {\doibase
  10.1103/PhysRevB.96.144432} {\bibfield  {journal} {\bibinfo  {journal} {Phys.
  Rev. B}\ }\textbf {\bibinfo {volume} {96}},\ \bibinfo {pages} {144432}
  (\bibinfo {year} {2017})}\BibitemShut {NoStop}%
\bibitem [{\citenamefont {Ohtsuki}\ and\ \citenamefont
  {Ohtsuki}(2016)}]{Ohtsuki16JPSJ}%
  \BibitemOpen
  \bibfield  {author} {\bibinfo {author} {\bibfnamefont {T.}~\bibnamefont
  {Ohtsuki}}\ and\ \bibinfo {author} {\bibfnamefont {T.}~\bibnamefont
  {Ohtsuki}},\ }\bibfield  {title} {\bibinfo {title} {\emph {Deep Learning the
  Quantum Phase Transitions in Random Two-Dimensional Electron Systems}},\
  }\href {\doibase 10.7566/JPSJ.85.123706} {\bibfield  {journal} {\bibinfo
  {journal} {Journal of the Physical Society of Japan}\ }\textbf {\bibinfo
  {volume} {85}},\ \bibinfo {pages} {123706} (\bibinfo {year}
  {2016})}\BibitemShut {NoStop}%
\bibitem [{\citenamefont {Kottmann}\ \emph {et~al.}(2020)\citenamefont
  {Kottmann}, \citenamefont {Huembeli}, \citenamefont {Lewenstein},\ and\
  \citenamefont {Ac{\'{i}}n}}]{kottmann2020unsupervised}%
  \BibitemOpen
  \bibfield  {author} {\bibinfo {author} {\bibfnamefont {K.}~\bibnamefont
  {Kottmann}}, \bibinfo {author} {\bibfnamefont {P.}~\bibnamefont {Huembeli}},
  \bibinfo {author} {\bibfnamefont {M.}~\bibnamefont {Lewenstein}}, \ and\
  \bibinfo {author} {\bibfnamefont {A.}~\bibnamefont {Ac{\'{i}}n}},\ }\bibfield
   {title} {\bibinfo {title} {\emph {{Unsupervised Phase Discovery with Deep
  Anomaly Detection}}},\ }\href {\doibase 10.1103/PhysRevLett.125.170603}
  {\bibfield  {journal} {\bibinfo  {journal} {Physical Review Letters}\
  }\textbf {\bibinfo {volume} {125}},\ \bibinfo {pages} {170603} (\bibinfo
  {year} {2020})}\BibitemShut {NoStop}%
\bibitem [{\citenamefont {Huembeli}\ \emph {et~al.}(2018)\citenamefont
  {Huembeli}, \citenamefont {Dauphin},\ and\ \citenamefont
  {Wittek}}]{Huembeli18PRB}%
  \BibitemOpen
  \bibfield  {author} {\bibinfo {author} {\bibfnamefont {P.}~\bibnamefont
  {Huembeli}}, \bibinfo {author} {\bibfnamefont {A.}~\bibnamefont {Dauphin}}, \
  and\ \bibinfo {author} {\bibfnamefont {P.}~\bibnamefont {Wittek}},\
  }\bibfield  {title} {\bibinfo {title} {\emph {{Identifying quantum phase
  transitions with adversarial neural networks}}},\ }\href {\doibase
  10.1103/PhysRevB.97.134109} {\bibfield  {journal} {\bibinfo  {journal} {Phys.
  Rev. B}\ }\textbf {\bibinfo {volume} {97}},\ \bibinfo {pages} {134109}
  (\bibinfo {year} {2018})},\ \Eprint {http://arxiv.org/abs/1710.08382}
  {arXiv:1710.08382} \BibitemShut {NoStop}%
\bibitem [{\citenamefont {Huembeli}\ \emph {et~al.}(2019)\citenamefont
  {Huembeli}, \citenamefont {Dauphin}, \citenamefont {Wittek},\ and\
  \citenamefont {Gogolin}}]{Huembeli19}%
  \BibitemOpen
  \bibfield  {author} {\bibinfo {author} {\bibfnamefont {P.}~\bibnamefont
  {Huembeli}}, \bibinfo {author} {\bibfnamefont {A.}~\bibnamefont {Dauphin}},
  \bibinfo {author} {\bibfnamefont {P.}~\bibnamefont {Wittek}}, \ and\ \bibinfo
  {author} {\bibfnamefont {C.}~\bibnamefont {Gogolin}},\ }\bibfield  {title}
  {\bibinfo {title} {\emph {Automated discovery of characteristic features of
  phase transitions in many-body localization}},\ }\href {\doibase
  10.1103/PhysRevB.99.104106} {\bibfield  {journal} {\bibinfo  {journal} {Phys.
  Rev. B}\ }\textbf {\bibinfo {volume} {99}},\ \bibinfo {pages} {104106}
  (\bibinfo {year} {2019})}\BibitemShut {NoStop}%
\bibitem [{\citenamefont {Zhang}\ \emph {et~al.}(2019)\citenamefont {Zhang},
  \citenamefont {Mesaros}, \citenamefont {Fujita}, \citenamefont {Edkins},
  \citenamefont {Hamidian}, \citenamefont {Ch'ng}, \citenamefont {Eisaki},
  \citenamefont {Uchida}, \citenamefont {Davis}, \citenamefont {Khatami},\ and\
  \citenamefont {Kim}}]{Zhang19Nature}%
  \BibitemOpen
  \bibfield  {author} {\bibinfo {author} {\bibfnamefont {Y.}~\bibnamefont
  {Zhang}}, \bibinfo {author} {\bibfnamefont {A.}~\bibnamefont {Mesaros}},
  \bibinfo {author} {\bibfnamefont {K.}~\bibnamefont {Fujita}}, \bibinfo
  {author} {\bibfnamefont {S.~D.}\ \bibnamefont {Edkins}}, \bibinfo {author}
  {\bibfnamefont {M.~H.}\ \bibnamefont {Hamidian}}, \bibinfo {author}
  {\bibfnamefont {K.}~\bibnamefont {Ch'ng}}, \bibinfo {author} {\bibfnamefont
  {H.}~\bibnamefont {Eisaki}}, \bibinfo {author} {\bibfnamefont
  {S.}~\bibnamefont {Uchida}}, \bibinfo {author} {\bibfnamefont {J.~C.}\
  \bibnamefont {Davis}}, \bibinfo {author} {\bibfnamefont {E.}~\bibnamefont
  {Khatami}}, \ and\ \bibinfo {author} {\bibfnamefont {E.~A.}\ \bibnamefont
  {Kim}},\ }\bibfield  {title} {\bibinfo {title} {\emph {{Machine learning in
  electronic-quantum-matter imaging experiments}}},\ }\href {\doibase
  10.1038/s41586-019-1319-8} {\bibfield  {journal} {\bibinfo  {journal}
  {Nature}\ }\textbf {\bibinfo {volume} {570}},\ \bibinfo {pages} {484}
  (\bibinfo {year} {2019})}\BibitemShut {NoStop}%
\bibitem [{\citenamefont {Ziatdinov}\ \emph {et~al.}(2016)\citenamefont
  {Ziatdinov}, \citenamefont {Maksov}, \citenamefont {Li}, \citenamefont
  {Sefat}, \citenamefont {Maksymovych},\ and\ \citenamefont
  {Kalinin}}]{Ziatdinov16Nanotechnology}%
  \BibitemOpen
  \bibfield  {author} {\bibinfo {author} {\bibfnamefont {M.}~\bibnamefont
  {Ziatdinov}}, \bibinfo {author} {\bibfnamefont {A.}~\bibnamefont {Maksov}},
  \bibinfo {author} {\bibfnamefont {L.}~\bibnamefont {Li}}, \bibinfo {author}
  {\bibfnamefont {A.~S.}\ \bibnamefont {Sefat}}, \bibinfo {author}
  {\bibfnamefont {P.}~\bibnamefont {Maksymovych}}, \ and\ \bibinfo {author}
  {\bibfnamefont {S.~V.}\ \bibnamefont {Kalinin}},\ }\bibfield  {title}
  {\bibinfo {title} {\emph {{Deep data mining in a real space: Separation of
  intertwined electronic responses in a lightly doped BaFe2As2}}},\ }\href
  {\doibase 10.1088/0957-4484/27/47/475706} {\bibfield  {journal} {\bibinfo
  {journal} {Nanotechnology}\ }\textbf {\bibinfo {volume} {27}} (\bibinfo
  {year} {2016}),\ 10.1088/0957-4484/27/47/475706}\BibitemShut {NoStop}%
\bibitem [{\citenamefont {Samarakoon}\ \emph {et~al.}(2020)\citenamefont
  {Samarakoon}, \citenamefont {Barros}, \citenamefont {Li}, \citenamefont
  {Eisenbach}, \citenamefont {Zhang}, \citenamefont {Ye}, \citenamefont
  {Sharma}, \citenamefont {Dun}, \citenamefont {Zhou}, \citenamefont {Grigera},
  \citenamefont {Batista},\ and\ \citenamefont
  {Tennant}}]{Samarakoon20NatComm}%
  \BibitemOpen
  \bibfield  {author} {\bibinfo {author} {\bibfnamefont {A.~M.}\ \bibnamefont
  {Samarakoon}}, \bibinfo {author} {\bibfnamefont {K.}~\bibnamefont {Barros}},
  \bibinfo {author} {\bibfnamefont {Y.~W.}\ \bibnamefont {Li}}, \bibinfo
  {author} {\bibfnamefont {M.}~\bibnamefont {Eisenbach}}, \bibinfo {author}
  {\bibfnamefont {Q.}~\bibnamefont {Zhang}}, \bibinfo {author} {\bibfnamefont
  {F.}~\bibnamefont {Ye}}, \bibinfo {author} {\bibfnamefont {V.}~\bibnamefont
  {Sharma}}, \bibinfo {author} {\bibfnamefont {Z.~L.}\ \bibnamefont {Dun}},
  \bibinfo {author} {\bibfnamefont {H.}~\bibnamefont {Zhou}}, \bibinfo {author}
  {\bibfnamefont {S.~A.}\ \bibnamefont {Grigera}}, \bibinfo {author}
  {\bibfnamefont {C.~D.}\ \bibnamefont {Batista}}, \ and\ \bibinfo {author}
  {\bibfnamefont {D.~A.}\ \bibnamefont {Tennant}},\ }\bibfield  {title}
  {\bibinfo {title} {\emph {{Machine-learning-assisted insight into spin ice
  Dy2Ti2O7}}},\ }\href {\doibase 10.1038/s41467-020-14660-y} {\bibfield
  {journal} {\bibinfo  {journal} {Nat. Comm.}\ }\textbf {\bibinfo {volume}
  {11}},\ \bibinfo {pages} {1} (\bibinfo {year} {2020})}\BibitemShut {NoStop}%
\bibitem [{\citenamefont {Rem}\ \emph {et~al.}(2019)\citenamefont {Rem},
  \citenamefont {K{\"{a}}ming}, \citenamefont {Tarnowski}, \citenamefont
  {Asteria}, \citenamefont {Fl{\"{a}}schner}, \citenamefont {Becker},
  \citenamefont {Sengstock},\ and\ \citenamefont {Weitenberg}}]{Rem19NatPhys}%
  \BibitemOpen
  \bibfield  {author} {\bibinfo {author} {\bibfnamefont {B.~S.}\ \bibnamefont
  {Rem}}, \bibinfo {author} {\bibfnamefont {N.}~\bibnamefont {K{\"{a}}ming}},
  \bibinfo {author} {\bibfnamefont {M.}~\bibnamefont {Tarnowski}}, \bibinfo
  {author} {\bibfnamefont {L.}~\bibnamefont {Asteria}}, \bibinfo {author}
  {\bibfnamefont {N.}~\bibnamefont {Fl{\"{a}}schner}}, \bibinfo {author}
  {\bibfnamefont {C.}~\bibnamefont {Becker}}, \bibinfo {author} {\bibfnamefont
  {K.}~\bibnamefont {Sengstock}}, \ and\ \bibinfo {author} {\bibfnamefont
  {C.}~\bibnamefont {Weitenberg}},\ }\bibfield  {title} {\bibinfo {title}
  {\emph {{Identifying quantum phase transitions using artificial neural
  networks on experimental data}}},\ }\href {\doibase
  10.1038/s41567-019-0554-0} {\bibfield  {journal} {\bibinfo  {journal} {Nat.
  Phys.}\ }\textbf {\bibinfo {volume} {15}},\ \bibinfo {pages} {917} (\bibinfo
  {year} {2019})},\ \Eprint {http://arxiv.org/abs/1809.05519}
  {arXiv:1809.05519} \BibitemShut {NoStop}%
\bibitem [{\citenamefont {Bohrdt}\ \emph {et~al.}(2019)\citenamefont {Bohrdt},
  \citenamefont {Chiu}, \citenamefont {Ji}, \citenamefont {Xu}, \citenamefont
  {Greif}, \citenamefont {Greiner}, \citenamefont {Demler}, \citenamefont
  {Grusdt},\ and\ \citenamefont {Knap}}]{Bohrdt19NatPhys}%
  \BibitemOpen
  \bibfield  {author} {\bibinfo {author} {\bibfnamefont {A.}~\bibnamefont
  {Bohrdt}}, \bibinfo {author} {\bibfnamefont {C.~S.}\ \bibnamefont {Chiu}},
  \bibinfo {author} {\bibfnamefont {G.}~\bibnamefont {Ji}}, \bibinfo {author}
  {\bibfnamefont {M.}~\bibnamefont {Xu}}, \bibinfo {author} {\bibfnamefont
  {D.}~\bibnamefont {Greif}}, \bibinfo {author} {\bibfnamefont
  {M.}~\bibnamefont {Greiner}}, \bibinfo {author} {\bibfnamefont
  {E.}~\bibnamefont {Demler}}, \bibinfo {author} {\bibfnamefont
  {F.}~\bibnamefont {Grusdt}}, \ and\ \bibinfo {author} {\bibfnamefont
  {M.}~\bibnamefont {Knap}},\ }\bibfield  {title} {\bibinfo {title} {\emph
  {Classifying snapshots of the doped Hubbard model with machine learning}},\
  }\href {\doibase 10.1038/s41567-019-0565-x} {\bibfield  {journal} {\bibinfo
  {journal} {Nat. Phys.}\ }\textbf {\bibinfo {volume} {15}},\ \bibinfo {pages}
  {921} (\bibinfo {year} {2019})}\BibitemShut {NoStop}%
\bibitem [{\citenamefont {Khatami}\ \emph {et~al.}(2020)\citenamefont
  {Khatami}, \citenamefont {Guardado-Sanchez}, \citenamefont {Spar},
  \citenamefont {Carrasquilla}, \citenamefont {Bakr},\ and\ \citenamefont
  {Scalettar}}]{Khatami20PRA}%
  \BibitemOpen
  \bibfield  {author} {\bibinfo {author} {\bibfnamefont {E.}~\bibnamefont
  {Khatami}}, \bibinfo {author} {\bibfnamefont {E.}~\bibnamefont
  {Guardado-Sanchez}}, \bibinfo {author} {\bibfnamefont {B.~M.}\ \bibnamefont
  {Spar}}, \bibinfo {author} {\bibfnamefont {J.~F.}\ \bibnamefont
  {Carrasquilla}}, \bibinfo {author} {\bibfnamefont {W.~S.}\ \bibnamefont
  {Bakr}}, \ and\ \bibinfo {author} {\bibfnamefont {R.~T.}\ \bibnamefont
  {Scalettar}},\ }\bibfield  {title} {\bibinfo {title} {\emph {Visualizing
  strange metallic correlations in the two-dimensional Fermi-Hubbard model with
  artificial intelligence}},\ }\href {\doibase 10.1103/PhysRevA.102.033326}
  {\bibfield  {journal} {\bibinfo  {journal} {Phys. Rev. A}\ }\textbf {\bibinfo
  {volume} {102}},\ \bibinfo {pages} {033326} (\bibinfo {year}
  {2020})}\BibitemShut {NoStop}%
\bibitem [{\citenamefont {Miles}\ \emph {et~al.}(2020)\citenamefont {Miles},
  \citenamefont {Bohrdt}, \citenamefont {Wu}, \citenamefont {Chiu},
  \citenamefont {Xu}, \citenamefont {Ji}, \citenamefont {Greiner},
  \citenamefont {Weinberger}, \citenamefont {Demler},\ and\ \citenamefont
  {Kim}}]{Miles20arXiv}%
  \BibitemOpen
  \bibfield  {author} {\bibinfo {author} {\bibfnamefont {C.}~\bibnamefont
  {Miles}}, \bibinfo {author} {\bibfnamefont {A.}~\bibnamefont {Bohrdt}},
  \bibinfo {author} {\bibfnamefont {R.}~\bibnamefont {Wu}}, \bibinfo {author}
  {\bibfnamefont {C.}~\bibnamefont {Chiu}}, \bibinfo {author} {\bibfnamefont
  {M.}~\bibnamefont {Xu}}, \bibinfo {author} {\bibfnamefont {G.}~\bibnamefont
  {Ji}}, \bibinfo {author} {\bibfnamefont {M.}~\bibnamefont {Greiner}},
  \bibinfo {author} {\bibfnamefont {K.~Q.}\ \bibnamefont {Weinberger}},
  \bibinfo {author} {\bibfnamefont {E.}~\bibnamefont {Demler}}, \ and\ \bibinfo
  {author} {\bibfnamefont {E.-A.}\ \bibnamefont {Kim}},\ }\bibfield  {title}
  {\bibinfo {title} {\emph {Correlator Convolutional Neural Networks: An
  Interpretable Architecture for Image-like Quantum Matter Data}},\ }\href@noop
  {} {\bibfield  {journal} {\bibinfo  {journal} {arXiv}\ } (\bibinfo {year}
  {2020})},\ \Eprint {http://arxiv.org/abs/2011.03474} {arXiv:2011.03474
  [cond-mat.str-el]} \BibitemShut {NoStop}%
\bibitem [{\citenamefont {Broecker}\ \emph
  {et~al.}(2017{\natexlab{b}})\citenamefont {Broecker}, \citenamefont
  {Assaad},\ and\ \citenamefont {Trebst}}]{Broecker17arxiv}%
  \BibitemOpen
  \bibfield  {author} {\bibinfo {author} {\bibfnamefont {P.}~\bibnamefont
  {Broecker}}, \bibinfo {author} {\bibfnamefont {F.~F.}\ \bibnamefont
  {Assaad}}, \ and\ \bibinfo {author} {\bibfnamefont {S.}~\bibnamefont
  {Trebst}},\ }\bibfield  {title} {\bibinfo {title} {\emph {{Quantum phase
  recognition via unsupervised machine learning}}},\ }\href {\doibase
  10.1016/j.ptsp.2011.01.003} {\bibfield  {journal} {\bibinfo  {journal}
  {arXiv:1707.00663}\ } (\bibinfo {year} {2017}{\natexlab{b}}),\
  10.1016/j.ptsp.2011.01.003},\ \Eprint {http://arxiv.org/abs/1707.00663}
  {arXiv:1707.00663} \BibitemShut {NoStop}%
\bibitem [{\citenamefont {Wetzel}(2017)}]{Wetzel17PRE}%
  \BibitemOpen
  \bibfield  {author} {\bibinfo {author} {\bibfnamefont {S.~J.}\ \bibnamefont
  {Wetzel}},\ }\bibfield  {title} {\bibinfo {title} {\emph {{Unsupervised
  learning of phase transitions: From principal component analysis to
  variational autoencoders}}},\ }\href {\doibase 10.1103/PhysRevE.96.022140}
  {\bibfield  {journal} {\bibinfo  {journal} {Physical Review E}\ }\textbf
  {\bibinfo {volume} {96}},\ \bibinfo {pages} {1} (\bibinfo {year} {2017})},\
  \Eprint {http://arxiv.org/abs/1703.02435} {arXiv:1703.02435} \BibitemShut
  {NoStop}%
\bibitem [{\citenamefont {Ch'ng}\ \emph {et~al.}(2018)\citenamefont {Ch'ng},
  \citenamefont {Vazquez},\ and\ \citenamefont {Khatami}}]{Chng18PRE}%
  \BibitemOpen
  \bibfield  {author} {\bibinfo {author} {\bibfnamefont {K.}~\bibnamefont
  {Ch'ng}}, \bibinfo {author} {\bibfnamefont {N.}~\bibnamefont {Vazquez}}, \
  and\ \bibinfo {author} {\bibfnamefont {E.}~\bibnamefont {Khatami}},\
  }\bibfield  {title} {\bibinfo {title} {\emph {Unsupervised machine learning
  account of magnetic transitions in the Hubbard model}},\ }\href {\doibase
  10.1103/PhysRevE.97.013306} {\bibfield  {journal} {\bibinfo  {journal} {Phys.
  Rev. E}\ }\textbf {\bibinfo {volume} {97}},\ \bibinfo {pages} {013306}
  (\bibinfo {year} {2018})}\BibitemShut {NoStop}%
\bibitem [{\citenamefont {Greplova}\ \emph {et~al.}(2020)\citenamefont
  {Greplova}, \citenamefont {Valenti}, \citenamefont {Boschung}, \citenamefont
  {Sch{\"{a}}fer}, \citenamefont {L{\"{o}}rch},\ and\ \citenamefont
  {Huber}}]{Greplova20NJP}%
  \BibitemOpen
  \bibfield  {author} {\bibinfo {author} {\bibfnamefont {E.}~\bibnamefont
  {Greplova}}, \bibinfo {author} {\bibfnamefont {A.}~\bibnamefont {Valenti}},
  \bibinfo {author} {\bibfnamefont {G.}~\bibnamefont {Boschung}}, \bibinfo
  {author} {\bibfnamefont {F.}~\bibnamefont {Sch{\"{a}}fer}}, \bibinfo {author}
  {\bibfnamefont {N.}~\bibnamefont {L{\"{o}}rch}}, \ and\ \bibinfo {author}
  {\bibfnamefont {S.~D.}\ \bibnamefont {Huber}},\ }\bibfield  {title} {\bibinfo
  {title} {\emph {{Unsupervised identification of topological phase transitions
  using predictive models}}},\ }\href {\doibase 10.1088/1367-2630/ab7771}
  {\bibfield  {journal} {\bibinfo  {journal} {New Journal of Physics}\ }\textbf
  {\bibinfo {volume} {22}},\ \bibinfo {pages} {045003} (\bibinfo {year}
  {2020})}\BibitemShut {NoStop}%
\bibitem [{\citenamefont {Lidiak}\ and\ \citenamefont
  {Gong}(2020)}]{Lidiak20PRL}%
  \BibitemOpen
  \bibfield  {author} {\bibinfo {author} {\bibfnamefont {A.}~\bibnamefont
  {Lidiak}}\ and\ \bibinfo {author} {\bibfnamefont {Z.}~\bibnamefont {Gong}},\
  }\bibfield  {title} {\bibinfo {title} {\emph {Unsupervised Machine Learning
  of Quantum Phase Transitions Using Diffusion Maps}},\ }\href {\doibase
  10.1103/PhysRevLett.125.225701} {\bibfield  {journal} {\bibinfo  {journal}
  {Phys. Rev. Lett.}\ }\textbf {\bibinfo {volume} {125}},\ \bibinfo {pages}
  {225701} (\bibinfo {year} {2020})}\BibitemShut {NoStop}%
\bibitem [{\citenamefont {Arnold}\ \emph {et~al.}(2020)\citenamefont {Arnold},
  \citenamefont {Sch}, \citenamefont {Zonda},\ and\ \citenamefont
  {Lode}}]{Arnold20arXiv}%
  \BibitemOpen
  \bibfield  {author} {\bibinfo {author} {\bibfnamefont {J.}~\bibnamefont
  {Arnold}}, \bibinfo {author} {\bibfnamefont {F.}~\bibnamefont {Sch}},
  \bibinfo {author} {\bibfnamefont {M.}~\bibnamefont {Zonda}}, \ and\ \bibinfo
  {author} {\bibfnamefont {A.~U.~J.}\ \bibnamefont {Lode}},\ }\bibfield
  {title} {\bibinfo {title} {\emph {{Interpretable and unsupervised phase
  classification}}},\ }\href@noop {} {\ ,\ \bibinfo {pages} {1} (\bibinfo
  {year} {2020})},\ \Eprint {http://arxiv.org/abs/arXiv:2010.04730v1}
  {arXiv:arXiv:2010.04730v1} \BibitemShut {NoStop}%
\bibitem [{\citenamefont {Casert}\ \emph {et~al.}(2020)\citenamefont {Casert},
  \citenamefont {Mills}, \citenamefont {Vieijra}, \citenamefont {Ryckebusch},\
  and\ \citenamefont {Tamblyn}}]{Casert20arXiv}%
  \BibitemOpen
  \bibfield  {author} {\bibinfo {author} {\bibfnamefont {C.}~\bibnamefont
  {Casert}}, \bibinfo {author} {\bibfnamefont {K.}~\bibnamefont {Mills}},
  \bibinfo {author} {\bibfnamefont {T.}~\bibnamefont {Vieijra}}, \bibinfo
  {author} {\bibfnamefont {J.}~\bibnamefont {Ryckebusch}}, \ and\ \bibinfo
  {author} {\bibfnamefont {I.}~\bibnamefont {Tamblyn}},\ }\bibfield  {title}
  {\bibinfo {title} {\emph {Optical lattice experiments at unobserved
  conditions and scales through generative adversarial deep learning}},\ }\href
  {https://arxiv.org/abs/2002.07055} {\bibfield  {journal} {\bibinfo  {journal}
  {arXiv:2002.07055}\ } (\bibinfo {year} {2020})}\BibitemShut {NoStop}%
\bibitem [{\citenamefont {Lu}\ \emph {et~al.}(2020)\citenamefont {Lu},
  \citenamefont {Kim},\ and\ \citenamefont {Solja\ifmmode \check{c}\else
  \v{c}\fi{}i\ifmmode~\acute{c}\else \'{c}\fi{}}}]{Lu20PRX}%
  \BibitemOpen
  \bibfield  {author} {\bibinfo {author} {\bibfnamefont {P.~Y.}\ \bibnamefont
  {Lu}}, \bibinfo {author} {\bibfnamefont {S.}~\bibnamefont {Kim}}, \ and\
  \bibinfo {author} {\bibfnamefont {M.}~\bibnamefont {Solja\ifmmode
  \check{c}\else \v{c}\fi{}i\ifmmode~\acute{c}\else \'{c}\fi{}}},\ }\bibfield
  {title} {\bibinfo {title} {\emph {Extracting Interpretable Physical
  Parameters from Spatiotemporal Systems Using Unsupervised Learning}},\ }\href
  {\doibase 10.1103/PhysRevX.10.031056} {\bibfield  {journal} {\bibinfo
  {journal} {Phys. Rev. X}\ }\textbf {\bibinfo {volume} {10}},\ \bibinfo
  {pages} {031056} (\bibinfo {year} {2020})}\BibitemShut {NoStop}%
\bibitem [{\citenamefont {Iten}\ \emph {et~al.}(2020)\citenamefont {Iten},
  \citenamefont {Metger}, \citenamefont {Wilming}, \citenamefont {del Rio},\
  and\ \citenamefont {Renner}}]{Iten20PRL}%
  \BibitemOpen
  \bibfield  {author} {\bibinfo {author} {\bibfnamefont {R.}~\bibnamefont
  {Iten}}, \bibinfo {author} {\bibfnamefont {T.}~\bibnamefont {Metger}},
  \bibinfo {author} {\bibfnamefont {H.}~\bibnamefont {Wilming}}, \bibinfo
  {author} {\bibfnamefont {L.}~\bibnamefont {del Rio}}, \ and\ \bibinfo
  {author} {\bibfnamefont {R.}~\bibnamefont {Renner}},\ }\bibfield  {title}
  {\bibinfo {title} {\emph {Discovering Physical Concepts with Neural
  Networks}},\ }\href {\doibase 10.1103/PhysRevLett.124.010508} {\bibfield
  {journal} {\bibinfo  {journal} {Phys. Rev. Lett.}\ }\textbf {\bibinfo
  {volume} {124}},\ \bibinfo {pages} {010508} (\bibinfo {year}
  {2020})}\BibitemShut {NoStop}%
\bibitem [{\citenamefont {Ponte}\ and\ \citenamefont
  {Melko}(2017)}]{Ponte17PRB}%
  \BibitemOpen
  \bibfield  {author} {\bibinfo {author} {\bibfnamefont {P.}~\bibnamefont
  {Ponte}}\ and\ \bibinfo {author} {\bibfnamefont {R.~G.}\ \bibnamefont
  {Melko}},\ }\bibfield  {title} {\bibinfo {title} {\emph {{Kernel methods for
  interpretable machine learning of order parameters}}},\ }\href {\doibase
  10.1103/PhysRevB.96.205146} {\bibfield  {journal} {\bibinfo  {journal}
  {Physical Review B}\ }\textbf {\bibinfo {volume} {96}},\ \bibinfo {pages}
  {205146} (\bibinfo {year} {2017})},\ \Eprint
  {http://arxiv.org/abs/1704.05848} {arXiv:1704.05848} \BibitemShut {NoStop}%
\bibitem [{\citenamefont {Greitemann}\ \emph {et~al.}(2019)\citenamefont
  {Greitemann}, \citenamefont {Liu},\ and\ \citenamefont
  {Pollet}}]{Greitemann19PRB}%
  \BibitemOpen
  \bibfield  {author} {\bibinfo {author} {\bibfnamefont {J.}~\bibnamefont
  {Greitemann}}, \bibinfo {author} {\bibfnamefont {K.}~\bibnamefont {Liu}}, \
  and\ \bibinfo {author} {\bibfnamefont {L.}~\bibnamefont {Pollet}},\
  }\bibfield  {title} {\bibinfo {title} {\emph {{Probing hidden spin order with
  interpretable machine learning}}},\ }\href {\doibase
  10.1103/PhysRevB.99.060404} {\bibfield  {journal} {\bibinfo  {journal}
  {Physical Review B}\ }\textbf {\bibinfo {volume} {99}},\ \bibinfo {pages}
  {060404(R)} (\bibinfo {year} {2019})},\ \Eprint
  {http://arxiv.org/abs/1804.08557} {arXiv:1804.08557} \BibitemShut {NoStop}%
\bibitem [{\citenamefont {Dawid}\ \emph {et~al.}(2020)\citenamefont {Dawid},
  \citenamefont {Huembeli}, \citenamefont {Tomza}, \citenamefont {Lewenstein},\
  and\ \citenamefont {Dauphin}}]{Dawid20NJP}%
  \BibitemOpen
  \bibfield  {author} {\bibinfo {author} {\bibfnamefont {A.}~\bibnamefont
  {Dawid}}, \bibinfo {author} {\bibfnamefont {P.}~\bibnamefont {Huembeli}},
  \bibinfo {author} {\bibfnamefont {M.}~\bibnamefont {Tomza}}, \bibinfo
  {author} {\bibfnamefont {M.}~\bibnamefont {Lewenstein}}, \ and\ \bibinfo
  {author} {\bibfnamefont {A.}~\bibnamefont {Dauphin}},\ }\bibfield  {title}
  {\bibinfo {title} {\emph {Phase Detection with Neural Networks: Interpreting
  the Black Box}},\ }\href {\doibase 10.1088/1367-2630/abc463} {\bibfield
  {journal} {\bibinfo  {journal} {New J. Phys.}\ }\textbf {\bibinfo {volume}
  {22}},\ \bibinfo {pages} {115001} (\bibinfo {year} {2020})},\ \Eprint
  {http://arxiv.org/abs/2004.04711} {2004.04711} \BibitemShut {NoStop}%
\bibitem [{\citenamefont {Zhang}\ \emph {et~al.}(2020)\citenamefont {Zhang},
  \citenamefont {Ginsparg},\ and\ \citenamefont {Kim}}]{Zhang20PRR}%
  \BibitemOpen
  \bibfield  {author} {\bibinfo {author} {\bibfnamefont {Y.}~\bibnamefont
  {Zhang}}, \bibinfo {author} {\bibfnamefont {P.}~\bibnamefont {Ginsparg}}, \
  and\ \bibinfo {author} {\bibfnamefont {E.-A.}\ \bibnamefont {Kim}},\
  }\bibfield  {title} {\bibinfo {title} {\emph {{Interpreting machine learning
  of topological quantum phase transitions}}},\ }\href {\doibase
  10.1103/physrevresearch.2.023283} {\bibfield  {journal} {\bibinfo  {journal}
  {Physical Review Research}\ }\textbf {\bibinfo {volume} {2}},\ \bibinfo
  {pages} {23283} (\bibinfo {year} {2020})},\ \Eprint
  {http://arxiv.org/abs/1912.10057} {arXiv:1912.10057} \BibitemShut {NoStop}%
\bibitem [{\citenamefont {Wetzel}\ \emph {et~al.}(2020)\citenamefont {Wetzel},
  \citenamefont {Melko}, \citenamefont {Scott}, \citenamefont {Panju},\ and\
  \citenamefont {Ganesh}}]{Wetzel20PRR}%
  \BibitemOpen
  \bibfield  {author} {\bibinfo {author} {\bibfnamefont {S.~J.}\ \bibnamefont
  {Wetzel}}, \bibinfo {author} {\bibfnamefont {R.~G.}\ \bibnamefont {Melko}},
  \bibinfo {author} {\bibfnamefont {J.}~\bibnamefont {Scott}}, \bibinfo
  {author} {\bibfnamefont {M.}~\bibnamefont {Panju}}, \ and\ \bibinfo {author}
  {\bibfnamefont {V.}~\bibnamefont {Ganesh}},\ }\bibfield  {title} {\bibinfo
  {title} {\emph {{Discovering symmetry invariants and conserved quantities by
  interpreting siamese neural networks}}},\ }\href {\doibase
  10.1103/physrevresearch.2.033499} {\bibfield  {journal} {\bibinfo  {journal}
  {Physical Review Research}\ }\textbf {\bibinfo {volume} {2}},\ \bibinfo
  {pages} {033499} (\bibinfo {year} {2020})},\ \Eprint
  {http://arxiv.org/abs/2003.04299} {arXiv:2003.04299} \BibitemShut {NoStop}%
\bibitem [{\citenamefont {Lewenstein}\ \emph {et~al.}(2012)\citenamefont
  {Lewenstein}, \citenamefont {Sanpera},\ and\ \citenamefont
  {Ahufinger}}]{Lewenstein12Oxford}%
  \BibitemOpen
  \bibfield  {author} {\bibinfo {author} {\bibfnamefont {M.}~\bibnamefont
  {Lewenstein}}, \bibinfo {author} {\bibfnamefont {A.}~\bibnamefont {Sanpera}},
  \ and\ \bibinfo {author} {\bibfnamefont {V.}~\bibnamefont {Ahufinger}},\
  }\href@noop {} {\emph {\bibinfo {title} {{Ultracold atoms in optical lattice:
  simulating quantum many-body systems}}}}\ (\bibinfo  {publisher} {Oxford
  Univ. Press},\ \bibinfo {year} {2012})\BibitemShut {NoStop}%
\bibitem [{\citenamefont {Dalibard}\ \emph {et~al.}(2011)\citenamefont
  {Dalibard}, \citenamefont {Gerbier}, \citenamefont {Juzeliunas},\ and\
  \citenamefont {{\"{O}}hberg}}]{Dalibard11RMP}%
  \BibitemOpen
  \bibfield  {author} {\bibinfo {author} {\bibfnamefont {J.}~\bibnamefont
  {Dalibard}}, \bibinfo {author} {\bibfnamefont {F.}~\bibnamefont {Gerbier}},
  \bibinfo {author} {\bibfnamefont {G.}~\bibnamefont {Juzeliunas}}, \ and\
  \bibinfo {author} {\bibfnamefont {P.}~\bibnamefont {{\"{O}}hberg}},\
  }\bibfield  {title} {\bibinfo {title} {\emph {{Colloquium: Artificial gauge
  potentials for neutral atoms}}},\ }\href@noop {} {\bibfield  {journal}
  {\bibinfo  {journal} {Rev. Mod. Phys.}\ }\textbf {\bibinfo {volume} {83}},\
  \bibinfo {pages} {1523} (\bibinfo {year} {2011})}\BibitemShut {NoStop}%
\bibitem [{\citenamefont {Cooper}\ \emph
  {et~al.}(2019{\natexlab{a}})\citenamefont {Cooper}, \citenamefont
  {Dalibard},\ and\ \citenamefont {Spielman}}]{Cooper19RMP}%
  \BibitemOpen
  \bibfield  {author} {\bibinfo {author} {\bibfnamefont {N.~R.}\ \bibnamefont
  {Cooper}}, \bibinfo {author} {\bibfnamefont {J.}~\bibnamefont {Dalibard}}, \
  and\ \bibinfo {author} {\bibfnamefont {I.~B.}\ \bibnamefont {Spielman}},\
  }\bibfield  {title} {\bibinfo {title} {\emph {{Topological bands for
  ultracold atoms}}},\ }\href {\doibase 10.1103/RevModPhys.91.015005}
  {\bibfield  {journal} {\bibinfo  {journal} {Reviews of Modern Physics}\
  }\textbf {\bibinfo {volume} {91}},\ \bibinfo {pages} {015005} (\bibinfo
  {year} {2019}{\natexlab{a}})},\ \Eprint {http://arxiv.org/abs/1803.00249}
  {arXiv:1803.00249} \BibitemShut {NoStop}%
\bibitem [{\citenamefont {Bukov}\ \emph {et~al.}(2015)\citenamefont {Bukov},
  \citenamefont {D'Alessio},\ and\ \citenamefont
  {Polkovnikov}}]{Bukov15AdvPhys}%
  \BibitemOpen
  \bibfield  {author} {\bibinfo {author} {\bibfnamefont {M.}~\bibnamefont
  {Bukov}}, \bibinfo {author} {\bibfnamefont {L.}~\bibnamefont {D'Alessio}}, \
  and\ \bibinfo {author} {\bibfnamefont {A.}~\bibnamefont {Polkovnikov}},\
  }\bibfield  {title} {\bibinfo {title} {\emph {{Universal High-Frequency
  Behavior of Periodically Driven Systems: from Dynamical Stabilization to
  Floquet Engineering}}},\ }\href {\doibase 10.1080/00018732.2015.1055918}
  {\bibfield  {journal} {\bibinfo  {journal} {Advances in Physics}\ }\textbf
  {\bibinfo {volume} {64}},\ \bibinfo {pages} {139} (\bibinfo {year}
  {2015})}\BibitemShut {NoStop}%
\bibitem [{\citenamefont {Eckardt}(2017)}]{Eckardt17RMP}%
  \BibitemOpen
  \bibfield  {author} {\bibinfo {author} {\bibfnamefont {A.}~\bibnamefont
  {Eckardt}},\ }\bibfield  {title} {\bibinfo {title} {\emph {{Colloquium:
  Atomic quantum gases in periodically driven optical lattices}}},\ }\href
  {\doibase 10.1103/RevModPhys.89.011004} {\bibfield  {journal} {\bibinfo
  {journal} {Rev. Mod. Phys.}\ }\textbf {\bibinfo {volume} {89}},\ \bibinfo
  {pages} {011004} (\bibinfo {year} {2017})}\BibitemShut {NoStop}%
\bibitem [{\citenamefont {Cooper}\ \emph
  {et~al.}(2019{\natexlab{b}})\citenamefont {Cooper}, \citenamefont
  {Dalibard},\ and\ \citenamefont {Spielman}}]{Cooper19}%
  \BibitemOpen
  \bibfield  {author} {\bibinfo {author} {\bibfnamefont {N.~R.}\ \bibnamefont
  {Cooper}}, \bibinfo {author} {\bibfnamefont {J.}~\bibnamefont {Dalibard}}, \
  and\ \bibinfo {author} {\bibfnamefont {I.~B.}\ \bibnamefont {Spielman}},\
  }\bibfield  {title} {\bibinfo {title} {\emph {Topological bands for ultracold
  atoms}},\ }\href {\doibase 10.1103/RevModPhys.91.015005} {\bibfield
  {journal} {\bibinfo  {journal} {Rev. Mod. Phys.}\ }\textbf {\bibinfo {volume}
  {91}},\ \bibinfo {pages} {015005} (\bibinfo {year}
  {2019}{\natexlab{b}})}\BibitemShut {NoStop}%
\bibitem [{\citenamefont {Zhang}\ and\ \citenamefont {Kim}(2017)}]{Zhang17PRL}%
  \BibitemOpen
  \bibfield  {author} {\bibinfo {author} {\bibfnamefont {Y.}~\bibnamefont
  {Zhang}}\ and\ \bibinfo {author} {\bibfnamefont {E.-A.}\ \bibnamefont
  {Kim}},\ }\bibfield  {title} {\bibinfo {title} {\emph {Quantum Loop
  Topography for Machine Learning}},\ }\href
  {https://link.aps.org/doi/10.1103/PhysRevLett.118.216401} {\bibfield
  {journal} {\bibinfo  {journal} {Phys. Rev. Lett.}\ }\textbf {\bibinfo
  {volume} {118}},\ \bibinfo {pages} {216401} (\bibinfo {year}
  {2017})}\BibitemShut {NoStop}%
\bibitem [{\citenamefont {Deng}\ \emph {et~al.}(2017)\citenamefont {Deng},
  \citenamefont {Li},\ and\ \citenamefont {Das~Sarma}}]{Deng17PRB}%
  \BibitemOpen
  \bibfield  {author} {\bibinfo {author} {\bibfnamefont {D.-L.}\ \bibnamefont
  {Deng}}, \bibinfo {author} {\bibfnamefont {X.}~\bibnamefont {Li}}, \ and\
  \bibinfo {author} {\bibfnamefont {S.}~\bibnamefont {Das~Sarma}},\ }\bibfield
  {title} {\bibinfo {title} {\emph {Machine learning topological states}},\
  }\href {\doibase 10.1103/PhysRevB.96.195145} {\bibfield  {journal} {\bibinfo
  {journal} {Phys. Rev. B}\ }\textbf {\bibinfo {volume} {96}},\ \bibinfo
  {pages} {195145} (\bibinfo {year} {2017})}\BibitemShut {NoStop}%
\bibitem [{\citenamefont {Zhang}\ \emph {et~al.}(2018)\citenamefont {Zhang},
  \citenamefont {Shen},\ and\ \citenamefont {Zhai}}]{Zhang18PRL}%
  \BibitemOpen
  \bibfield  {author} {\bibinfo {author} {\bibfnamefont {P.}~\bibnamefont
  {Zhang}}, \bibinfo {author} {\bibfnamefont {H.}~\bibnamefont {Shen}}, \ and\
  \bibinfo {author} {\bibfnamefont {H.}~\bibnamefont {Zhai}},\ }\bibfield
  {title} {\bibinfo {title} {\emph {Machine Learning Topological Invariants
  with Neural Networks}},\ }\href {\doibase 10.1103/PhysRevLett.120.066401}
  {\bibfield  {journal} {\bibinfo  {journal} {Phys. Rev. Lett.}\ }\textbf
  {\bibinfo {volume} {120}},\ \bibinfo {pages} {066401} (\bibinfo {year}
  {2018})}\BibitemShut {NoStop}%
\bibitem [{\citenamefont {Carvalho}\ \emph {et~al.}(2018)\citenamefont
  {Carvalho}, \citenamefont {Garc{\'{i}}a-Mart{\'{i}}nez}, \citenamefont
  {Lado},\ and\ \citenamefont {Fern{\'{a}}ndez-Rossier}}]{Carvalho18PRB}%
  \BibitemOpen
  \bibfield  {author} {\bibinfo {author} {\bibfnamefont {D.}~\bibnamefont
  {Carvalho}}, \bibinfo {author} {\bibfnamefont {N.~A.}\ \bibnamefont
  {Garc{\'{i}}a-Mart{\'{i}}nez}}, \bibinfo {author} {\bibfnamefont {J.~L.}\
  \bibnamefont {Lado}}, \ and\ \bibinfo {author} {\bibfnamefont
  {J.}~\bibnamefont {Fern{\'{a}}ndez-Rossier}},\ }\bibfield  {title} {\bibinfo
  {title} {\emph {{Real-space mapping of topological invariants using
  artificial neural networks}}},\ }\href {\doibase 10.1103/PhysRevB.97.115453}
  {\bibfield  {journal} {\bibinfo  {journal} {Physical Review B}\ }\textbf
  {\bibinfo {volume} {97}},\ \bibinfo {pages} {115453} (\bibinfo {year}
  {2018})},\ \Eprint {http://arxiv.org/abs/1801.09655} {arXiv:1801.09655}
  \BibitemShut {NoStop}%
\bibitem [{\citenamefont {Beach}\ \emph {et~al.}(2018)\citenamefont {Beach},
  \citenamefont {Golubeva},\ and\ \citenamefont {Melko}}]{Beach2018PRB}%
  \BibitemOpen
  \bibfield  {author} {\bibinfo {author} {\bibfnamefont {M.~J.~S.}\
  \bibnamefont {Beach}}, \bibinfo {author} {\bibfnamefont {A.}~\bibnamefont
  {Golubeva}}, \ and\ \bibinfo {author} {\bibfnamefont {R.~G.}\ \bibnamefont
  {Melko}},\ }\bibfield  {title} {\bibinfo {title} {\emph {Machine learning
  vortices at the Kosterlitz-Thouless transition}},\ }\href {\doibase
  10.1103/PhysRevB.97.045207} {\bibfield  {journal} {\bibinfo  {journal} {Phys.
  Rev. B}\ }\textbf {\bibinfo {volume} {97}},\ \bibinfo {pages} {045207}
  (\bibinfo {year} {2018})}\BibitemShut {NoStop}%
\bibitem [{\citenamefont {Rodriguez-Nieva}\ and\ \citenamefont
  {Scheurer}(2019)}]{Rodriguez-Nieva19NatPhys}%
  \BibitemOpen
  \bibfield  {author} {\bibinfo {author} {\bibfnamefont {J.~F.}\ \bibnamefont
  {Rodriguez-Nieva}}\ and\ \bibinfo {author} {\bibfnamefont {M.~S.}\
  \bibnamefont {Scheurer}},\ }\bibfield  {title} {\bibinfo {title} {\emph
  {{Identifying topological order through unsupervised machine learning}}},\
  }\href {\doibase 10.1038/s41567-019-0512-x} {\bibfield  {journal} {\bibinfo
  {journal} {Nat. Phys.}\ }\textbf {\bibinfo {volume} {15}},\ \bibinfo {pages}
  {790} (\bibinfo {year} {2019})},\ \Eprint {http://arxiv.org/abs/1805.05961}
  {arXiv:1805.05961} \BibitemShut {NoStop}%
\bibitem [{\citenamefont {Holanda}\ and\ \citenamefont
  {Griffith}(2020)}]{Holanda20PRB}%
  \BibitemOpen
  \bibfield  {author} {\bibinfo {author} {\bibfnamefont {N.~L.}\ \bibnamefont
  {Holanda}}\ and\ \bibinfo {author} {\bibfnamefont {M.~A.}\ \bibnamefont
  {Griffith}},\ }\bibfield  {title} {\bibinfo {title} {\emph {{Machine learning
  topological phases in real space}}},\ }\href {\doibase
  10.1103/PhysRevB.102.054107} {\bibfield  {journal} {\bibinfo  {journal}
  {Physical Review B}\ }\textbf {\bibinfo {volume} {102}},\ \bibinfo {pages}
  {054107} (\bibinfo {year} {2020})},\ \Eprint
  {http://arxiv.org/abs/1901.01963} {arXiv:1901.01963} \BibitemShut {NoStop}%
\bibitem [{\citenamefont {Long}\ \emph {et~al.}(2020)\citenamefont {Long},
  \citenamefont {Ren},\ and\ \citenamefont {Chen}}]{Long20PRL}%
  \BibitemOpen
  \bibfield  {author} {\bibinfo {author} {\bibfnamefont {Y.}~\bibnamefont
  {Long}}, \bibinfo {author} {\bibfnamefont {J.}~\bibnamefont {Ren}}, \ and\
  \bibinfo {author} {\bibfnamefont {H.}~\bibnamefont {Chen}},\ }\bibfield
  {title} {\bibinfo {title} {\emph {Unsupervised Manifold Clustering of
  Topological Phononics}},\ }\href {\doibase 10.1103/PhysRevLett.124.185501}
  {\bibfield  {journal} {\bibinfo  {journal} {Phys. Rev. Lett.}\ }\textbf
  {\bibinfo {volume} {124}},\ \bibinfo {pages} {185501} (\bibinfo {year}
  {2020})}\BibitemShut {NoStop}%
\bibitem [{\citenamefont {Scheurer}\ and\ \citenamefont
  {Slager}(2020)}]{Scheurer20PRL}%
  \BibitemOpen
  \bibfield  {author} {\bibinfo {author} {\bibfnamefont {M.~S.}\ \bibnamefont
  {Scheurer}}\ and\ \bibinfo {author} {\bibfnamefont {R.-J.}\ \bibnamefont
  {Slager}},\ }\bibfield  {title} {\bibinfo {title} {\emph {Unsupervised
  Machine Learning and Band Topology}},\ }\href {\doibase
  10.1103/PhysRevLett.124.226401} {\bibfield  {journal} {\bibinfo  {journal}
  {Phys. Rev. Lett.}\ }\textbf {\bibinfo {volume} {124}},\ \bibinfo {pages}
  {226401} (\bibinfo {year} {2020})}\BibitemShut {NoStop}%
\bibitem [{\citenamefont {Price}\ and\ \citenamefont {Cooper}(2012)}]{price12}%
  \BibitemOpen
  \bibfield  {author} {\bibinfo {author} {\bibfnamefont {H.~M.}\ \bibnamefont
  {Price}}\ and\ \bibinfo {author} {\bibfnamefont {N.~R.}\ \bibnamefont
  {Cooper}},\ }\bibfield  {title} {\bibinfo {title} {\emph {Mapping the Berry
  curvature from semiclassical dynamics in optical lattices}},\ }\href
  {\doibase 10.1103/PhysRevA.85.033620} {\bibfield  {journal} {\bibinfo
  {journal} {Phys. Rev. A}\ }\textbf {\bibinfo {volume} {85}},\ \bibinfo
  {pages} {033620} (\bibinfo {year} {2012})}\BibitemShut {NoStop}%
\bibitem [{\citenamefont {Dauphin}\ and\ \citenamefont
  {Goldman}(2013)}]{dauphin13}%
  \BibitemOpen
  \bibfield  {author} {\bibinfo {author} {\bibfnamefont {A.}~\bibnamefont
  {Dauphin}}\ and\ \bibinfo {author} {\bibfnamefont {N.}~\bibnamefont
  {Goldman}},\ }\bibfield  {title} {\bibinfo {title} {\emph {Extracting the
  Chern Number from the Dynamics of a Fermi Gas: Implementing a Quantum Hall
  Bar for Cold Atoms}},\ }\href {\doibase 10.1103/PhysRevLett.111.135302}
  {\bibfield  {journal} {\bibinfo  {journal} {Phys. Rev. Lett.}\ }\textbf
  {\bibinfo {volume} {111}},\ \bibinfo {pages} {135302} (\bibinfo {year}
  {2013})}\BibitemShut {NoStop}%
\bibitem [{\citenamefont {Jotzu}\ \emph {et~al.}(2014)\citenamefont {Jotzu},
  \citenamefont {Messer}, \citenamefont {Desbuquois}, \citenamefont {Lebrat},
  \citenamefont {Uehlinger}, \citenamefont {Greif},\ and\ \citenamefont
  {Esslinger}}]{Jotzu14Nature}%
  \BibitemOpen
  \bibfield  {author} {\bibinfo {author} {\bibfnamefont {G.}~\bibnamefont
  {Jotzu}}, \bibinfo {author} {\bibfnamefont {M.}~\bibnamefont {Messer}},
  \bibinfo {author} {\bibfnamefont {R.}~\bibnamefont {Desbuquois}}, \bibinfo
  {author} {\bibfnamefont {M.}~\bibnamefont {Lebrat}}, \bibinfo {author}
  {\bibfnamefont {T.}~\bibnamefont {Uehlinger}}, \bibinfo {author}
  {\bibfnamefont {D.}~\bibnamefont {Greif}}, \ and\ \bibinfo {author}
  {\bibfnamefont {T.}~\bibnamefont {Esslinger}},\ }\bibfield  {title} {\bibinfo
  {title} {\emph {{Experimental realisation of the topological Haldane model
  with ultracold fermions}}},\ }\href {\doibase 10.1038/nature13915} {\bibfield
   {journal} {\bibinfo  {journal} {Nature}\ }\textbf {\bibinfo {volume}
  {515}},\ \bibinfo {pages} {237} (\bibinfo {year} {2014})}\BibitemShut
  {NoStop}%
\bibitem [{\citenamefont {Aidelsburger}\ \emph {et~al.}(2015)\citenamefont
  {Aidelsburger}, \citenamefont {Lohse}, \citenamefont {Schweizer},
  \citenamefont {Atala}, \citenamefont {Barreiro}, \citenamefont
  {Nascimb{\`{e}}ne}, \citenamefont {Cooper}, \citenamefont {Bloch},\ and\
  \citenamefont {Goldman}}]{Aidelsburger15NatPhys}%
  \BibitemOpen
  \bibfield  {author} {\bibinfo {author} {\bibfnamefont {M.}~\bibnamefont
  {Aidelsburger}}, \bibinfo {author} {\bibfnamefont {M.}~\bibnamefont {Lohse}},
  \bibinfo {author} {\bibfnamefont {C.}~\bibnamefont {Schweizer}}, \bibinfo
  {author} {\bibfnamefont {M.}~\bibnamefont {Atala}}, \bibinfo {author}
  {\bibfnamefont {J.~T.}\ \bibnamefont {Barreiro}}, \bibinfo {author}
  {\bibfnamefont {S.}~\bibnamefont {Nascimb{\`{e}}ne}}, \bibinfo {author}
  {\bibfnamefont {N.~R.}\ \bibnamefont {Cooper}}, \bibinfo {author}
  {\bibfnamefont {I.}~\bibnamefont {Bloch}}, \ and\ \bibinfo {author}
  {\bibfnamefont {N.}~\bibnamefont {Goldman}},\ }\bibfield  {title} {\bibinfo
  {title} {\emph {{Measuring the Chern number of Hofstadter bands with
  ultracold bosonic atoms}}},\ }\href {\doibase 10.1038/nphys3171} {\bibfield
  {journal} {\bibinfo  {journal} {Nature Physics}\ }\textbf {\bibinfo {volume}
  {11}},\ \bibinfo {pages} {162} (\bibinfo {year} {2015})},\ \Eprint
  {http://arxiv.org/abs/1407.4205} {arXiv:1407.4205} \BibitemShut {NoStop}%
\bibitem [{\citenamefont {Duca}\ \emph {et~al.}(2015)\citenamefont {Duca},
  \citenamefont {Li}, \citenamefont {Reitter}, \citenamefont {Bloch},
  \citenamefont {Schleier-Smith},\ and\ \citenamefont
  {Schneider}}]{Duca15Science}%
  \BibitemOpen
  \bibfield  {author} {\bibinfo {author} {\bibfnamefont {L.}~\bibnamefont
  {Duca}}, \bibinfo {author} {\bibfnamefont {T.}~\bibnamefont {Li}}, \bibinfo
  {author} {\bibfnamefont {M.}~\bibnamefont {Reitter}}, \bibinfo {author}
  {\bibfnamefont {I.}~\bibnamefont {Bloch}}, \bibinfo {author} {\bibfnamefont
  {M.}~\bibnamefont {Schleier-Smith}}, \ and\ \bibinfo {author} {\bibfnamefont
  {U.}~\bibnamefont {Schneider}},\ }\bibfield  {title} {\bibinfo {title} {\emph
  {{An Aharonov-Bohm interferometer for determining Bloch band topology}}},\
  }\href {\doibase 10.1126/science.1259052} {\bibfield  {journal} {\bibinfo
  {journal} {Science}\ }\textbf {\bibinfo {volume} {347}},\ \bibinfo {pages}
  {288} (\bibinfo {year} {2015})},\ \Eprint {http://arxiv.org/abs/1407.5635}
  {arXiv:1407.5635} \BibitemShut {NoStop}%
\bibitem [{\citenamefont {Tran}\ \emph {et~al.}(2017)\citenamefont {Tran},
  \citenamefont {Dauphin}, \citenamefont {Grushin}, \citenamefont {Zoller},\
  and\ \citenamefont {Goldman}}]{Tran17SciAdv}%
  \BibitemOpen
  \bibfield  {author} {\bibinfo {author} {\bibfnamefont {D.~T.}\ \bibnamefont
  {Tran}}, \bibinfo {author} {\bibfnamefont {A.}~\bibnamefont {Dauphin}},
  \bibinfo {author} {\bibfnamefont {A.~G.}\ \bibnamefont {Grushin}}, \bibinfo
  {author} {\bibfnamefont {P.}~\bibnamefont {Zoller}}, \ and\ \bibinfo {author}
  {\bibfnamefont {N.}~\bibnamefont {Goldman}},\ }\bibfield  {title} {\bibinfo
  {title} {\emph {{Probing topology by "heating": Quantized circular dichroism
  in ultracold atoms}}},\ }\href {\doibase 10.1126/sciadv.1701207} {\bibfield
  {journal} {\bibinfo  {journal} {Science Advances}\ }\textbf {\bibinfo
  {volume} {3}} (\bibinfo {year} {2017}),\ 10.1126/sciadv.1701207},\ \Eprint
  {http://arxiv.org/abs/1704.01990} {arXiv:1704.01990} \BibitemShut {NoStop}%
\bibitem [{\citenamefont {Asteria}\ \emph {et~al.}(2019)\citenamefont
  {Asteria}, \citenamefont {Tran}, \citenamefont {Ozawa}, \citenamefont
  {Tarnowski}, \citenamefont {Rem}, \citenamefont {Fl{\"{a}}schner},
  \citenamefont {Sengstock}, \citenamefont {Goldman},\ and\ \citenamefont
  {Weitenberg}}]{Asteria19NatPhys}%
  \BibitemOpen
  \bibfield  {author} {\bibinfo {author} {\bibfnamefont {L.}~\bibnamefont
  {Asteria}}, \bibinfo {author} {\bibfnamefont {D.~T.}\ \bibnamefont {Tran}},
  \bibinfo {author} {\bibfnamefont {T.}~\bibnamefont {Ozawa}}, \bibinfo
  {author} {\bibfnamefont {M.}~\bibnamefont {Tarnowski}}, \bibinfo {author}
  {\bibfnamefont {B.~S.}\ \bibnamefont {Rem}}, \bibinfo {author} {\bibfnamefont
  {N.}~\bibnamefont {Fl{\"{a}}schner}}, \bibinfo {author} {\bibfnamefont
  {K.}~\bibnamefont {Sengstock}}, \bibinfo {author} {\bibfnamefont
  {N.}~\bibnamefont {Goldman}}, \ and\ \bibinfo {author} {\bibfnamefont
  {C.}~\bibnamefont {Weitenberg}},\ }\bibfield  {title} {\bibinfo {title}
  {\emph {{Measuring quantized circular dichroism in ultracold topological
  matter}}},\ }\href {http://arxiv.org/abs/1805.11077} {\bibfield  {journal}
  {\bibinfo  {journal} {Nature Physics}\ }\textbf {\bibinfo {volume} {15}},\
  \bibinfo {pages} {449} (\bibinfo {year} {2019})},\ \Eprint
  {http://arxiv.org/abs/1805.11077} {arXiv:1805.11077} \BibitemShut {NoStop}%
\bibitem [{\citenamefont {Alba}\ \emph {et~al.}(2011)\citenamefont {Alba},
  \citenamefont {Fernandez-Gonzalvo}, \citenamefont {Mur-Petit}, \citenamefont
  {Pachos},\ and\ \citenamefont {Garcia-Ripoll}}]{Alba11PRL}%
  \BibitemOpen
  \bibfield  {author} {\bibinfo {author} {\bibfnamefont {E.}~\bibnamefont
  {Alba}}, \bibinfo {author} {\bibfnamefont {X.}~\bibnamefont
  {Fernandez-Gonzalvo}}, \bibinfo {author} {\bibfnamefont {J.}~\bibnamefont
  {Mur-Petit}}, \bibinfo {author} {\bibfnamefont {J.~K.}\ \bibnamefont
  {Pachos}}, \ and\ \bibinfo {author} {\bibfnamefont {J.~J.}\ \bibnamefont
  {Garcia-Ripoll}},\ }\bibfield  {title} {\bibinfo {title} {\emph {{Seeing
  topological order in time-of-flight measurements}}},\ }\href {\doibase
  10.1103/PhysRevLett.107.235301} {\bibfield  {journal} {\bibinfo  {journal}
  {Phys. Rev. Lett.}\ }\textbf {\bibinfo {volume} {107}},\ \bibinfo {pages}
  {235301} (\bibinfo {year} {2011})}\BibitemShut {NoStop}%
\bibitem [{\citenamefont {Hauke}\ \emph {et~al.}(2014)\citenamefont {Hauke},
  \citenamefont {Lewenstein},\ and\ \citenamefont {Eckardt}}]{Hauke14PRL}%
  \BibitemOpen
  \bibfield  {author} {\bibinfo {author} {\bibfnamefont {P.}~\bibnamefont
  {Hauke}}, \bibinfo {author} {\bibfnamefont {M.}~\bibnamefont {Lewenstein}}, \
  and\ \bibinfo {author} {\bibfnamefont {A.}~\bibnamefont {Eckardt}},\
  }\bibfield  {title} {\bibinfo {title} {\emph {{Tomography of band insulators
  from quench dynamics}}},\ }\href {\doibase 10.1103/PhysRevLett.113.045303}
  {\bibfield  {journal} {\bibinfo  {journal} {Phys. Rev. Lett.}\ }\textbf
  {\bibinfo {volume} {113}},\ \bibinfo {pages} {045303} (\bibinfo {year}
  {2014})}\BibitemShut {NoStop}%
\bibitem [{\citenamefont {Fl{\"{a}}schner}\ \emph {et~al.}(2016)\citenamefont
  {Fl{\"{a}}schner}, \citenamefont {Rem}, \citenamefont {Tarnowski},
  \citenamefont {Vogel}, \citenamefont {L{\"{u}}hmann}, \citenamefont
  {Sengstock},\ and\ \citenamefont {Weitenberg}}]{Flaschner16Science}%
  \BibitemOpen
  \bibfield  {author} {\bibinfo {author} {\bibfnamefont {N.}~\bibnamefont
  {Fl{\"{a}}schner}}, \bibinfo {author} {\bibfnamefont {B.~S.}\ \bibnamefont
  {Rem}}, \bibinfo {author} {\bibfnamefont {M.}~\bibnamefont {Tarnowski}},
  \bibinfo {author} {\bibfnamefont {D.}~\bibnamefont {Vogel}}, \bibinfo
  {author} {\bibfnamefont {D.-S.}\ \bibnamefont {L{\"{u}}hmann}}, \bibinfo
  {author} {\bibfnamefont {K.}~\bibnamefont {Sengstock}}, \ and\ \bibinfo
  {author} {\bibfnamefont {C.}~\bibnamefont {Weitenberg}},\ }\bibfield  {title}
  {\bibinfo {title} {\emph {{Experimental reconstruction of the Berry curvature
  in a Floquet Bloch band}}},\ }\href {\doibase 10.1126/science.aad4568}
  {\bibfield  {journal} {\bibinfo  {journal} {Science}\ }\textbf {\bibinfo
  {volume} {352}},\ \bibinfo {pages} {1091} (\bibinfo {year}
  {2016})}\BibitemShut {NoStop}%
\bibitem [{\citenamefont {Fl{\"{a}}schner}\ \emph {et~al.}(2018)\citenamefont
  {Fl{\"{a}}schner}, \citenamefont {Vogel}, \citenamefont {Tarnowski},
  \citenamefont {Rem}, \citenamefont {L{\"{u}}hmann}, \citenamefont {Heyl},
  \citenamefont {Budich}, \citenamefont {Mathey}, \citenamefont {Sengstock},\
  and\ \citenamefont {Weitenberg}}]{Flaschner18NatPhys}%
  \BibitemOpen
  \bibfield  {author} {\bibinfo {author} {\bibfnamefont {N.}~\bibnamefont
  {Fl{\"{a}}schner}}, \bibinfo {author} {\bibfnamefont {D.}~\bibnamefont
  {Vogel}}, \bibinfo {author} {\bibfnamefont {M.}~\bibnamefont {Tarnowski}},
  \bibinfo {author} {\bibfnamefont {B.~S.}\ \bibnamefont {Rem}}, \bibinfo
  {author} {\bibfnamefont {D.~S.}\ \bibnamefont {L{\"{u}}hmann}}, \bibinfo
  {author} {\bibfnamefont {M.}~\bibnamefont {Heyl}}, \bibinfo {author}
  {\bibfnamefont {J.~C.}\ \bibnamefont {Budich}}, \bibinfo {author}
  {\bibfnamefont {L.}~\bibnamefont {Mathey}}, \bibinfo {author} {\bibfnamefont
  {K.}~\bibnamefont {Sengstock}}, \ and\ \bibinfo {author} {\bibfnamefont
  {C.}~\bibnamefont {Weitenberg}},\ }\bibfield  {title} {\bibinfo {title}
  {\emph {{Observation of dynamical vortices after quenches in a system with
  topology}}},\ }\href {\doibase 10.1038/s41567-017-0013-8} {\bibfield
  {journal} {\bibinfo  {journal} {Nature Physics}\ }\textbf {\bibinfo {volume}
  {14}},\ \bibinfo {pages} {265} (\bibinfo {year} {2018})}\BibitemShut
  {NoStop}%
\bibitem [{\citenamefont {Tarnowski}\ \emph {et~al.}(2019)\citenamefont
  {Tarnowski}, \citenamefont {{\"{U}}nal}, \citenamefont {Fl{\"{a}}schner},
  \citenamefont {Rem}, \citenamefont {Eckardt}, \citenamefont {Sengstock},\
  and\ \citenamefont {Weitenberg}}]{Tarnowski19NatComm}%
  \BibitemOpen
  \bibfield  {author} {\bibinfo {author} {\bibfnamefont {M.}~\bibnamefont
  {Tarnowski}}, \bibinfo {author} {\bibfnamefont {F.~N.}\ \bibnamefont
  {{\"{U}}nal}}, \bibinfo {author} {\bibfnamefont {N.}~\bibnamefont
  {Fl{\"{a}}schner}}, \bibinfo {author} {\bibfnamefont {B.~S.}\ \bibnamefont
  {Rem}}, \bibinfo {author} {\bibfnamefont {A.}~\bibnamefont {Eckardt}},
  \bibinfo {author} {\bibfnamefont {K.}~\bibnamefont {Sengstock}}, \ and\
  \bibinfo {author} {\bibfnamefont {C.}~\bibnamefont {Weitenberg}},\ }\bibfield
   {title} {\bibinfo {title} {\emph {{Measuring topology from dynamics by
  obtaining the Chern number from a linking number}}},\ }\href {\doibase
  10.1038/s41467-019-09668-y} {\bibfield  {journal} {\bibinfo  {journal}
  {Nature Communications}\ }\textbf {\bibinfo {volume} {10}},\ \bibinfo {pages}
  {1728} (\bibinfo {year} {2019})}\BibitemShut {NoStop}%
\bibitem [{\citenamefont {Haldane}(1988)}]{Haldane88PRL}%
  \BibitemOpen
  \bibfield  {author} {\bibinfo {author} {\bibfnamefont {F.~D.~M.}\
  \bibnamefont {Haldane}},\ }\bibfield  {title} {\bibinfo {title} {\emph {Model
  for a Quantum Hall Effect without Landau Levels: Condensed-Matter Realization
  of the "Parity Anomaly"}},\ }\href {\doibase 10.1103/PhysRevLett.61.2015}
  {\bibfield  {journal} {\bibinfo  {journal} {Phys. Rev. Lett.}\ }\textbf
  {\bibinfo {volume} {61}},\ \bibinfo {pages} {2015} (\bibinfo {year}
  {1988})}\BibitemShut {NoStop}%
\bibitem [{\citenamefont {Hu}\ \emph {et~al.}(2017)\citenamefont {Hu},
  \citenamefont {Singh},\ and\ \citenamefont {Scalettar}}]{Hu17}%
  \BibitemOpen
  \bibfield  {author} {\bibinfo {author} {\bibfnamefont {W.}~\bibnamefont
  {Hu}}, \bibinfo {author} {\bibfnamefont {R.~R.~P.}\ \bibnamefont {Singh}}, \
  and\ \bibinfo {author} {\bibfnamefont {R.~T.}\ \bibnamefont {Scalettar}},\
  }\bibfield  {title} {\bibinfo {title} {\emph {Discovering phases, phase
  transitions, and crossovers through unsupervised machine learning: A critical
  examination}},\ }\href {\doibase 10.1103/PhysRevE.95.062122} {\bibfield
  {journal} {\bibinfo  {journal} {Phys. Rev. E}\ }\textbf {\bibinfo {volume}
  {95}},\ \bibinfo {pages} {062122} (\bibinfo {year} {2017})}\BibitemShut
  {NoStop}%
\bibitem [{\citenamefont {Ming}\ \emph {et~al.}(2019)\citenamefont {Ming},
  \citenamefont {Lin}, \citenamefont {Bartlett},\ and\ \citenamefont
  {Zhang}}]{Ming19npj}%
  \BibitemOpen
  \bibfield  {author} {\bibinfo {author} {\bibfnamefont {Y.}~\bibnamefont
  {Ming}}, \bibinfo {author} {\bibfnamefont {C.~T.}\ \bibnamefont {Lin}},
  \bibinfo {author} {\bibfnamefont {S.~D.}\ \bibnamefont {Bartlett}}, \ and\
  \bibinfo {author} {\bibfnamefont {W.~W.}\ \bibnamefont {Zhang}},\ }\bibfield
  {title} {\bibinfo {title} {\emph {{Quantum topology identification with deep
  neural networks and quantum walks}}},\ }\href {\doibase
  10.1038/s41524-019-0224-x} {\bibfield  {journal} {\bibinfo  {journal} {npj
  Computational Materials}\ }\textbf {\bibinfo {volume} {5}},\ \bibinfo {pages}
  {88} (\bibinfo {year} {2019})},\ \Eprint {http://arxiv.org/abs/1811.12630}
  {arXiv:1811.12630} \BibitemShut {NoStop}%
\bibitem [{\citenamefont {Rosson}\ \emph {et~al.}(2020)\citenamefont {Rosson},
  \citenamefont {Kiffner}, \citenamefont {Mur-Petit},\ and\ \citenamefont
  {Jaksch}}]{Rosson20PRA}%
  \BibitemOpen
  \bibfield  {author} {\bibinfo {author} {\bibfnamefont {P.}~\bibnamefont
  {Rosson}}, \bibinfo {author} {\bibfnamefont {M.}~\bibnamefont {Kiffner}},
  \bibinfo {author} {\bibfnamefont {J.}~\bibnamefont {Mur-Petit}}, \ and\
  \bibinfo {author} {\bibfnamefont {D.}~\bibnamefont {Jaksch}},\ }\bibfield
  {title} {\bibinfo {title} {\emph {Characterizing the phase diagram of
  finite-size dipolar Bose-Hubbard systems}},\ }\href {\doibase
  10.1103/PhysRevA.101.013616} {\bibfield  {journal} {\bibinfo  {journal}
  {Phys. Rev. A}\ }\textbf {\bibinfo {volume} {101}},\ \bibinfo {pages}
  {013616} (\bibinfo {year} {2020})}\BibitemShut {NoStop}%
\bibitem [{\citenamefont {Oka}\ and\ \citenamefont {Aoki}(2009)}]{Oka09PRB}%
  \BibitemOpen
  \bibfield  {author} {\bibinfo {author} {\bibfnamefont {T.}~\bibnamefont
  {Oka}}\ and\ \bibinfo {author} {\bibfnamefont {H.}~\bibnamefont {Aoki}},\
  }\bibfield  {title} {\bibinfo {title} {\emph {Photovoltaic Hall effect in
  graphene}},\ }\href {\doibase 10.1103/PhysRevB.79.081406} {\bibfield
  {journal} {\bibinfo  {journal} {Phys. Rev. B}\ }\textbf {\bibinfo {volume}
  {79}},\ \bibinfo {pages} {081406} (\bibinfo {year} {2009})}\BibitemShut
  {NoStop}%
\bibitem [{\citenamefont {Rechtsman}\ \emph {et~al.}(2013)\citenamefont
  {Rechtsman}, \citenamefont {Zeuner}, \citenamefont {Plotnik}, \citenamefont
  {Lumer}, \citenamefont {Podolsky}, \citenamefont {Dreisow}, \citenamefont
  {Nolte}, \citenamefont {Segev},\ and\ \citenamefont
  {Szameit}}]{Rechtsman13Nature}%
  \BibitemOpen
  \bibfield  {author} {\bibinfo {author} {\bibfnamefont {M.~C.}\ \bibnamefont
  {Rechtsman}}, \bibinfo {author} {\bibfnamefont {J.~M.}\ \bibnamefont
  {Zeuner}}, \bibinfo {author} {\bibfnamefont {Y.}~\bibnamefont {Plotnik}},
  \bibinfo {author} {\bibfnamefont {Y.}~\bibnamefont {Lumer}}, \bibinfo
  {author} {\bibfnamefont {D.}~\bibnamefont {Podolsky}}, \bibinfo {author}
  {\bibfnamefont {F.}~\bibnamefont {Dreisow}}, \bibinfo {author} {\bibfnamefont
  {S.}~\bibnamefont {Nolte}}, \bibinfo {author} {\bibfnamefont
  {M.}~\bibnamefont {Segev}}, \ and\ \bibinfo {author} {\bibfnamefont
  {A.}~\bibnamefont {Szameit}},\ }\bibfield  {title} {\bibinfo {title} {\emph
  {{Photonic Floquet topological insulators}}},\ }\href {\doibase
  10.1002/pssr.201206451} {\bibfield  {journal} {\bibinfo  {journal} {Nature}\
  }\textbf {\bibinfo {volume} {496}},\ \bibinfo {pages} {196} (\bibinfo {year}
  {2013})}\BibitemShut {NoStop}%
\bibitem [{\citenamefont {Kitagawa}\ \emph {et~al.}(2010)\citenamefont
  {Kitagawa}, \citenamefont {Berg}, \citenamefont {Rudner},\ and\ \citenamefont
  {Demler}}]{Kitagawa_2010}%
  \BibitemOpen
  \bibfield  {author} {\bibinfo {author} {\bibfnamefont {T.}~\bibnamefont
  {Kitagawa}}, \bibinfo {author} {\bibfnamefont {E.}~\bibnamefont {Berg}},
  \bibinfo {author} {\bibfnamefont {M.}~\bibnamefont {Rudner}}, \ and\ \bibinfo
  {author} {\bibfnamefont {E.}~\bibnamefont {Demler}},\ }\bibfield  {title}
  {\bibinfo {title} {\emph {Topological characterization of periodically driven
  quantum systems}},\ }\href {\doibase 10.1103/PhysRevB.82.235114} {\bibfield
  {journal} {\bibinfo  {journal} {Phys. Rev. B}\ }\textbf {\bibinfo {volume}
  {82}},\ \bibinfo {pages} {235114} (\bibinfo {year} {2010})}\BibitemShut
  {NoStop}%
\bibitem [{\citenamefont {Rudner}\ \emph {et~al.}(2013)\citenamefont {Rudner},
  \citenamefont {Lindner}, \citenamefont {Berg},\ and\ \citenamefont
  {Levin}}]{Rudner13PRX}%
  \BibitemOpen
  \bibfield  {author} {\bibinfo {author} {\bibfnamefont {M.~S.}\ \bibnamefont
  {Rudner}}, \bibinfo {author} {\bibfnamefont {N.~H.}\ \bibnamefont {Lindner}},
  \bibinfo {author} {\bibfnamefont {E.}~\bibnamefont {Berg}}, \ and\ \bibinfo
  {author} {\bibfnamefont {M.}~\bibnamefont {Levin}},\ }\bibfield  {title}
  {\bibinfo {title} {\emph {Anomalous Edge States and the Bulk-Edge
  Correspondence for Periodically Driven Two-Dimensional Systems}},\ }\href
  {\doibase 10.1103/PhysRevX.3.031005} {\bibfield  {journal} {\bibinfo
  {journal} {Phys. Rev. X}\ }\textbf {\bibinfo {volume} {3}},\ \bibinfo {pages}
  {031005} (\bibinfo {year} {2013})}\BibitemShut {NoStop}%
\bibitem [{\citenamefont {Dauphin}\ \emph {et~al.}(2017)\citenamefont
  {Dauphin}, \citenamefont {Tran}, \citenamefont {Lewenstein},\ and\
  \citenamefont {Goldman}}]{Dauphin_2017}%
  \BibitemOpen
  \bibfield  {author} {\bibinfo {author} {\bibfnamefont {A.}~\bibnamefont
  {Dauphin}}, \bibinfo {author} {\bibfnamefont {D.-T.}\ \bibnamefont {Tran}},
  \bibinfo {author} {\bibfnamefont {M.}~\bibnamefont {Lewenstein}}, \ and\
  \bibinfo {author} {\bibfnamefont {N.}~\bibnamefont {Goldman}},\ }\bibfield
  {title} {\bibinfo {title} {\emph {{Loading ultracold gases in topological
  Floquet bands: the fate of current and center-of-mass responses}}},\ }\href
  {\doibase 10.1088/2053-1583/aa6a3b} {\bibfield  {journal} {\bibinfo
  {journal} {2D Mater.}\ }\textbf {\bibinfo {volume} {4}},\ \bibinfo {pages}
  {024010} (\bibinfo {year} {2017})}\BibitemShut {NoStop}%
\bibitem [{\citenamefont {Kumar}\ \emph {et~al.}(2020)\citenamefont {Kumar},
  \citenamefont {Rodriguez-Vega}, \citenamefont {Pereg-Barnea},\ and\
  \citenamefont {Seradjeh}}]{Kumar_2020}%
  \BibitemOpen
  \bibfield  {author} {\bibinfo {author} {\bibfnamefont {A.}~\bibnamefont
  {Kumar}}, \bibinfo {author} {\bibfnamefont {M.}~\bibnamefont
  {Rodriguez-Vega}}, \bibinfo {author} {\bibfnamefont {T.}~\bibnamefont
  {Pereg-Barnea}}, \ and\ \bibinfo {author} {\bibfnamefont {B.}~\bibnamefont
  {Seradjeh}},\ }\bibfield  {title} {\bibinfo {title} {\emph {{Linear response
  theory and optical conductivity of Floquet topological insulators}}},\ }\href
  {\doibase 10.1103/PhysRevB.101.174314} {\bibfield  {journal} {\bibinfo
  {journal} {Phys. Rev. B}\ }\textbf {\bibinfo {volume} {101}},\ \bibinfo
  {pages} {174314} (\bibinfo {year} {2020})}\BibitemShut {NoStop}%
\bibitem [{\citenamefont {Goodfellow}\ \emph {et~al.}(2016)\citenamefont
  {Goodfellow}, \citenamefont {Bengio},\ and\ \citenamefont
  {Courville}}]{Goodfellow16}%
  \BibitemOpen
  \bibfield  {author} {\bibinfo {author} {\bibfnamefont {I.}~\bibnamefont
  {Goodfellow}}, \bibinfo {author} {\bibfnamefont {Y.}~\bibnamefont {Bengio}},
  \ and\ \bibinfo {author} {\bibfnamefont {A.}~\bibnamefont {Courville}},\
  }\href@noop {} {\emph {\bibinfo {title} {Deep Learning}}}\ (\bibinfo
  {publisher} {MIT Press},\ \bibinfo {year} {2016})\ \bibinfo {note}
  {\url{http://www.deeplearningbook.org}}\BibitemShut {NoStop}%
\bibitem [{\citenamefont {Maas}(2013)}]{Maas2013}%
  \BibitemOpen
  \bibfield  {author} {\bibinfo {author} {\bibfnamefont {A.~L.}\ \bibnamefont
  {Maas}},\ }\bibfield  {title} {\bibinfo {title} {\emph {Rectifier
  Nonlinearities Improve Neural Network Acoustic Models}}\ }(\bibinfo {year}
  {2013})\BibitemShut {NoStop}%
\bibitem [{\citenamefont {Kingma}\ and\ \citenamefont {Ba}(2015)}]{Kingma2015}%
  \BibitemOpen
  \bibfield  {author} {\bibinfo {author} {\bibfnamefont {D.~P.}\ \bibnamefont
  {Kingma}}\ and\ \bibinfo {author} {\bibfnamefont {J.~L.}\ \bibnamefont
  {Ba}},\ }\bibfield  {title} {\bibinfo {title} {\emph {{Adam: A method for
  stochastic optimization}}},\ }in\ \href {https://arxiv.org/abs/1412.6980v9}
  {\emph {\bibinfo {booktitle} {3rd International Conference on Learning
  Representations, ICLR 2015 - Conference Track Proceedings}}}\ (\bibinfo
  {publisher} {International Conference on Learning Representations, ICLR},\
  \bibinfo {year} {2015})\ \Eprint {http://arxiv.org/abs/1412.6980}
  {arXiv:1412.6980} \BibitemShut {NoStop}%
\bibitem [{\citenamefont {Lecun}(1987)}]{LeCun1987PhD}%
  \BibitemOpen
  \bibfield  {author} {\bibinfo {author} {\bibfnamefont {Y.}~\bibnamefont
  {Lecun}},\ }\href@noop {} {\emph {\bibinfo {title} {PhD thesis: Modeles
  connexionnistes de l'apprentissage (connectionist learning models)}}}\
  (\bibinfo  {publisher} {Universite P. et M. Curie (Paris 6)},\ \bibinfo
  {year} {1987})\BibitemShut {NoStop}%
\bibitem [{\citenamefont {Bourlard}\ and\ \citenamefont
  {Kamp}(1988)}]{Bourlard1988}%
  \BibitemOpen
  \bibfield  {author} {\bibinfo {author} {\bibfnamefont {H.}~\bibnamefont
  {Bourlard}}\ and\ \bibinfo {author} {\bibfnamefont {Y.}~\bibnamefont
  {Kamp}},\ }\bibfield  {title} {\bibinfo {title} {\emph {Auto-association by
  multilayer perceptrons and singular value decomposition}},\ }\href {\doibase
  10.1007/bf00332918} {\bibfield  {journal} {\bibinfo  {journal} {Biological
  Cybernetics}\ }\textbf {\bibinfo {volume} {59}},\ \bibinfo {pages} {291}
  (\bibinfo {year} {1988})}\BibitemShut {NoStop}%
\bibitem [{\citenamefont {Hinton}\ and\ \citenamefont
  {Zemel}(1993)}]{Hinton1993}%
  \BibitemOpen
  \bibfield  {author} {\bibinfo {author} {\bibfnamefont {G.~E.}\ \bibnamefont
  {Hinton}}\ and\ \bibinfo {author} {\bibfnamefont {R.~S.}\ \bibnamefont
  {Zemel}},\ }\bibfield  {title} {\bibinfo {title} {\emph {Autoencoders,
  Minimum Description Length and Helmholtz Free Energy}},\ }in\ \href@noop {}
  {\emph {\bibinfo {booktitle} {Proceedings of the 6th International Conference
  on Neural Information Processing Systems}}},\ \bibinfo {series and number}
  {NIPS'93}\ (\bibinfo  {publisher} {Morgan Kaufmann Publishers Inc.},\
  \bibinfo {address} {San Francisco, CA, USA},\ \bibinfo {year} {1993})\ pp.\
  \bibinfo {pages} {3--10}\BibitemShut {NoStop}%
\bibitem [{\citenamefont {Vincent}\ \emph {et~al.}(2008)\citenamefont
  {Vincent}, \citenamefont {Larochelle}, \citenamefont {Bengio},\ and\
  \citenamefont {Manzagol}}]{Vincent2008ICML}%
  \BibitemOpen
  \bibfield  {author} {\bibinfo {author} {\bibfnamefont {P.}~\bibnamefont
  {Vincent}}, \bibinfo {author} {\bibfnamefont {H.}~\bibnamefont {Larochelle}},
  \bibinfo {author} {\bibfnamefont {Y.}~\bibnamefont {Bengio}}, \ and\ \bibinfo
  {author} {\bibfnamefont {P.-A.}\ \bibnamefont {Manzagol}},\ }\bibfield
  {title} {\bibinfo {title} {\emph {Extracting and Composing Robust Features
  with Denoising Autoencoders}},\ }in\ \href {\doibase 10.1145/1390156.1390294}
  {\emph {\bibinfo {booktitle} {Proceedings of the 25th International
  Conference on Machine Learning}}},\ \bibinfo {series and number} {ICML '08}\
  (\bibinfo  {publisher} {Association for Computing Machinery},\ \bibinfo
  {address} {New York, NY, USA},\ \bibinfo {year} {2008})\ p.\ \bibinfo {pages}
  {1096–1103}\BibitemShut {NoStop}%
\bibitem [{\citenamefont {Xie}\ \emph {et~al.}(2012)\citenamefont {Xie},
  \citenamefont {Xu},\ and\ \citenamefont {Chen}}]{Xie2012NIPS}%
  \BibitemOpen
  \bibfield  {author} {\bibinfo {author} {\bibfnamefont {J.}~\bibnamefont
  {Xie}}, \bibinfo {author} {\bibfnamefont {L.}~\bibnamefont {Xu}}, \ and\
  \bibinfo {author} {\bibfnamefont {E.}~\bibnamefont {Chen}},\ }\bibfield
  {title} {\bibinfo {title} {\emph {Image Denoising and Inpainting with Deep
  Neural Networks}},\ }in\ \href
  {https://proceedings.neurips.cc/paper/2012/file/6cdd60ea0045eb7a6ec44c54d29ed402-Paper.pdf}
  {\emph {\bibinfo {booktitle} {Advances in Neural Information Processing
  Systems}}},\ Vol.~\bibinfo {volume} {25},\ \bibinfo {editor} {edited by\
  \bibinfo {editor} {\bibfnamefont {F.}~\bibnamefont {Pereira}}, \bibinfo
  {editor} {\bibfnamefont {C.~J.~C.}\ \bibnamefont {Burges}}, \bibinfo {editor}
  {\bibfnamefont {L.}~\bibnamefont {Bottou}}, \ and\ \bibinfo {editor}
  {\bibfnamefont {K.~Q.}\ \bibnamefont {Weinberger}}}\ (\bibinfo  {publisher}
  {Curran Associates, Inc.},\ \bibinfo {year} {2012})\ pp.\ \bibinfo {pages}
  {341--349}\BibitemShut {NoStop}%
\bibitem [{\citenamefont {Baldassarre}\ \emph {et~al.}(2017)\citenamefont
  {Baldassarre}, \citenamefont {Mor{\'{\i}}n},\ and\ \citenamefont
  {Rod{\'{e}}s{-}Guirao}}]{Baldassarre2017arXiv}%
  \BibitemOpen
  \bibfield  {author} {\bibinfo {author} {\bibfnamefont {F.}~\bibnamefont
  {Baldassarre}}, \bibinfo {author} {\bibfnamefont {D.~G.}\ \bibnamefont
  {Mor{\'{\i}}n}}, \ and\ \bibinfo {author} {\bibfnamefont {L.}~\bibnamefont
  {Rod{\'{e}}s{-}Guirao}},\ }\bibfield  {title} {\bibinfo {title} {\emph {Deep
  Koalarization: Image Colorization using CNNs and Inception-ResNet-v2}},\
  }\href {http://arxiv.org/abs/1712.03400} {\bibfield  {journal} {\bibinfo
  {journal} {CoRR}\ }\textbf {\bibinfo {volume} {abs/1712.03400}} (\bibinfo
  {year} {2017})},\ \Eprint {http://arxiv.org/abs/1712.03400}
  {arXiv:1712.03400} \BibitemShut {NoStop}%
\bibitem [{\citenamefont {Kingma}\ and\ \citenamefont
  {Welling}(2014)}]{kingma2014autoencoding}%
  \BibitemOpen
  \bibfield  {author} {\bibinfo {author} {\bibfnamefont {D.~P.}\ \bibnamefont
  {Kingma}}\ and\ \bibinfo {author} {\bibfnamefont {M.}~\bibnamefont
  {Welling}},\ }\href@noop {} {\bibinfo {title} {\emph {Auto-Encoding
  Variational Bayes}}} (\bibinfo {year} {2014}),\ \Eprint
  {http://arxiv.org/abs/1312.6114} {arXiv:1312.6114 [stat.ML]} \BibitemShut
  {NoStop}%
\bibitem [{\citenamefont {Rezende}\ \emph {et~al.}(2014)\citenamefont
  {Rezende}, \citenamefont {Mohamed},\ and\ \citenamefont
  {Wierstra}}]{rezende2014stochastic}%
  \BibitemOpen
  \bibfield  {author} {\bibinfo {author} {\bibfnamefont {D.~J.}\ \bibnamefont
  {Rezende}}, \bibinfo {author} {\bibfnamefont {S.}~\bibnamefont {Mohamed}}, \
  and\ \bibinfo {author} {\bibfnamefont {D.}~\bibnamefont {Wierstra}},\
  }\href@noop {} {\bibinfo {title} {\emph {Stochastic Backpropagation and
  Approximate Inference in Deep Generative Models}}} (\bibinfo {year} {2014}),\
  \Eprint {http://arxiv.org/abs/1401.4082} {arXiv:1401.4082 [stat.ML]}
  \BibitemShut {NoStop}%
\bibitem [{\citenamefont {Doersch}(2016)}]{doersch2016tutorial}%
  \BibitemOpen
  \bibfield  {author} {\bibinfo {author} {\bibfnamefont {C.}~\bibnamefont
  {Doersch}},\ }\href@noop {} {\bibinfo {title} {\emph {Tutorial on Variational
  Autoencoders}}} (\bibinfo {year} {2016}),\ \Eprint
  {http://arxiv.org/abs/1606.05908} {arXiv:1606.05908 [stat.ML]} \BibitemShut
  {NoStop}%
\bibitem [{\citenamefont {Cook}(1977)}]{Cook77}%
  \BibitemOpen
  \bibfield  {author} {\bibinfo {author} {\bibfnamefont {R.~D.}\ \bibnamefont
  {Cook}},\ }\bibfield  {title} {\bibinfo {title} {\emph {Detection of
  Influential Observation in Linear Regression}},\ }\href {\doibase
  10.2307/1268249} {\bibfield  {journal} {\bibinfo  {journal} {Technometrics}\
  }\textbf {\bibinfo {volume} {19}},\ \bibinfo {pages} {15} (\bibinfo {year}
  {1977})}\BibitemShut {NoStop}%
\bibitem [{\citenamefont {Koh}\ and\ \citenamefont {Liang}(2017)}]{Koh17}%
  \BibitemOpen
  \bibfield  {author} {\bibinfo {author} {\bibfnamefont {P.~W.}\ \bibnamefont
  {Koh}}\ and\ \bibinfo {author} {\bibfnamefont {P.}~\bibnamefont {Liang}},\
  }\bibfield  {title} {\bibinfo {title} {\emph {Understanding Black-box
  Predictions via Influence Functions}},\ }in\ \href
  {http://proceedings.mlr.press/v70/koh17a.html} {\emph {\bibinfo {booktitle}
  {{Proceedings of the 34th International Conference} on {Machine Learning}}}}\
  (\bibinfo {year} {2017})\BibitemShut {NoStop}%
\bibitem [{\citenamefont {{van der Walt}}\ \emph {et~al.}(2011)\citenamefont
  {{van der Walt}}, \citenamefont {{Colbert}},\ and\ \citenamefont
  {{Varoquaux}}}]{NumPy}%
  \BibitemOpen
  \bibfield  {author} {\bibinfo {author} {\bibfnamefont {S.}~\bibnamefont {{van
  der Walt}}}, \bibinfo {author} {\bibfnamefont {S.~C.}\ \bibnamefont
  {{Colbert}}}, \ and\ \bibinfo {author} {\bibfnamefont {G.}~\bibnamefont
  {{Varoquaux}}},\ }\bibfield  {title} {\bibinfo {title} {\emph {The NumPy
  Array: A Structure for Efficient Numerical Computation}},\ }\href@noop {}
  {\bibfield  {journal} {\bibinfo  {journal} {Computing in Science
  Engineering}\ }\textbf {\bibinfo {volume} {13}},\ \bibinfo {pages} {22}
  (\bibinfo {year} {2011})}\BibitemShut {NoStop}%
\bibitem [{\citenamefont {Paszke}\ \emph {et~al.}(2017)\citenamefont {Paszke},
  \citenamefont {Gross}, \citenamefont {Chintala}, \citenamefont {Chanan},
  \citenamefont {Yang}, \citenamefont {DeVito}, \citenamefont {Lin},
  \citenamefont {Desmaison}, \citenamefont {Antiga},\ and\ \citenamefont
  {Lerer}}]{pytorch}%
  \BibitemOpen
  \bibfield  {author} {\bibinfo {author} {\bibfnamefont {A.}~\bibnamefont
  {Paszke}}, \bibinfo {author} {\bibfnamefont {S.}~\bibnamefont {Gross}},
  \bibinfo {author} {\bibfnamefont {S.}~\bibnamefont {Chintala}}, \bibinfo
  {author} {\bibfnamefont {G.}~\bibnamefont {Chanan}}, \bibinfo {author}
  {\bibfnamefont {E.}~\bibnamefont {Yang}}, \bibinfo {author} {\bibfnamefont
  {Z.}~\bibnamefont {DeVito}}, \bibinfo {author} {\bibfnamefont
  {Z.}~\bibnamefont {Lin}}, \bibinfo {author} {\bibfnamefont {A.}~\bibnamefont
  {Desmaison}}, \bibinfo {author} {\bibfnamefont {L.}~\bibnamefont {Antiga}}, \
  and\ \bibinfo {author} {\bibfnamefont {A.}~\bibnamefont {Lerer}},\ }\bibfield
   {title} {\bibinfo {title} {\emph {Automatic differentiation in PyTorch}},\
  }in\ \href@noop {} {\emph {\bibinfo {booktitle} {NIPS-W}}}\ (\bibinfo {year}
  {2017})\BibitemShut {NoStop}%
\bibitem [{\citenamefont {Abadi}\ \emph {et~al.}(2015)\citenamefont {Abadi},
  \citenamefont {Agarwal}, \citenamefont {Barham}, \citenamefont {Brevdo},
  \citenamefont {Chen}, \citenamefont {Citro}, \citenamefont {Corrado},
  \citenamefont {Davis}, \citenamefont {Dean}, \citenamefont {Devin},
  \citenamefont {Ghemawat}, \citenamefont {Goodfellow}, \citenamefont {Harp},
  \citenamefont {Irving}, \citenamefont {Isard}, \citenamefont {Jia},
  \citenamefont {Jozefowicz}, \citenamefont {Kaiser}, \citenamefont {Kudlur},
  \citenamefont {Levenberg}, \citenamefont {Man\'{e}}, \citenamefont {Monga},
  \citenamefont {Moore}, \citenamefont {Murray}, \citenamefont {Olah},
  \citenamefont {Schuster}, \citenamefont {Shlens}, \citenamefont {Steiner},
  \citenamefont {Sutskever}, \citenamefont {Talwar}, \citenamefont {Tucker},
  \citenamefont {Vanhoucke}, \citenamefont {Vasudevan}, \citenamefont
  {Vi\'{e}gas}, \citenamefont {Vinyals}, \citenamefont {Warden}, \citenamefont
  {Wattenberg}, \citenamefont {Wicke}, \citenamefont {Yu},\ and\ \citenamefont
  {Zheng}}]{tensorflow15}%
  \BibitemOpen
  \bibfield  {author} {\bibinfo {author} {\bibfnamefont {M.}~\bibnamefont
  {Abadi}}, \bibinfo {author} {\bibfnamefont {A.}~\bibnamefont {Agarwal}},
  \bibinfo {author} {\bibfnamefont {P.}~\bibnamefont {Barham}}, \bibinfo
  {author} {\bibfnamefont {E.}~\bibnamefont {Brevdo}}, \bibinfo {author}
  {\bibfnamefont {Z.}~\bibnamefont {Chen}}, \bibinfo {author} {\bibfnamefont
  {C.}~\bibnamefont {Citro}}, \bibinfo {author} {\bibfnamefont {G.~S.}\
  \bibnamefont {Corrado}}, \bibinfo {author} {\bibfnamefont {A.}~\bibnamefont
  {Davis}}, \bibinfo {author} {\bibfnamefont {J.}~\bibnamefont {Dean}},
  \bibinfo {author} {\bibfnamefont {M.}~\bibnamefont {Devin}}, \bibinfo
  {author} {\bibfnamefont {S.}~\bibnamefont {Ghemawat}}, \bibinfo {author}
  {\bibfnamefont {I.}~\bibnamefont {Goodfellow}}, \bibinfo {author}
  {\bibfnamefont {A.}~\bibnamefont {Harp}}, \bibinfo {author} {\bibfnamefont
  {G.}~\bibnamefont {Irving}}, \bibinfo {author} {\bibfnamefont
  {M.}~\bibnamefont {Isard}}, \bibinfo {author} {\bibfnamefont
  {Y.}~\bibnamefont {Jia}}, \bibinfo {author} {\bibfnamefont {R.}~\bibnamefont
  {Jozefowicz}}, \bibinfo {author} {\bibfnamefont {L.}~\bibnamefont {Kaiser}},
  \bibinfo {author} {\bibfnamefont {M.}~\bibnamefont {Kudlur}}, \bibinfo
  {author} {\bibfnamefont {J.}~\bibnamefont {Levenberg}}, \bibinfo {author}
  {\bibfnamefont {D.}~\bibnamefont {Man\'{e}}}, \bibinfo {author}
  {\bibfnamefont {R.}~\bibnamefont {Monga}}, \bibinfo {author} {\bibfnamefont
  {S.}~\bibnamefont {Moore}}, \bibinfo {author} {\bibfnamefont
  {D.}~\bibnamefont {Murray}}, \bibinfo {author} {\bibfnamefont
  {C.}~\bibnamefont {Olah}}, \bibinfo {author} {\bibfnamefont {M.}~\bibnamefont
  {Schuster}}, \bibinfo {author} {\bibfnamefont {J.}~\bibnamefont {Shlens}},
  \bibinfo {author} {\bibfnamefont {B.}~\bibnamefont {Steiner}}, \bibinfo
  {author} {\bibfnamefont {I.}~\bibnamefont {Sutskever}}, \bibinfo {author}
  {\bibfnamefont {K.}~\bibnamefont {Talwar}}, \bibinfo {author} {\bibfnamefont
  {P.}~\bibnamefont {Tucker}}, \bibinfo {author} {\bibfnamefont
  {V.}~\bibnamefont {Vanhoucke}}, \bibinfo {author} {\bibfnamefont
  {V.}~\bibnamefont {Vasudevan}}, \bibinfo {author} {\bibfnamefont
  {F.}~\bibnamefont {Vi\'{e}gas}}, \bibinfo {author} {\bibfnamefont
  {O.}~\bibnamefont {Vinyals}}, \bibinfo {author} {\bibfnamefont
  {P.}~\bibnamefont {Warden}}, \bibinfo {author} {\bibfnamefont
  {M.}~\bibnamefont {Wattenberg}}, \bibinfo {author} {\bibfnamefont
  {M.}~\bibnamefont {Wicke}}, \bibinfo {author} {\bibfnamefont
  {Y.}~\bibnamefont {Yu}}, \ and\ \bibinfo {author} {\bibfnamefont
  {X.}~\bibnamefont {Zheng}},\ }\href {http://tensorflow.org/} {\bibinfo
  {title} {\emph {{TensorFlow}: Large-Scale Machine Learning on Heterogeneous
  Systems}}} (\bibinfo {year} {2015}),\ \bibinfo {note} {software available
  from tensorflow.org}\BibitemShut {NoStop}%
\bibitem [{\citenamefont {Käming}\ \emph {et~al.}(2021)\citenamefont
  {Käming}, \citenamefont {Dawid}, \citenamefont {Kottmann}, \citenamefont
  {Lewenstein}, \citenamefont {Sengstock}, \citenamefont {Dauphin},\ and\
  \citenamefont {Weitenberg}}]{notebooks}%
  \BibitemOpen
  \bibfield  {author} {\bibinfo {author} {\bibfnamefont {N.}~\bibnamefont
  {Käming}}, \bibinfo {author} {\bibfnamefont {A.}~\bibnamefont {Dawid}},
  \bibinfo {author} {\bibfnamefont {K.}~\bibnamefont {Kottmann}}, \bibinfo
  {author} {\bibfnamefont {M.}~\bibnamefont {Lewenstein}}, \bibinfo {author}
  {\bibfnamefont {K.}~\bibnamefont {Sengstock}}, \bibinfo {author}
  {\bibfnamefont {A.}~\bibnamefont {Dauphin}}, \ and\ \bibinfo {author}
  {\bibfnamefont {C.}~\bibnamefont {Weitenberg}},\ }\bibfield  {title}
  {\bibinfo {title} {\emph {Code and Data for Unsupervised machine learning of
  topological phase transitions from experimental data}},\ }\href {\doibase
  10.5281/zenodo.4459311} {\  (\bibinfo {year} {2021}),\
  10.5281/zenodo.4459311}\BibitemShut {NoStop}%
\bibitem [{\citenamefont {Hinton}\ and\ \citenamefont
  {Salakhutdinov}(2006)}]{Hinton06Science}%
  \BibitemOpen
  \bibfield  {author} {\bibinfo {author} {\bibfnamefont {G.~E.}\ \bibnamefont
  {Hinton}}\ and\ \bibinfo {author} {\bibfnamefont {R.~R.}\ \bibnamefont
  {Salakhutdinov}},\ }\bibfield  {title} {\bibinfo {title} {\emph {Reducing the
  Dimensionality of Data with Neural Networks}},\ }\href {\doibase
  10.1126/science.1127647} {\bibfield  {journal} {\bibinfo  {journal}
  {Science}\ }\textbf {\bibinfo {volume} {313}},\ \bibinfo {pages} {504}
  (\bibinfo {year} {2006})}\BibitemShut {NoStop}%
\bibitem [{\citenamefont {{Fitzgibbon}}\ \emph {et~al.}(1999)\citenamefont
  {{Fitzgibbon}}, \citenamefont {{Pilu}},\ and\ \citenamefont
  {{Fisher}}}]{FitzgibbonIEEE}%
  \BibitemOpen
  \bibfield  {author} {\bibinfo {author} {\bibfnamefont {A.}~\bibnamefont
  {{Fitzgibbon}}}, \bibinfo {author} {\bibfnamefont {M.}~\bibnamefont
  {{Pilu}}}, \ and\ \bibinfo {author} {\bibfnamefont {R.~B.}\ \bibnamefont
  {{Fisher}}},\ }\bibfield  {title} {\bibinfo {title} {\emph {Direct least
  square fitting of ellipses}},\ }\href {\doibase 10.1109/34.765658} {\bibfield
   {journal} {\bibinfo  {journal} {IEEE Transactions on Pattern Analysis and
  Machine Intelligence}\ }\textbf {\bibinfo {volume} {21}},\ \bibinfo {pages}
  {476} (\bibinfo {year} {1999})}\BibitemShut {NoStop}%
\bibitem [{\citenamefont {Ockeloen}\ \emph {et~al.}(2010)\citenamefont
  {Ockeloen}, \citenamefont {Tauschinsky}, \citenamefont {Spreeuw},\ and\
  \citenamefont {Whitlock}}]{Oeckeloen10PRA}%
  \BibitemOpen
  \bibfield  {author} {\bibinfo {author} {\bibfnamefont {C.~F.}\ \bibnamefont
  {Ockeloen}}, \bibinfo {author} {\bibfnamefont {A.~F.}\ \bibnamefont
  {Tauschinsky}}, \bibinfo {author} {\bibfnamefont {R.~J.~C.}\ \bibnamefont
  {Spreeuw}}, \ and\ \bibinfo {author} {\bibfnamefont {S.}~\bibnamefont
  {Whitlock}},\ }\bibfield  {title} {\bibinfo {title} {\emph {Detection of
  small atom numbers through image processing}},\ }\href {\doibase
  10.1103/PhysRevA.82.061606} {\bibfield  {journal} {\bibinfo  {journal} {Phys.
  Rev. A}\ }\textbf {\bibinfo {volume} {82}},\ \bibinfo {pages} {061606}
  (\bibinfo {year} {2010})}\BibitemShut {NoStop}%
\bibitem [{\citenamefont {Fung}\ \emph {et~al.}(2016)\citenamefont {Fung},
  \citenamefont {Hanna}, \citenamefont {Vendrell}, \citenamefont {Ramakrishna},
  \citenamefont {Seideman}, \citenamefont {Santra},\ and\ \citenamefont
  {Ourmazd}}]{Fung16Nature}%
  \BibitemOpen
  \bibfield  {author} {\bibinfo {author} {\bibfnamefont {R.}~\bibnamefont
  {Fung}}, \bibinfo {author} {\bibfnamefont {A.~M.}\ \bibnamefont {Hanna}},
  \bibinfo {author} {\bibfnamefont {O.}~\bibnamefont {Vendrell}}, \bibinfo
  {author} {\bibfnamefont {S.}~\bibnamefont {Ramakrishna}}, \bibinfo {author}
  {\bibfnamefont {T.}~\bibnamefont {Seideman}}, \bibinfo {author}
  {\bibfnamefont {R.}~\bibnamefont {Santra}}, \ and\ \bibinfo {author}
  {\bibfnamefont {A.}~\bibnamefont {Ourmazd}},\ }\bibfield  {title} {\bibinfo
  {title} {\emph {{Dynamics from noisy data with extreme timing
  uncertainty}}},\ }\href {\doibase 10.1038/nature17627} {\bibfield  {journal}
  {\bibinfo  {journal} {Nature}\ }\textbf {\bibinfo {volume} {532}},\ \bibinfo
  {pages} {471} (\bibinfo {year} {2016})}\BibitemShut {NoStop}%
\bibitem [{\citenamefont {Akiba}\ \emph {et~al.}(2019)\citenamefont {Akiba},
  \citenamefont {Sano}, \citenamefont {Yanase}, \citenamefont {Ohta},\ and\
  \citenamefont {Koyama}}]{optuna19}%
  \BibitemOpen
  \bibfield  {author} {\bibinfo {author} {\bibfnamefont {T.}~\bibnamefont
  {Akiba}}, \bibinfo {author} {\bibfnamefont {S.}~\bibnamefont {Sano}},
  \bibinfo {author} {\bibfnamefont {T.}~\bibnamefont {Yanase}}, \bibinfo
  {author} {\bibfnamefont {T.}~\bibnamefont {Ohta}}, \ and\ \bibinfo {author}
  {\bibfnamefont {M.}~\bibnamefont {Koyama}},\ }\bibfield  {title} {\bibinfo
  {title} {\emph {Optuna: A Next-generation Hyperparameter Optimization
  Framework}},\ }in\ \href@noop {} {\emph {\bibinfo {booktitle} {Proceedings of
  the 25rd {ACM} {SIGKDD} International Conference on Knowledge Discovery and
  Data Mining}}}\ (\bibinfo {year} {2019})\BibitemShut {NoStop}%
\bibitem [{\citenamefont {{Zhou Wang}}\ \emph {et~al.}(2004)\citenamefont
  {{Zhou Wang}}, \citenamefont {{Bovik}}, \citenamefont {{Sheikh}},\ and\
  \citenamefont {{Simoncelli}}}]{wang14ssim}%
  \BibitemOpen
  \bibfield  {author} {\bibinfo {author} {\bibnamefont {{Zhou Wang}}}, \bibinfo
  {author} {\bibfnamefont {A.~C.}\ \bibnamefont {{Bovik}}}, \bibinfo {author}
  {\bibfnamefont {H.~R.}\ \bibnamefont {{Sheikh}}}, \ and\ \bibinfo {author}
  {\bibfnamefont {E.~P.}\ \bibnamefont {{Simoncelli}}},\ }\bibfield  {title}
  {\bibinfo {title} {\emph {Image quality assessment: from error visibility to
  structural similarity}},\ }\href {\doibase 10.1109/TIP.2003.819861}
  {\bibfield  {journal} {\bibinfo  {journal} {IEEE Transactions on Image
  Processing}\ }\textbf {\bibinfo {volume} {13}},\ \bibinfo {pages} {600}
  (\bibinfo {year} {2004})}\BibitemShut {NoStop}%
\bibitem [{\citenamefont {Pedregosa}\ \emph {et~al.}(2011)\citenamefont
  {Pedregosa}, \citenamefont {Varoquaux}, \citenamefont {Gramfort},
  \citenamefont {Michel}, \citenamefont {Thirion}, \citenamefont {Grisel},
  \citenamefont {Blondel}, \citenamefont {Prettenhofer}, \citenamefont {Weiss},
  \citenamefont {Dubourg}, \citenamefont {Vanderplas}, \citenamefont {Passos},
  \citenamefont {Cournapeau}, \citenamefont {Brucher}, \citenamefont {Perrot},\
  and\ \citenamefont {Duchesnay}}]{scikit-learn}%
  \BibitemOpen
  \bibfield  {author} {\bibinfo {author} {\bibfnamefont {F.}~\bibnamefont
  {Pedregosa}}, \bibinfo {author} {\bibfnamefont {G.}~\bibnamefont
  {Varoquaux}}, \bibinfo {author} {\bibfnamefont {A.}~\bibnamefont {Gramfort}},
  \bibinfo {author} {\bibfnamefont {V.}~\bibnamefont {Michel}}, \bibinfo
  {author} {\bibfnamefont {B.}~\bibnamefont {Thirion}}, \bibinfo {author}
  {\bibfnamefont {O.}~\bibnamefont {Grisel}}, \bibinfo {author} {\bibfnamefont
  {M.}~\bibnamefont {Blondel}}, \bibinfo {author} {\bibfnamefont
  {P.}~\bibnamefont {Prettenhofer}}, \bibinfo {author} {\bibfnamefont
  {R.}~\bibnamefont {Weiss}}, \bibinfo {author} {\bibfnamefont
  {V.}~\bibnamefont {Dubourg}}, \bibinfo {author} {\bibfnamefont
  {J.}~\bibnamefont {Vanderplas}}, \bibinfo {author} {\bibfnamefont
  {A.}~\bibnamefont {Passos}}, \bibinfo {author} {\bibfnamefont
  {D.}~\bibnamefont {Cournapeau}}, \bibinfo {author} {\bibfnamefont
  {M.}~\bibnamefont {Brucher}}, \bibinfo {author} {\bibfnamefont
  {M.}~\bibnamefont {Perrot}}, \ and\ \bibinfo {author} {\bibfnamefont
  {E.}~\bibnamefont {Duchesnay}},\ }\bibfield  {title} {\bibinfo {title} {\emph
  {Scikit-learn: Machine Learning in {P}ython}},\ }\href@noop {} {\bibfield
  {journal} {\bibinfo  {journal} {Journal of Machine Learning Research}\
  }\textbf {\bibinfo {volume} {12}},\ \bibinfo {pages} {2825} (\bibinfo {year}
  {2011})}\BibitemShut {NoStop}%
\end{thebibliography}
%

\newpage
\appendix

\section{Center of Mass and Micromotion}
\label{app:com_micromotion}

\begin{figure}[h]
\begin{center}
\includegraphics[width=.9\columnwidth]{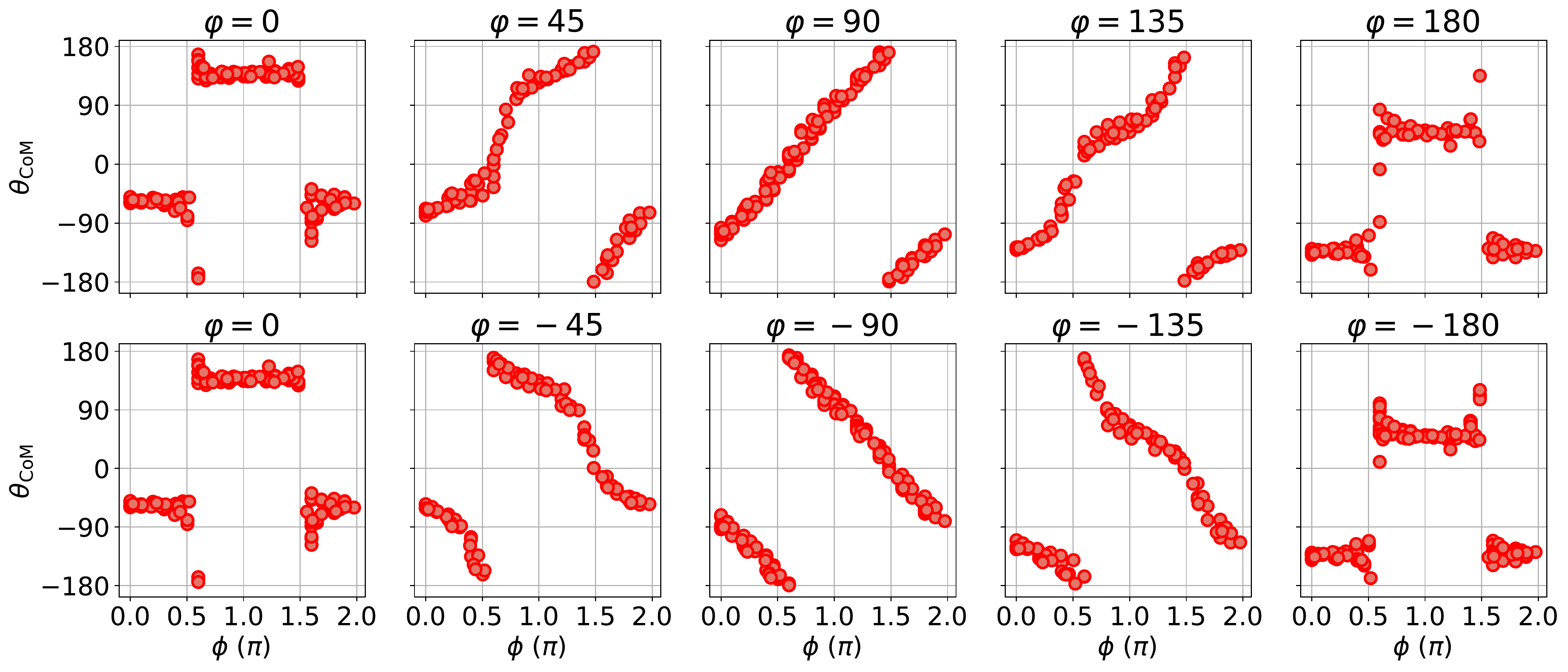}
\end{center}
\caption{Dependence of the azimuthal coordinate $\theta_{\mathrm{CoM}}$ of the center of mass and the micromotion phase $\phi$. The dependence can be explained by the elliptical shaking. For the shaking phases $\varphi=0,\pm180$ the shaking is linear, thus the cloud can only be displaced in k-space along the direction of the shaking. For a shaking phase of $\varphi=\pm90$ the shaking is circular and thus the dependence is linear. The sign of $\varphi$ decides on the direction of shaking which is encoded in the phase jump and the direction of the slope.}
\label{fig:micromotion_com}
\end{figure}

In figure \ref{fig:micromotion_com} we show the dependence of the azimuthal coordinate of the center of mass on the micromotion phase. For circular shaking, i.e. a shaking phase of $\pm 90^{\circ}$, the center of mass moves in a circular fashion yielding a linear dependence between the azimuthal coordinate of the center of mass and the micromotion phase. For linear shaking, i.e. a shaking phase of $0^{\circ}$ and $180^\circ$, the center of mass moves along a diagonal line yielding a constant azimuthal coordinate of the center of mass at $\pm 45^{\circ}$, with only an irrelevant jump by $180^{\circ}$ for certain micromotion phases. Other shaking phases interpolate between these two behaviors. In conclusion, the movement of the center of mass of the momentum distribution follows the shaking trajectories as expected. 

\section{PCA Analysis}
\label{app:pca}
\begin{figure}[h]
\begin{center}
\includegraphics[width=.9\columnwidth]{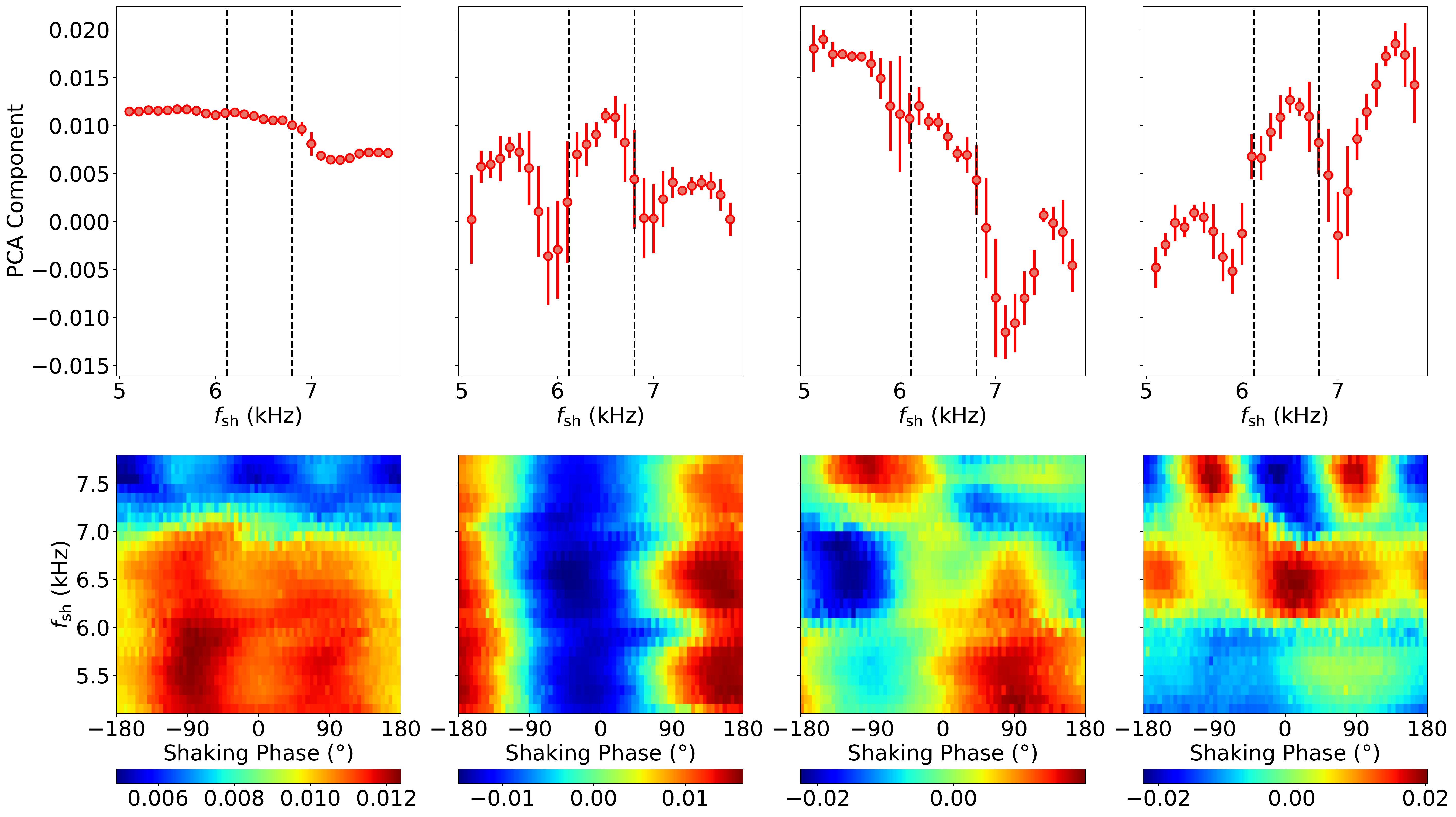}
\end{center}
\caption{PCA Analysis of the ToF images rephased to a fixed micromotion phase. From left to right: PCA component 1, 2, 6 and 8 which have been selected arbitrarily. To plot all other components see the notebooks in \cite{notebooks}. In the upper row a cut along a shaking phase of $90^\circ$ is plotted. The error is the standard deviation of the different values for the component for the given shaking frequency. The theoretical predictions for the phase transitions are given by the dashed lines. In the lower row, the averaged components are plotted in the Haldane phase diagram fashion.}
\label{fig:pca}
\end{figure}

In addition to the methods in the main text we also used principal component analysis on the processed data. In figure \ref{fig:pca} we selected four components of this analysis. The different principle components clearly show features that correspond well with the theoretical predictions such as sharp local minima at the expected phase transitions. However, the data also shows features that are not related to phase transitions such as strong dependence on the shaking phase in the trivial regions. Therefore the data does not provide a clear recipe for identifying the topological phase transitions in a completely unsupervised way. This is true in particular for the choice of the components to be analysed.

\section{Anomaly detection in phase direction.}
\label{appendix:anomaly-detection}

\begin{figure}[h]
\begin{center}
\includegraphics[width=.9\columnwidth]{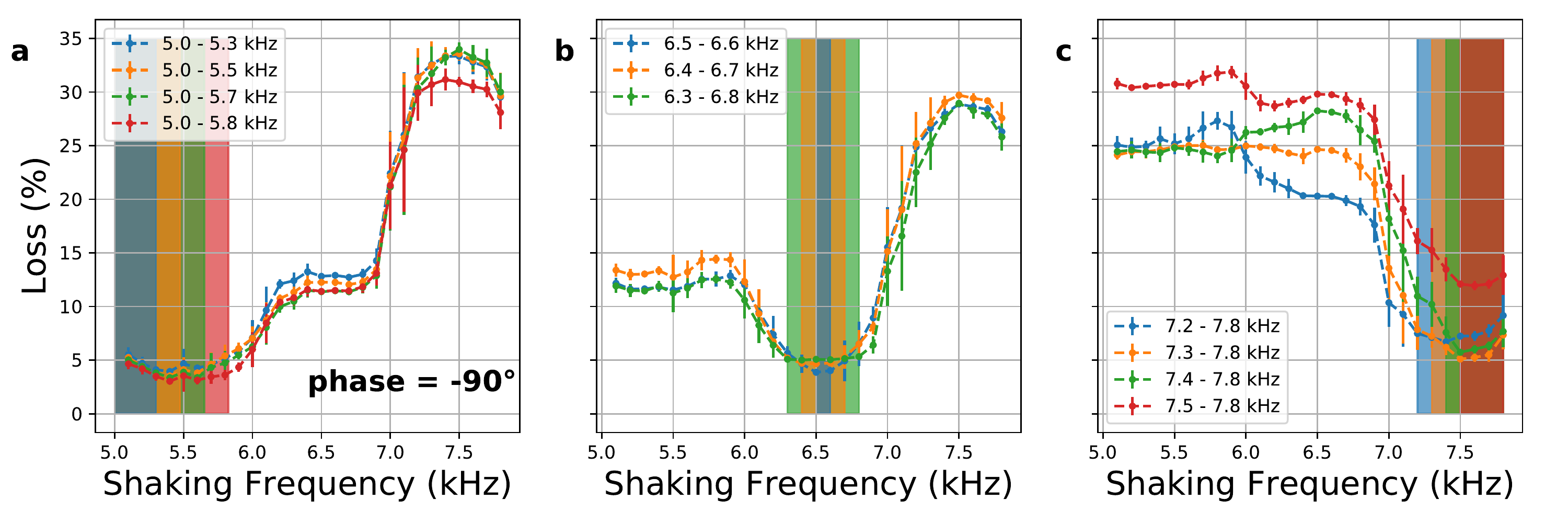}
\end{center}
\caption{Consistency checks for the obtained phase boundaries by varying the size of the training region for all three phases (\textbf{a}-\textbf{c}). We see that the the results are consistent and changing the training region size does not shift the obtained boundaries.}
\label{fig:boundary_consistency}
\end{figure}

\begin{figure}[h]
\begin{center}
\includegraphics[width=.9\columnwidth]{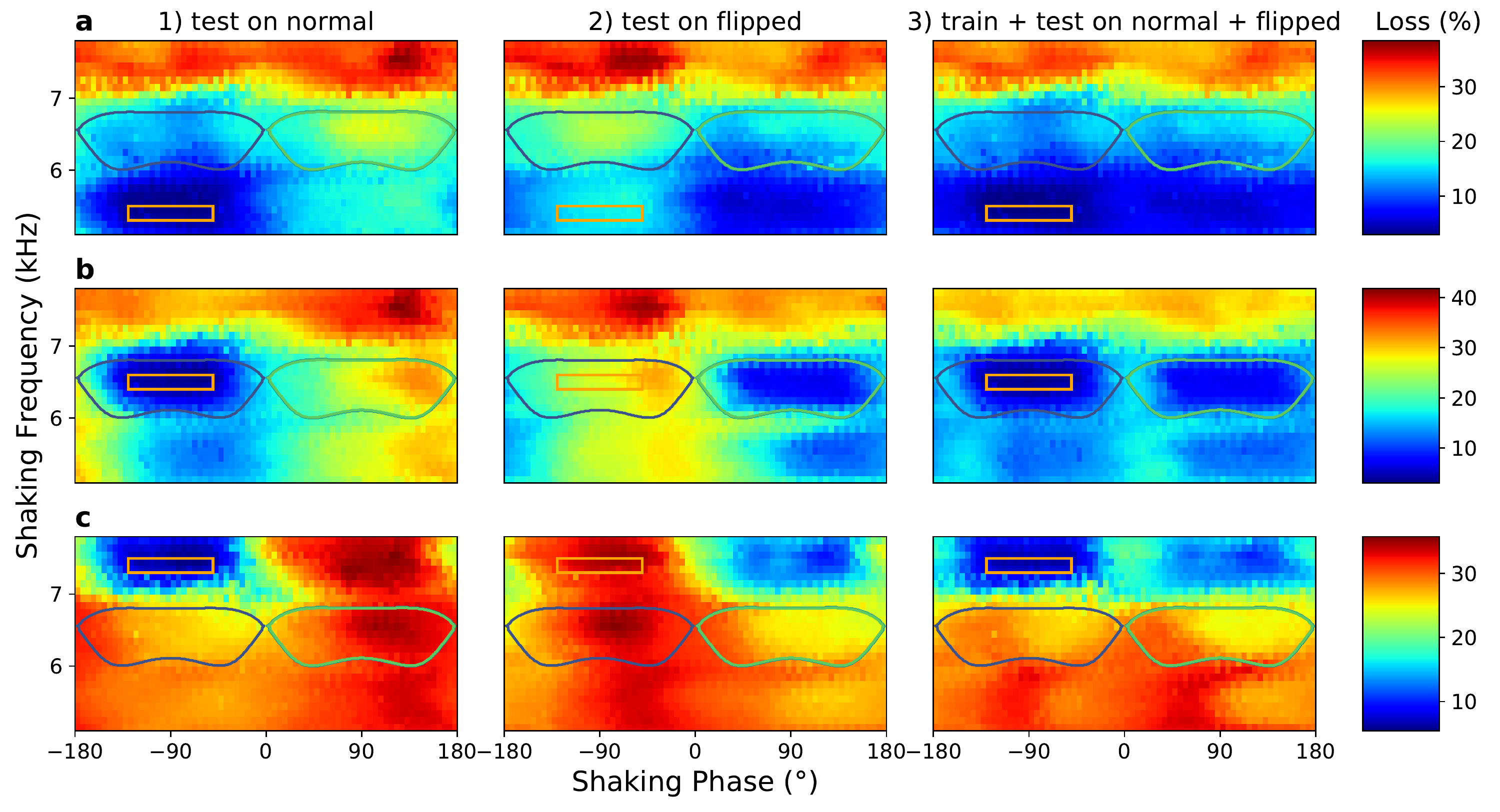}
\end{center}
\caption{1\textbf{a},\textbf{b},\textbf{c}) Training with small box and testing on the normal images. The autoencoder does not generalize well within the same phase in the shaking phase parameter. 2\textbf{a},\textbf{b},\textbf{c}) Same AE as in 1 but now tested on flipped images. The loss map is qualitatively mirrored around the phase $= 0^\circ$ axis. 3\textbf{a},\textbf{b},\textbf{c}) Now both training in testing are performed on a dataset consisting of both the normal and flipped images.}
\label{fig:flipped-business}
\end{figure}

We perform a consistency check to confirm the method is well behaved in the frequency parameter. For this, we train with different sizes of the training region in the respective three phases. As seen in \cref{fig:boundary_consistency}, the obtained results are consistent and we are confident about the three obtained boundaries presented in the main text. In panel c) there are discrepancies inside the blue-detuned trivial phase where we perform the training. However, the onset of the plateau for the topologically non-trivial phase is consistent for all four training region sizes.
We find that the method does not generalize well when performing the same analysis in the phase parameter. We show in \cref{fig:flipped-business} 1\textbf{a}-\textbf{c} how for smaller boxes in the phase-parameter, the network has problems reproducing images from the same phase but with different and unseen shaking phase parameters. This is the reason we are not able to differentiate between the non-trivial topological phases with Chern number $+1$ and $-1$. We use the same network and test it on flipped images in \cref{fig:flipped-business} 2\textbf{a}-\textbf{c}. Flipping here corresponds to the phase-space transformation $(k_x,k_y) \rightarrow (-k_x,k_y)$. As expected, this operation physically corresponds to changing the shaking phase $\varphi$ to $-\varphi$ and should invert the sign of the Chern number. In 2\textbf{a}-\textbf{c} we train and evaluate the network on a dataset were we use both normal and flipped images. In 3\textbf{a} and \textbf{b} we see that now we obtain the expected behaviour, i.e. generalizing in the whole red-detuned trivial phase and separating the chern number $+1$ and $-1$ hases. However, in 3\textbf{c} the network still does not generalize well in the trivial blue-detuned phase. We therefore conclude that with this architecture we are not able to separate the two topologically non-trivial phases from each other.

\end{document}